\documentclass[12pt, oneside]{book}
\usepackage[top=1in, bottom=1in, left=1in, right=1in, textwidth=6.5in]{geometry}
\usepackage{microtype}
\usepackage{setspace}
\usepackage{amsmath}
\usepackage{tikz}
\usepackage{tikz-feynman}
\tikzfeynmanset{compat=1.1.0}
\usepackage{graphicx}
\usetikzlibrary{calc,positioning}
\usepackage{caption}
\usepackage{ragged2e}
\usepackage{subcaption}
\DeclareMathOperator{\sgn}{sgn}
\usepackage{dsfont}
\usepackage{gensymb}
\usepackage{braket}
\usepackage{slashed}
\usepackage{multicol}
\usepackage{soul}
\usepackage{bm}
\usepackage{mathtools}
\usepackage{esvect}
\usepackage{amssymb}
\usepackage[mathcal]{eucal}
\DeclareMathOperator{\Sym}{Sym}
\usepackage[T1]{fontenc}
\usepackage[OT2,T1]{fontenc}
\DeclareSymbolFont{cyrletters}{OT2}{wncyr}{m}{n}
\DeclareMathSymbol{\Sha}{\mathalpha}{cyrletters}{"58}

\usepackage{enumitem}

\usepackage[style=numeric, sorting=none, citestyle=numeric-comp, 
            maxnames=10, minnames=10]{biblatex}
\usepackage{xurl}
\AtBeginBibliography{%
    \setlength{\emergencystretch}{3em}%
}
\title{\textbf{Macroscopic Classical and Quantum Models of\\
Inverse Compton Scattering}}
\author{Emerson Penn Rogers\\
\small PhD Dissertation, Old Dominion University, August 2026}
\date{}

\begin{document}

\frontmatter
\maketitle

\chapter*{Abstract}
\addcontentsline{toc}{chapter}{Abstract}
Inverse Compton sources --- in which a relativistic electron beam scatters a laser
pulse to produce tunable, collimated, high-energy radiation --- have emerged as among
the most promising compact radiation sources, with applications ranging from nuclear
photonics to medical and nanoscale imaging and metrology. The most viable current
tabletop configuration couples laser-wakefield acceleration with inverse Compton
scattering, producing GeV-scale electron beams over millimeter distances. As laser
intensities increase and electron energies grow, the interaction enters the radiation
reaction regime, where the energy radiated by the electron becomes a significant
fraction of its kinetic energy. Predicting the scattered electron energy spectrum ---
the primary observable of these experiments --- from first principles has remained an
unsolved problem. Existing models are either classical equations of motion in which
spectra are aggregated via large-scale multiparticle simulation, or quantum models that
rely on approximations of uncertain validity and face the same simulation burden. None
directly produce a spectral prediction. This dissertation presents a novel framework
that resolves this deficiency. Building on a unified variational treatment of classical
and quantum electrodynamics, the laser pulse is represented as a coherent quantum state
of the electromagnetic field and the electron beam as a statistical quantum state
encoding its momentum distribution. This is the natural realization of scattering a
laser pulse by an electron beam within quantum electrodynamics, as opposed to the
traditional particle-particle scattering perspective. Moreover, it yields the scattered
spectrum as a primary analytic output. For a Gaussian laser pulse, a closed-form
expression for the scattered spectrum is derived that requires no simulation, no
approximation of the laser field profile, and no large particle ensembles. The
classical limit of the framework recovers Landau-Lifshitz dynamics --- the appropriate
self-consistent description of classical radiation reaction --- exactly, establishing
the coherent-state model as its quantum electrodynamic counterpart. The framework is
validated against existing experimental data, and the fundamental structural limitations
of all current modeling approaches are identified and analyzed.

\newpage
\thispagestyle{empty}
\vspace*{\fill}
\begin{center}
\textit{To my parents.}
\end{center}
\vspace*{\fill}
\newpage

\chapter*{Acknowledgements}
\addcontentsline{toc}{chapter}{Acknowledgements}
I am grateful for the generous support of several funding sources. I gratefully
acknowledge the National Science Foundation under Grant Nos.~1847771 and 2513760,
the Corey Sargent Memorial Scholarship (2023), the JSA/Jefferson Lab Graduate
Fellowship (2025--2026), and the Old Dominion University Graduate School Dissertation
Completion Award (Summer 2026). The Department of Physics at Old Dominion University
and Jefferson Lab Scientific Computing provided the institutional and computational
resources that made this research possible.

A great debt of gratitude is owed to my advisor and mentor, Professor Balsa Terzi\'{c},
whose guidance, patience, and intellectual generosity shaped this work in ways that are
difficult to overstate. The remaining members of my dissertation committee ---
Dr.~Geoffrey Krafft, Dr.~Alexandre Deur, Dr.~Lawrence Weinstein, and Dr.~Sookyung
Joo --- provided thoughtful feedback and a genuine willingness to engage with ideas
that did not always fit neatly into existing frameworks. Thanks are also due to
Elizabeth Breen, whose collaboration and intellectual partnership throughout this work
have been invaluable, and to the broader membership of the ODU Compton and ODU GRSI
research groups for stimulating discussions and ongoing collaboration. Ted Rogers
deserves special mention for his generous engagement regarding quantum field theory.

Finally, I would like to thank my friends and family who have helped support me more
than they know. Most importantly, I thank my loving partner Megan for always having
faith in me.

\tableofcontents
\listoffigures

\chapter*{Notation and Conventions}
\addcontentsline{toc}{chapter}{Notation and Conventions}

\noindent The following conventions are maintained throughout this dissertation unless
explicitly stated otherwise.

\vspace{0.5cm}
\noindent\textbf{Units.} The elementary electric charge is $e<0$. Heaviside-Lorentz units are used throughout. Expressions in SI units may be recovered via
\begin{align*}
    A^{\text{HL}}=\sqrt{\epsilon_{0}}A^{\text{SI}},
\end{align*}
and
\begin{align*}
    e^{\text{HL}}=\frac{e^{\text{SI}}}{\sqrt{\epsilon_{0}}}
\end{align*}
for the electromagnetic field and the elementary charge, respectively.

Natural units are also assumed:
\begin{align*}
    \hbar = c = 1.
\end{align*}
However, the factors $\hbar$ and $c$ are reintroduced explicitly where clarity or dimensional analysis demands it.

\vspace{0.5cm}
\noindent\textbf{Vectors and polarizations.} Spatial three-vectors will be denoted by bold Roman letters, e.g. $\bm{k}$ or $\bm{p}$. Letters $\bm{k}$ and $\bm{l}$ are respectively used to denote incident and scattered photon wave-vectors (also referred to as momenta). Analogously, letters $\bm{p}$, $\bm{q}$, and $\bm{r}$ (when necessary) denote incident and scattered electron momentum. The frequency (also referred to as energy) of the photon with wave-vector $\bm{k}$ is denoted $\omega_{\bm{k}}$, while the energy of the electron with momentum $\bm{p}$ is denoted $E_{\bm{p}}$. Lorentz four-vectors will be denoted by non-bold Roman letters, e.g. $k\equiv(\omega_{\bm{k}},\bm{k})$ or $p\equiv(E_{\bm{p}},\bm{p})$. Polarization and spin index will be denoted by Greek letters, e.g. $\kappa$, $\pi$. The polarization-wave-vector and spin-momentum couples will be denoted by capital letters (bold or non-bold), e.g. $\bm{K}\equiv(\kappa,\bm{k})$ or $K\equiv(\kappa, k )$. Finally, we also denote ordered $n$-tuples of, say, scattered electron spin-momenta as $\bm{Q}^{(n)}\equiv(\bm{Q}_{0},\dots,\bm{Q}_{n-1})$.

Specifically for a laser four-wavevector $k_L\equiv(\omega_{\bm{k}_{L}},\bm{k}_{L})$, we adopt the notations $\hat{\bm{k}}_{L}\equiv\frac{\bm{k}_{L}}{\omega_{\bm{k}_{L}}}$ and $\hat{k}_{L}\equiv\frac{k_{L}}{\omega_{\bm{k}_{L}}}=(1,\hat{\bm{k}}_{L})$.

\vspace{0.5cm}
\noindent\textbf{Metric.} The metric signature is $(+,-,-,-)$ and the Minkowski metric is denoted $\eta$, so that the Minkowski inner product of two four-vectors $a$ and $b$ is
\begin{align*}
    (ab) \equiv \eta_{\mu\nu}a^\mu b^\nu = a^0 b^0 - \bm{a}\cdot\bm{b}.
\end{align*}

\vspace{0.5cm}
\noindent\textbf{Lightfront coordinates.} The lightfront coordinates are defined by
\begin{align*}
    x^{\pm} \equiv x^{0} \pm x^{3},
    \qquad
    \bm{x}_\perp \equiv (x^1, x^2),
\end{align*}
with inverse relations $x^0 = (x^+ + x^-)/2$ and $x^3 = (x^+ - x^-)/2$.

\vspace{0.5cm}
\noindent\textbf{Momentum space measures.} We repeatedly make use of the following notations when integrating over momentum space:
\begin{subequations}
    \begin{align*}
        \int d^{3}\bm{k}
        &\Longleftrightarrow
        \sum_{\bm{k}_{T}}\int\Delta^{2}\bm{k}_{T}\,dk_{\parallel},\\[4pt]
        \int d^{3}\lambda_{\gamma}(\bm{k})
        &\Longleftrightarrow
        \sum_{\bm{k}_{T}}\int\frac{\Delta^{2}\bm{k}_{T}\,dk_{\parallel}}
        {(2\pi)^{3}\sqrt{2\omega_{\bm{k}}}},\\[4pt]
        \int d^{3}\mu_{\gamma}(\bm{k})
        &\Longleftrightarrow
        \sum_{\bm{k}_{T}}\int\frac{\Delta^{2}\bm{k}_{T}\,dk_{\parallel}}
        {(2\pi)^{3}(2\omega_{\bm{k}})},
    \end{align*}
\end{subequations}
where $\Delta^{2}\bm{k}_{T}=\frac{(2\pi)^{2}}{L^{2}}$ is the transverse momentum spacing. In the continuum limit $L\rightarrow\infty$, the left-hand notation remains unchanged. Furthermore, we absorb the polarization sum into the momentum integral by a deliberate abuse of notation, writing $d^{3}\bm{K}$, $d^{3}\lambda_{\gamma}(\bm{K})$, and $d^{3}\mu_{\gamma}(\bm{K})$ respectively. Analogous notations are used for electrons/positrons: $d^{3}\bm{P}$, $d^{3}\lambda_{e^{\pm}}(\bm{P})$, and $d^{3}\mu_{e^{\pm}}(\bm{P})$.

\vspace{0.5cm}
\noindent\textbf{Absorption/emission operators and particle states.} We denote an absorption operator, say, for a photon with polarization index $\kappa$ and wave-vector $\bm{k}$ as $a_{\gamma}(\bm{K})\equiv a_{\kappa}(\bm{k})$. The corresponding emission operator is the Hermitian conjugate, so that the single-photon state may be written 
\begin{align*}
    \ket{\bm{K}}=\sqrt{2\omega_{\bm{k}}}a_{\gamma}(\bm{K})^{\dagger}\ket{0},
\end{align*}
and analogously for electrons/positrons. We thus adopt the convention that kets have contravariant spin/polarization indices.

\vspace{0.5cm}
\noindent\textbf{Electromagnetic field.} The classical laser field is described by
the four-potential $A_{\text{cl}}^\mu$ in the Lorenz gauge $({\partial}A_{\text{cl}})
= 0$. The normalized vector potential amplitude is $\tilde{A}_{\text{cl}} \equiv
eA_{\text{cl}}/m$. For a plane-wave laser pulse, primes denote differentiation with respect to the lightfront
phase $(\hat{k}_L x)$.

\mainmatter

\chapter{Background and History of Inverse Compton Scattering and Radiation Reaction}

The development of Inverse Compton Sources (ICS) represents a paradigm shift in our ability to generate high-energy, collimated radiation within a laboratory environment. By facilitating the collision of relativistic electron beams with intense laser pulses, these compact sources provide a tunable and monochromatic alternative to traditional large-scale accelerator facilities, with cutting-edge designs now approaching tabletop dimensions. However, as laser intensities push toward strong-field limits, the interaction enters a regime where the standard classical description of particle dynamics is no longer sufficient. At these field-strengths, the radiation reaction---an effect arising when the exchange of energy-momentum between the electron and the electromagnetic field becomes non-negligible, resulting in a net depletion of electron energy-momentum---transitions from a minor perturbation to a major dynamical correction. As field strengths escalate further, profound quantum effects emerge, culminating in a fully non-perturbative regime of electron-photon scattering.

The ultimate objective of this research is the development of a comprehensive quantum model of inverse Compton scattering, one extendable to all intensity regimes and applicable to realistic experimental configurations. This dissertation establishes the foundational theoretical framework required for such a model. To date, the inability to accurately predict scattered spectra precludes the ability to tune and commercialize tabletop inverse Compton sources. For this reason, the primary motivation of this research is a fundamental framework from which the full scattered spectra can be derived as a single expression. We have done this encompassing both classical and quantum radiation reaction effects in the limiting case of a plane-wave laser pulse.

The remainder of this introductory chapter surveys the rapidly expanding applications of inverse Compton sources, followed by an examination of the historical evolution of the theory of the radiation reaction in parallel with major developments in electrodynamics.

Chapters 2 and 3 systematically develop the classical and quantum theories of electrodynamics from a modern, rigorous perspective, establishing the mathematical formalism and notation employed throughout the manuscript.

In Chapter 4, we introduce various models of Inverse Compton scattering. Following an overview of these alternative radiation reaction models including a novel spectral pushforward, we critically evaluate their theoretical shortcomings and compare their predictions with existing experimental data. This analysis serves as the primary physical motivation for the formulation of the subsequent fully quantum approach.

The focal point of this dissertation is detailed in Chapter 5, wherein we construct the fully quantum description of the radiation reaction designed to be scaled across all field regimes and experimental setups. We further demonstrate that, in the classical limit, this model successfully recovers the Landau-Lifshitz model, and we benchmark its accuracy by comparison with the spectral pushforward model fitted to the experimental spectral data.

We conclude in Chapter 6 by discussing the roadmap for extending our quantum model to even higher intensity regimes and more complex, realistic experimental architectures. There, we also explore the broader implications of this model for the commercialization of ICS technology and its crucial role in driving forthcoming scientific discovery.

\section{Applications of Inverse Compton Sources}

The transition of quality x-ray sources from massive, billion-dollar synchrotron installations to compact, tabletop-scale architectures has democratized access to high-fidelity radiation, unlocking a multitude of advanced applications. The defining advantage of ICS technology lies in its capacity to generate quasi-monoenergetic, highly collimated, and strictly energy-tunable X-ray and gamma-ray beams. By actively modulating the relativistic electron energy and the laser collision geometry, the spectral density and bandwidth of the emitted radiation can be precisely tailored to the specific atomic or nuclear resonances of a target material. This unprecedented tunability, coupled with the immense spatial coherence and ultrashort pulse duration of the beams, positions modern ICS as a transformative diagnostic and analytical engine across industrial, nuclear, and fundamental sciences.

\subsection{Preclinical Micro-CT, Medical Diagnostics, and Therapeutics}

The transition to tunable, monochromatic radiation heralds a major advancement not only in clinical oncology but also in high-resolution preclinical imaging. In the realm of biomedical research, the integration of inverse Compton sources with micro-computed tomography (micro-CT) has established a new standard for volumetric, non-destructive anatomical analysis. Conventional micro-CT systems are fundamentally constrained by the flux and focal spot size of standard bremsstrahlung X-ray tubes, which introduce beam-hardening artifacts and limit soft-tissue contrast. By leveraging the micron-scale source size and quasi-monoenergetic spectrum of an ICS, researchers can achieve unprecedented spatial resolution and phase-contrast sensitivity. This allows for true "virtual histology"---the three-dimensional, cellular-level visualization of complex soft-tissue architectures, such as pulmonary alveoli and vascular networks, without the need for destructive physical sectioning~\cite{Gunther:2020}.

\begin{figure}[h!]
    \centering
    \includegraphics[width=0.8\linewidth]{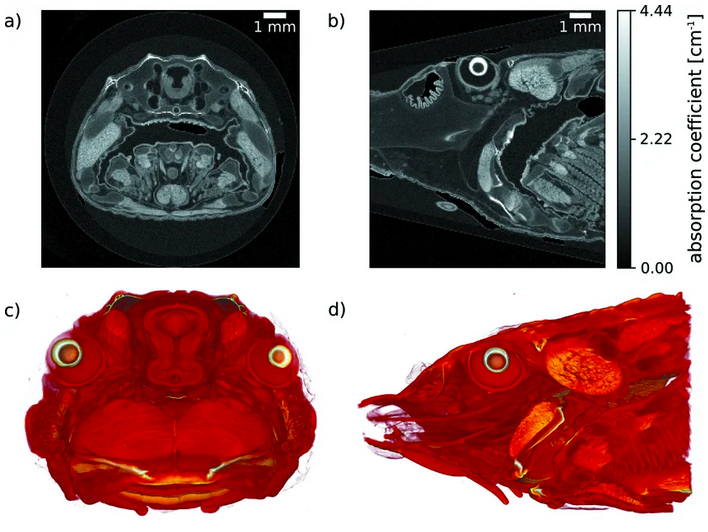}
    \caption[High-resolution X-ray micro-CT of a teleost fish head]{High-resolution X-ray micro-computed tomography (micro-CT) of front and side profiles of a teleost fish head, acquired utilizing a compact inverse Compton source. The quasi-monoenergetic and highly coherent nature of the ICS beam enables advanced phase-contrast imaging, providing exceptional soft-tissue differentiation and micron-scale resolution of the delicate cranial and branchial osteology. The tunable, narrow-bandwidth radiation mitigates the beam-hardening artifacts that fundamentally limit conventional bremsstrahlung X-ray tubes in complex biological imaging~\cite{Gunther:2020}.}
    \label{fig:microct}
\end{figure}

Scaling these capabilities to clinical radiology, the ICS beam mitigates the severe dose penalties associated with broad-spectrum sources, as no low-energy, non-image-forming photons are absorbed by the patient. Clinicians can fully exploit phase-contrast X-ray imaging (PCI) to acquire vastly superior soft-tissue resolution compared to traditional absorption radiography. Simultaneously, K-edge subtraction (KES) angiography leverages the strict energy tunability of the ICS to maximize the visibility of specific contrast agents, such as iodine or gadolinium, at substantially reduced patient doses~\cite{Paterno:2020}. Furthermore, in the therapeutic domain, ICS technology enables targeted Auger electron therapy. By tuning the incident beam to precisely match the K-edge of high-Z nanoclusters embedded exclusively within tumor cells, the ICS triggers the emission of highly localized, destructive Auger electrons, maximizing tumor cell death while completely sparing surrounding healthy tissue~\cite{Carroll:2002}.

\subsection{Structural Biology and Protein Crystallography}
In the realm of structural biology, unraveling the complex, three-dimensional architecture of macromolecular proteins is essential for rational drug design and understanding fundamental cellular functions. Historically, researchers have been required to compete for beamtime at massive, third-generation synchrotron facilities for empirical determination of these folded structures via X-ray crystallography. Compact inverse Compton sources disrupt this bottleneck by delivering synchrotron-quality, high-flux, monochromatic hard X-rays directly to the university or pharmaceutical laboratory. Because protein crystals are notoriously small and susceptible to radiation damage, the highly collimated and tunable nature of an ICS beam is critical. It allows researchers to seamlessly match the X-ray energy to the absorption edges of specific heavy atoms within the crystal—a requirement for Multi-wavelength Anomalous Dispersion (MAD) phasing techniques. This capability facilitates the rapid, high-resolution empirical reconstruction of folded protein structures, accelerating the pipeline for structure-based pharmaceutical development without the logistical constraints of a national laboratory~\cite{Graves:2009,Abendroth:2010}.

\begin{figure}[h!]
    \centering
    \begin{subfigure}{\textwidth}
        \centering
        \includegraphics[width=.35\linewidth]{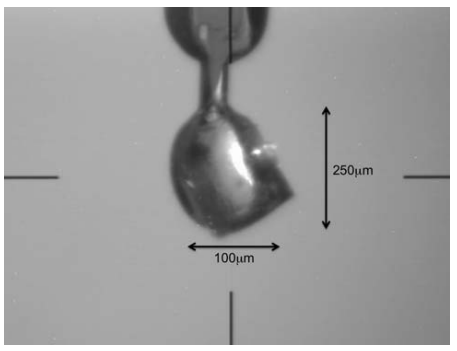}
        \hfill
        \includegraphics[width=.6\linewidth]{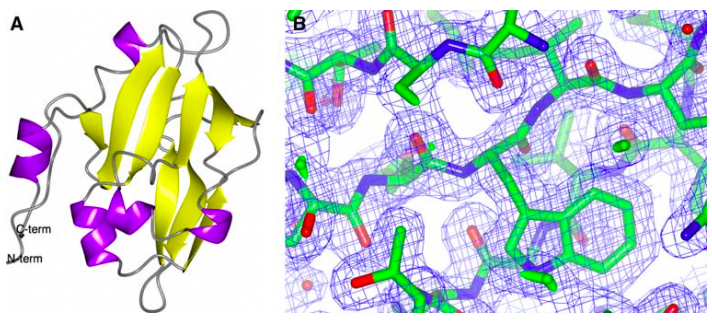}
    \end{subfigure}
    \caption[X-ray diffraction data from \textit{Mycobacterium tuberculosis}]{X-ray diffraction data from the Compact Light Source developed by Lyncean Technologies for crystals of the Glycine cleavage system protein H from \textit{Mycobacterium tuberculosis} was used to solve its protein structure \cite{Abendroth:2010}. Sample (left) and derived 3D protein structure (right).}
    \label{fig:proteinvisuals}
\end{figure}

There is an important distinction to make here: while computational models like AlphaFold famously predict protein folding, the empirical 3D structure of a folded protein must be inferred using X-ray crystallography. This is demonstrated in Fig.~\ref{fig:proteinvisuals}.

\subsection{Industrial and Materials Science}
In the domains of advanced manufacturing and materials science, the demand for high-resolution, non-destructive testing of dense components is essential. Traditional bremsstrahlung X-ray tubes inherently produce a broad, continuous energy spectrum, which introduces severe beam-hardening artifacts and fundamentally limits both penetration depth and image contrast when interrogating high-Z materials. Conversely, the tunable, high-energy gamma rays generated by an ICS cleanly penetrate massive, dense geometries—such as aerospace alloys, turbine blades, and complex additive manufacturing builds. When it is coupled with micro-focus computed tomography (CT) techniques, the micron-scale source size of an all-optical ICS enables the three-dimensional reconstruction of hyperfine internal structures. This allows for the precise detection of microscopic voids, stress fractures, and material inconsistencies without compromising the structural integrity of the component~\cite{Ma:2020}.

\subsection{Solid-State Physics and Advanced Microelectronics}
In the realm of condensed matter physics, the ultrashort pulse duration and high spectral brilliance of inverse Compton sources offer an unprecedented window into the ultrafast dynamics of quantum materials. For semiconductor metrology, the capability to perform time-resolved X-ray diffraction (TR-XRD) on a tabletop scale is invaluable. It enables the precise characterization of sub-nanometer lattice defects, transient strain dynamics, and carrier lifetimes, which are critical parameters for the development of next-generation microprocessors and extreme ultraviolet (EUV) lithography components. Furthermore, these tunable X-ray pulses are uniquely suited for probing the complex electronic architectures of superconductors. By employing femtosecond pump-probe methodologies, researchers utilize ICS beams to interrogate electron-phonon coupling, Cooper pair dynamics, and transient phase transitions in strongly correlated systems like cuprates and pnictides. This time-resolved structural analysis provides the empirical foundation necessary to unravel the enduring mysteries of high-$T_{c}$ superconductivity and to engineer novel quantum devices~\cite{Graves:2014}.

\subsection{Nuclear Security and Isotope Identification}
Moving beyond classical, attenuation-based radiography, the ultra-narrow energy bandwidth of inverse Compton sources facilitates highly advanced isotopic assay via Nuclear Resonance Fluorescence (NRF). NRF is a process wherein an incident photon is resonantly absorbed by a target nucleus, which subsequently de-excites by emitting characteristic gamma radiation. Because these resonant excitation levels are uniquely defined for every individual isotope, an ICS beam tuned to a specific resonant energy acts as an impeccable, non-destructive isotopic probe. This capability is revolutionary for nuclear security and safeguards; it allows for the precise quantification of spent nuclear fuel, the rapid detection of heavily shielded special nuclear materials (SNMs) like $^{235}\text{U}$ and $^{239}\text{Pu}$ within commercial cargo, and the verification of non-proliferation treaties~\cite{Kikuzawa:2009}. Furthermore, contemporary research has demonstrated that NRF-based inspection driven by an ICS can selectively identify the elemental ratios of illicit chemical compounds, narcotics, and explosives concealed behind thick steel shielding, cementing the technology's critical role in modern global security infrastructure~\cite{Lan:2021}.

\subsection{Fundamental Physics and Next-Generation Accelerators}
At the frontier of fundamental physics, high-intensity laser-electron interactions are poised to serve as critical probes of the quantum vacuum. In the strong-field QED regime, the collision between an ultra-relativistic electron bunch and an ultra-intense laser pulse triggers a cascade of non-perturbative phenomena. As the electrons are driven through the extreme optical field, they emit high-energy gamma rays via non-linear inverse Compton scattering. These emitted gamma rays can subsequently interact with the macroscopic laser field to generate electron-positron cascades through the multi-photon Breit-Wheeler process, or serve as probes for elusive vacuum polarization effects such as vacuum birefringence. The critical immediacy of this research is highlighted by the establishment of dedicated strong-field QED laboratories, such as the pioneering facility at the University of Rochester. These specialized environments are explicitly engineered to drive these exact laser-electron interactions deep into the fully non-perturbative quantum regime~\cite{Mironov:2025}. Interpreting the complex dynamics generated within such facilities requires the rigorous, scalable quantum models of the radiation reaction and non-linear scattering developed within this dissertation.

Furthermore, in the context of high-energy particle physics, ICS technology is a leading candidate for the generation of highly polarized positron beams required for next-generation lepton colliders, such as the proposed International Linear Collider (ILC). By directing circularly polarized laser light onto a relativistic electron beam, the resulting polarized gamma rays can undergo pair production in a high-Z target, yielding a dense, polarized positron cascade ~\cite{Omori:2006}. This directly bypasses the limitations of traditional bremsstrahlung-driven positron sources and is essential for precision tests of the Standard Model.

\subsection{The Necessity of a Comprehensive Model}
Successful deployment and subsequent commercialization of these diverse technologies hinges upon the precise control and prediction of the emitted radiation spectrum. This modeling challenge is most acutely pronounced in the pursuit of truly tabletop architectures, notably the Laser Wakefield-driven Inverse Compton Source (LWFA-ICS). By replacing traditional, large-scale radio-frequency linacs with millimeter-scale plasma wakefields, the LWFA-ICS represents the ultimate realization of a compact, high-flux light source. However, this extreme miniaturization introduces significant experimental complexities.

In an LWFA-ICS, the interaction dynamics are governed not by idealized plane waves, but by the collision of a realistic scattering laser pulse with a highly localized, dynamically evolving electron bunch. To achieve requisite luminosities within a microscopic interaction volume, the scattering laser is deliberately allowed to propagate past its focal point to maximize the volumetric overlap with the electron bunch. Thus, the electrons are subjected to a rapidly decaying optical field throughout the interaction region. Consequently, the beam dynamics are nontrivially complicated by a confluence of macroscopic and microscopic effects: the strict spatiotemporal evolution of these transient beams, inextricably coupled with the onset of the radiation reaction and profound strong-field quantum effects. Therefore, developing a comprehensive model that accurately captures both these fundamental high-field interactions and the macroscopic beam imperfections inherent to LWFA-ICS is not merely a theoretical exercise, but an essential prerequisite for the engineering, tuning, and optimization of next-generation, all-optical inverse Compton sources.

\section{A Brief Overview of Inverse Compton Scattering}
The generation of high-energy radiation via laser-electron collisions is governed by the fundamental kinematics of photon-electron scattering. To rigorously describe the extreme dynamics within a LWFA-ICS, it is necessary to first delineate the standard regimes of these interactions, tracing the conceptual evolution from classical elastic scattering to fully non-linear, quantum emission.

\subsection{Thomson and Compton Scattering}
The classical limit of photon-electron interaction is described by Thomson scattering. In this low-energy regime, an incident electromagnetic wave accelerates a stationary or slowly moving free electron, causing it to oscillate and subsequently emit dipole radiation at the same frequency $\omega$ as the incident wave. Because the energy of the incident photon is negligibly small compared to the rest energy of the electron $\hslash\omega\ll mc^{2}$ (with $\hslash$ the reduced Planck constant, $m$ the electron mass, and $c$ the speed of light), the recoil of the electron is practically zero. Consequently, the scattering process is elastic.

When the energy of the incident photon becomes comparable to or exceeds the electron's rest energy ($\hslash\omega\gtrsim mc^{2}$), the classical wave description breaks down, and the quantum corpuscular nature of light must be invoked. In this regime, known as Compton scattering, the photon transfers a portion of its energy and momentum to the electron, resulting in a scattered photon with a strictly lower energy and a correspondingly longer wavelength. The kinematics of this collision are captured by the Compton shift, which quantifies the change in wavelength $\Delta\lambda$ as a function of the scattering angle $\theta$:
\begin{align*}
    \Delta\lambda=\frac{h}{mc}(1-\cos{\theta})
\end{align*}
where $h$ is Planck's constant and $\frac{h}{mc}$ represents the Compton wavelength of the electron. This shift serves as the foundational signature of discrete quantum momentum transfer between radiation and matter.

\subsection{Inverse Scattering and the Governing Parameters}
While standard Compton and Thomson scattering describe a high-energy photon losing energy to a stationary electron, their inverse counterparts refer to the opposite effect in which a highly relativistic electron (relativistic factor $\gamma\gg 1$) collides head-on with a low-energy photon (such as an optical laser photon). Here, kinetic energy is transferred from the motion of the electron to the scattered photon, upscattering the energy of the incident light by a factor of $4\gamma^{2}$ into the X-ray or gamma-ray regime.

In a realistic LWFA-ICS environment, there are various intermediate regimes between simple inverse Thomson scattering and non-perturbative quantum emission. These regimes are governed by three distinct dimensionless parameters\footnote{These parameters technically only make sense for idealized laser pulse descriptions. While we will ultimately be concerned with realistic laser pulses, they constitute an effective scheme for categorizing inverse scattering phenomena and experiments.} that characterize the local electromagnetic field and the electron kinematics:
\begin{itemize}
    \item The Normalized Peak Field Strength ($a_{0}$): Defined as
    \begin{align}
    a_{0}=\frac{eE}{mc\omega},
    \label{NormalizedPeakFieldStrength}
    \end{align} where $e<0$ is the elementary charge of the electron and $m$, its mass, $E$ is the amplitude of the electric field of the laser envelope with laser frequency $\omega$, and $c$ is the speed of light. This parameter measures the classical kinematic nonlinearity of the interaction, representing the ratio of the electron's quiver momentum in the laser field to its rest mass momentum. It can alternatively be defined simply as the normalized amplitude of the electromagnetic field of the laser pulse.
    \item The Classical Radiation Parameter ($R_{c}$): Defined as 
    \begin{align}
    R_{c}=\alpha a_{0}^{2}\frac{\hslash (k_{L}p)}{(mc)^{2}},
    \label{ClassicalRadiationParameter}
    \end{align}
    where $\alpha$ is the fine-structure constant, $k_{L}$ is the four-wavevector of the laser, and $p$ is the four-momentum of the electron. This parameter quantifies the onset of macroscopic energy-momentum depletion, dictating when the electron's classical trajectory is significantly altered by its own emission.
    \item The Quantum Nonlinearity Parameter ($\chi$): Expressed invariantly as 
    \begin{align}
        \chi=\frac{\sqrt{-F_{\lambda\mu}F^{\lambda\nu}p^{\mu}p_{\nu}}}{mcE_{S}},
        \label{QuantumNonlinearityParameter}
    \end{align}
    where $F_{\mu\nu}$ is the laser field-strength tensor and $E_{S}$ is the critical Schwinger field. This parameter determines quantum recoil. It dictates when the discrete, stochastic nature of photon emission fundamentally breaks classical models of continuous radiation.
\end{itemize}

Using these parameters, we can cleanly categorize the interaction dynamics into four distinct inverse scattering regimes:
\begin{enumerate}
    \item Linear inverse Thomson/Compton scattering: ($a_{0}\ll 1$, $R_{c}\ll 1$, $\chi\ll 1$)\\
    In the perturbative limit, the transverse momentum imparted to the electron by the laser field remains strictly non-relativistic. The electron oscillates harmonically, acting as a classical dipole antenna moving at relativistic (longitudinal) bulk velocities. The interaction is characterized by the linear scattering of isolated, single photons, resulting in a scattered radiation spectrum with a single, sharp spectral peak at the fundamental frequency. Here, both radiation reaction effects and quantum recoil are entirely negligible. 
    \item Nonlinear inverse Compton scattering: ($a_{0}\approx1$, $R_{c}\ll 1$, $\chi \ll 1$)\\
    As the laser intensity increases and $a_{0}$ approaches unity, the linear approximation fails. In this regime, the electron interacts coherently with the macroscopic optical field, simultaneously absorbing multiple low-energy photons before emitting a single, highly energetic scattered photon. The Lorentz force drives the electron into an anharmonic figure-eight trajectory in its average rest frame. The scattered spectrum is fundamentally altered, exhibiting distinct signatures of classical nonlinearity: the emission of higher-order harmonics and the onset of ponderomotive broadening—a red-shifting of the spectral peaks due to the intensity-dependent effective mass of the electron within the strong optical field.
    \item Classical radiation reaction: ($a_{0}\gg 1$, $R_{c}\gtrsim 0.01$, $\chi \ll 1$)\\
    As the laser intensifies and at high initial electron energies, the interaction crosses a critical threshold where the exchange of energy-momentum between the electron and the electromagnetic field becomes non-negligible. The classical radiation parameter approaches or exceeds unity, indicating a severe, continuous net loss of energy-momentum for the electron throughout the interaction. While the scattered radiation spectrum becomes highly complex, noisy, and broadened, the physical signature of this regime is most profoundly observed in the scattered electron spectrum, which exhibits significant broadening and a distinct deceleration shift. Because $\chi$ remains relatively small, this macroscopic energy-momentum depletion can still be modeled classically via continuous formulations, such as the Landau-Lifshitz equation.
    \item The Quantum Strong-Field Regime ($a_{0}\gg 1$, $R_{c}\gg 0.01$, $\chi \gtrsim 1$)
    As the interaction pushes deeper into the strong-field limit and $\chi$ approaches unity, the assumption of continuous, classical radiation emission completely breaks down. The energy of the emitted photons becomes a substantial fraction of the electron's kinetic energy, resulting in severe and discrete quantum recoil. The scattered radiation spectrum becomes highly stochastic, while the electron spectrum undergoes profound quantum straggling—a variance in energy loss driven by the probabilistic nature of photon emission. In this fully non-perturbative regime, classical electrodynamics (CED) is entirely insufficient, necessitating the rigorous quantum electrodynamic (QED) framework developed in this dissertation.
\end{enumerate}

\begin{figure}[h!]
    \centering
    \includegraphics[width=0.8\linewidth]{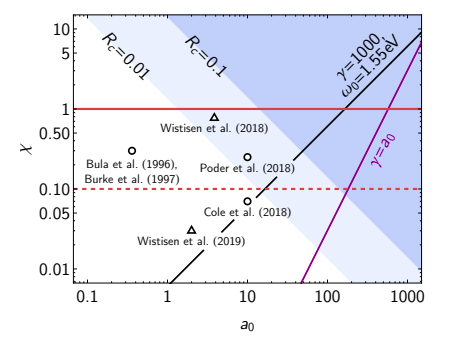}
    \caption[Characterization of radiation reaction experiments.]{The type of radiation reaction effects can be parametrized by the laser field strength (a.k.a. classical nonlinearity parameter) $a_{0}$ and the quantum nonlinearity parameter $\chi$. Classical radiation damping becomes strong when $R_{c}>0.01$ (light blue) and dominates when $R_{c} > 0.1$ (darker blue). Quantum corrections to the spectrum become necessary when $\chi > 0.1$. Electron-positron pair creation and QED cascades are important for $\chi > 1$ \cite{Blackburn:2020}. }
    \label{fig:rrexperiments}
\end{figure}

These regimes, along with inverse electron-photon scattering experiments are depicted succinctly in Fig.~\ref{fig:rrexperiments}.

\subsection{Spectral Signatures of the Radiation Reaction}

To fully appreciate the impact of the radiation reaction on beam dynamics, it is instructive to examine the prototypical scattered radiation spectra generated within the non-linear regime. By mathematically isolating the energy-momentum exchange, we can directly observe how the onset of these strong-field effects degrades the idealized emission profile.

\begin{figure}[h!]
    \centering
    \includegraphics[width=0.8\textwidth]{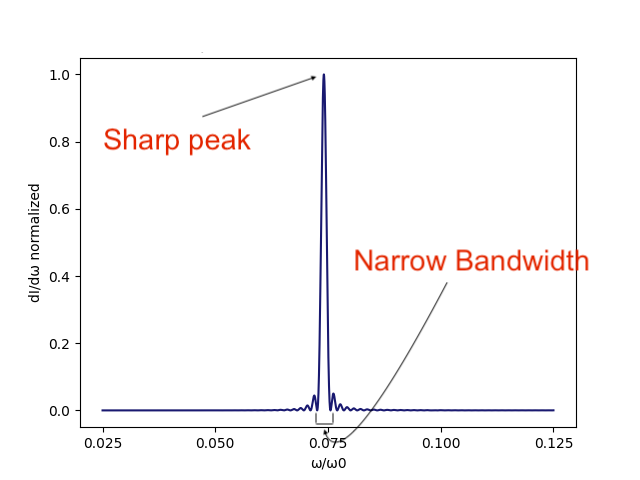}
    \caption[Scattered radiation spectrum from XSENSE]{Scattered radiation spectrum created with simulation code XSENSE~\cite{Ranjan:2018} is plotted for a rectangular envelope with $a_{0}=5.0$, wavelength $\lambda_{0}=1066\,\text{nm}$, and $\sigma=90$ optical cycles, scattering an electron with initial energy $511\,\text{MeV}$.}
    \label{fig:ruijterNORR}
\end{figure}

In an idealized non-linear interaction---where the normalized vector potential $a_{0}\approx 1$ but the radiation reaction is artificially suppressed---the scattered spectrum presents as a well-defined series of discrete harmonic peaks. Because the electron is driven into a stable, periodic figure-eight trajectory by the Lorentz force, the resulting radiation is highly coherent. The primary signature here is ponderomotive broadening: a predictable, intensity-dependent red-shift of the fundamental frequency and its harmonics caused by the increased effective mass of the electron as it traverses the intense optical field. However, because the electron's initial kinetic energy is strictly conserved in this idealized model, the spectral peaks remain sharp, as in Fig.~\ref{fig:ruijterNORR}. This strict monochromaticity is precisely the desirable quality of an inverse Compton source that enables the advanced diagnostic, therapeutic, and fundamental applications discussed previously in this chapter.

\begin{figure}[h!]
    \centering
    \includegraphics[width=0.8\textwidth]{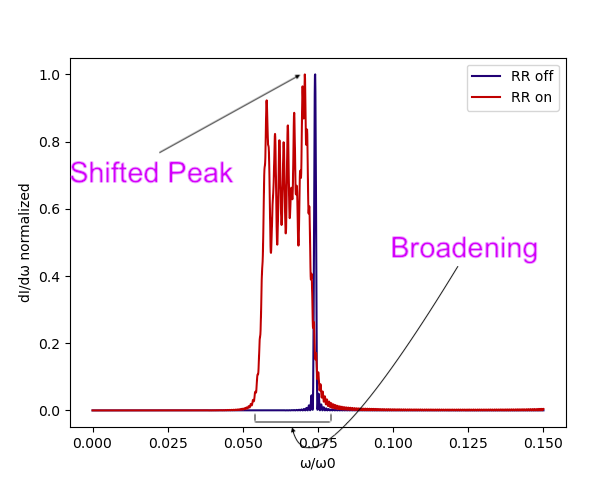}
    \caption[Scattered radiation spectrum from XSENSE including radiation reaction]{Scattered radiation spectrum created with simulation code XSENSE including classical radiation reaction is plotted for a rectangular envelope with $a_{0}=5.0$, wavelength $\lambda_{0}=1066\,\text{nm}$, and $\sigma=90$ optical cycles, scattering an electron with initial energy $511\,\text{MeV}$.}
    \label{fig:ruijterRR}
\end{figure}

Unfortunately, when we turn on the radiation reaction to accurately reflect a non-negligible net loss of energy-momentum for the electron, this ideal spectrum fundamentally degrades. The spectral qualities that make ICS technology so valuable are lost. When modeled classically (e.g., via the Landau-Lifshitz equation), the electron's kinetic energy is depleted continuously as it traverses the laser pulse. This continuous deceleration introduces further redshift, broadening, and noise to the scattered radiation spectrum, as in Fig.~\ref{fig:ruijterRR}.

As the interaction is pushed further into the strong-field regime and quantum effects emerge, the spectrum degrades even further. The discrete, probabilistic nature of photon emission introduces stochastic broadening of the electron energy distribution. Rather than experiencing smooth and continuous dampening, the electron undergoes stochastic recoil events. Consequently, the scattered radiation spectrum becomes highly noisy and erratic.

Because the scattered radiation spectrum becomes so complex in the radiation reaction regime, directly extracting predictive diagnostic information from the photons becomes exceedingly difficult. However, it is precisely at this threshold that the scattered electron spectrum begins to exhibit measurable macroscopic changes. Since the scattered electron and radiation spectra are inextricably linked, the electron distribution acts as a high-fidelity surrogate for the interaction dynamics. Therefore, while our ultimate concern remains the quality of the scattered radiation, our methodological focus shifts to its theoretical equivalent: rigorously modeling the scattered electron spectrum.

\subsubsection{The Predictive Failure of Current Models}
It is crucial to note that in reality---in a physical LWFA-ICS --- these fundamental spectral complexities are coupled with the macroscopic beam configurations. As the electrons traverse a rapidly decaying optical field, they experience dynamically shifting ranges of field-strengths. The resulting spectrum is a complex convolution of varying ponderomotive shifts, continuous classical energy depletion, and stochastic quantum recoil. Modeling and accurately predicting this convoluted spectral broadening forms the physical mandate for the remainder of this research.

Within the context of this dissertation, to "model" inverse Compton scattering refers to a grounded physical description of the underlying dynamics, while "simulating" an inverse Compton scattering event refers to a computational and numerical implementation of a specific model, designed to compute its scattered spectra. The scattered spectrum is the definitive physical fingerprint of the interaction, encoding the entire dynamic history of the electrodynamic system. Accurately predicting these scattered spectra is the strict operational prerequisite for any tunable inverse Compton source. Whether a facility is engineered for industrial non-destructive testing, clinical phase-contrast imaging, or probing the quantum vacuum, the ability to predictably tune the machine relies entirely upon theoretical models that can faithfully calculate the emitted spectrum across all intended operational regimes.

While analytical and numerical models have achieved remarkable success in predicting these scattered spectra within the linear and weakly non-linear regimes ($a_{0}\lesssim 1$), this predictive power collapses entirely as the interaction pushes into the radiation reaction and fully quantum regimes. In the strong-field regime, when the exchange of energy-momentum between the electron beam and the laser pulse becomes non-negligible, the resulting spectral smearing and stochastic quantum noise, coupled to realistic beam effects, overwhelm standard approaches.

To date, no existing theoretical model has successfully predicted the scattered spectra of a real, physical experiment operating within the radiation reaction regime or beyond. This glaring disconnect between theoretical formalism and experimental reality is a symptom of a deep-rooted historical pathology embedded within the foundations of strong-field electrodynamics. For nearly a century, standard theoretical frameworks have been tied to idealized descriptions of optical fields. While these approximations function elegantly on paper, they render the models fundamentally incapable of tolerating the transient, tightly focused, and rapidly decaying laser pulses characteristic of an actual LWFA-ICS. Therefore, accurately predicting the true spectral fingerprint of these next-generation sources requires finally abandoning this idealized paradigm and developing a rigorous theoretical framework grounded in reality.

\section{The Oldest Unsolved Problem in Electrodynamics}

\begin{figure}[htbp]
    \makebox[\textwidth][c]{
    
    \def\timelineWidth{15.5cm} 
    \def\offsetY{2.2cm}
    \begin{tikzpicture}[x=\timelineWidth, y=\offsetY]
        \pgfmathsetmacro{\startYear}{1885}
        \pgfmathsetmacro{\endYear}{2015}
        \draw[thick, -stealth] (0,0) -- (1,0);
        \foreach \tickYear in {1890, 1895, ..., 2010} {
            \pgfmathsetmacro{\tickPos}{(\tickYear - \startYear) / (\endYear - \startYear)}
            \draw (\tickPos, 3pt) -- (\tickPos, -3pt);
        }
        \foreach \tickYear in {1890, 1900, ..., 2010} {
            \pgfmathsetmacro{\tickPos}{(\tickYear - \startYear) / (\endYear - \startYear)}
            \draw[thick] (\tickPos, 5pt) -- (\tickPos, -5pt);
            \node[fill=white, inner sep=2pt, below=6pt] at (\tickPos, 0) {\small \tickYear};
        }
        \newcommand{\physicistPlot}[4]{
            \pgfmathsetmacro{\xPos}{(#1 - \startYear) / (\endYear - \startYear)}
            \pgfmathsetmacro{\yPos}{#4 * 2 - 1}
            \pgfmathsetmacro{\labelY}{\yPos + 0.68 * (2 * #4 - 1)}
            \draw[thick, gray] (\xPos, 0) -- (\xPos, \yPos);
            \node[
                draw=black!60,
                thick,
                rectangle,
                inner sep=0pt,
                fill=white
            ] at (\xPos,\yPos) {\includegraphics[height=2.2cm]{#2}};
            \node[font=\scriptsize] at (\xPos, \labelY) {#3};
        }
        \physicistPlot{1892}{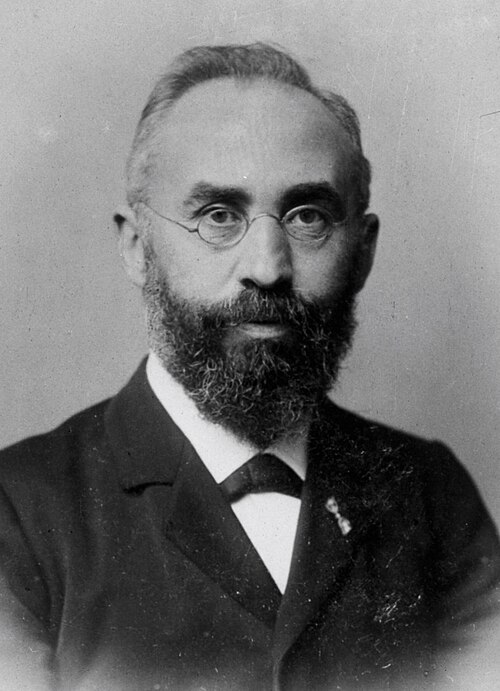}{Lorentz}{1}
        \physicistPlot{1903}{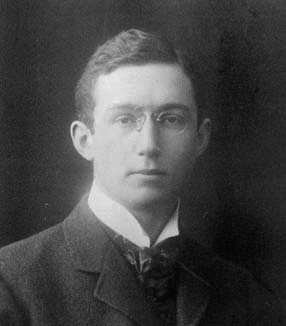}{Abraham}{0}
        \physicistPlot{1912}{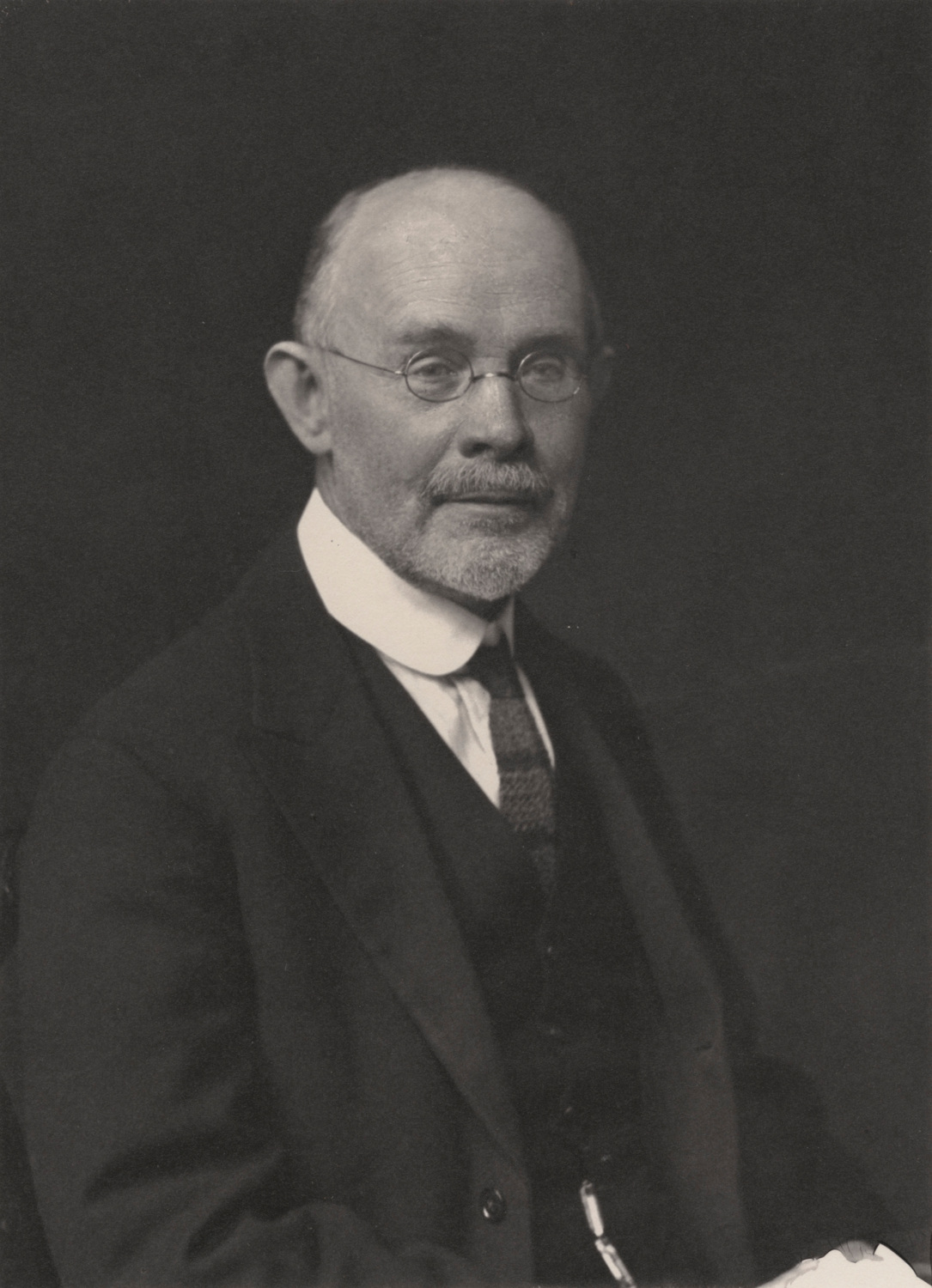}{Schott}{1}
        \physicistPlot{1935}{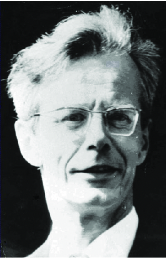}{Volkov}{0}
        \physicistPlot{1938}{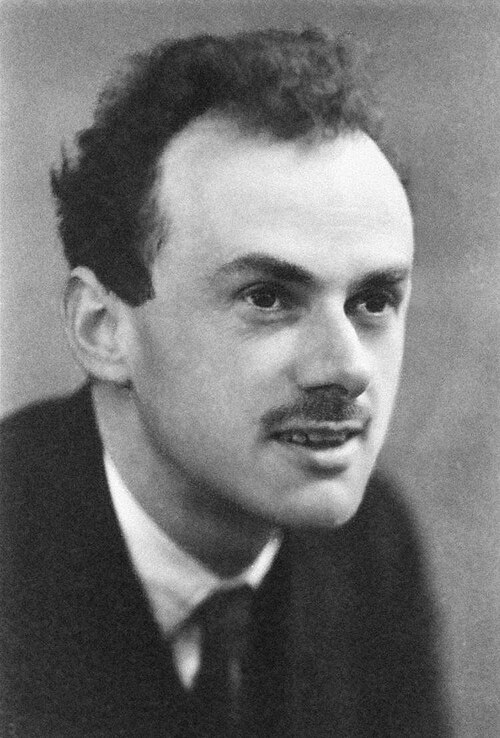}{Dirac}{1}
        \physicistPlot{1951}{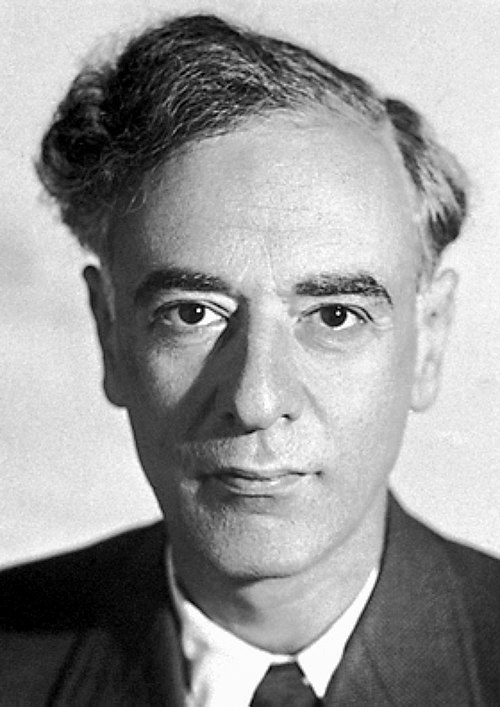}{Landau}{0}
        \physicistPlot{1951}{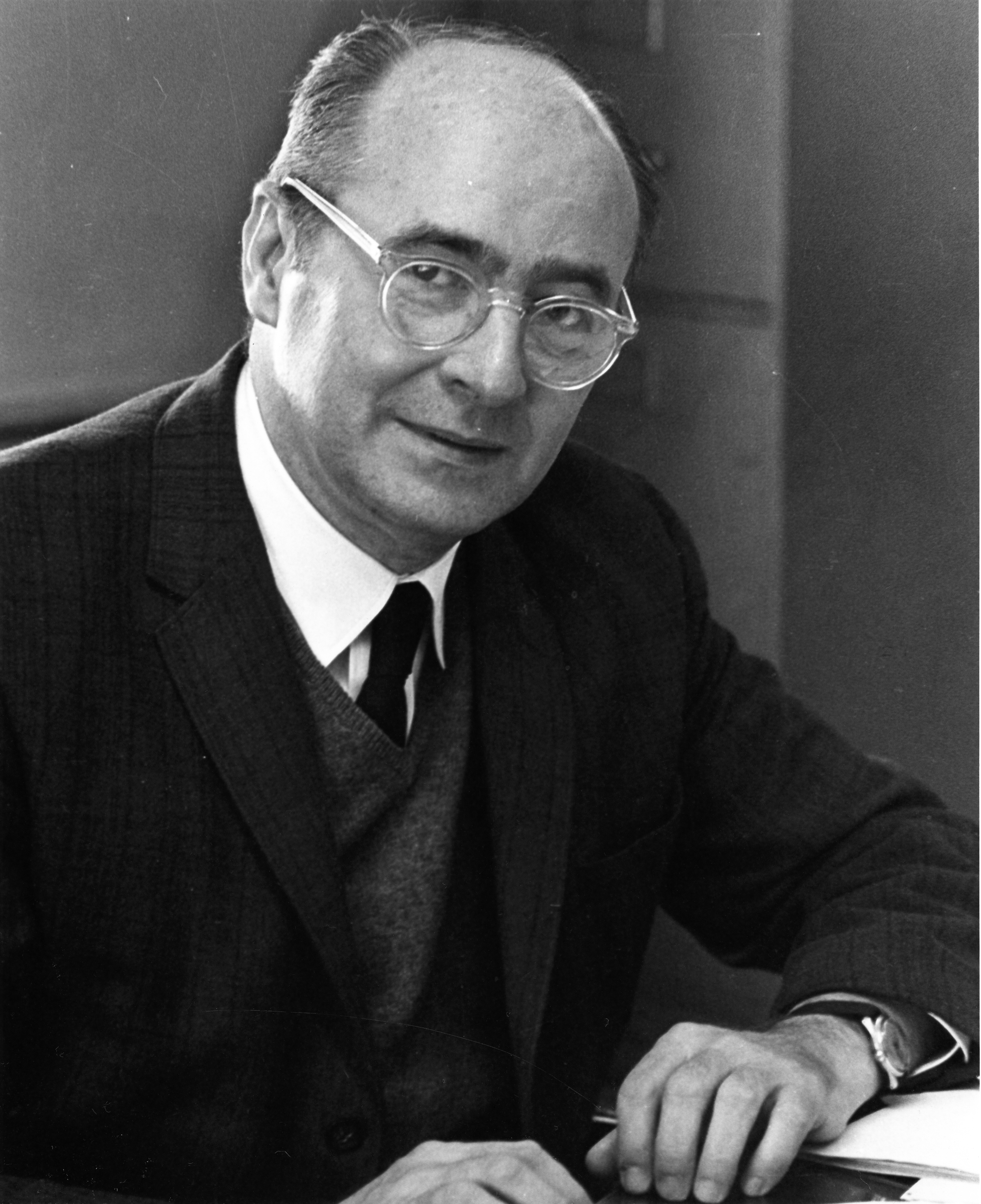}{Furry}{1}
        \physicistPlot{1975}{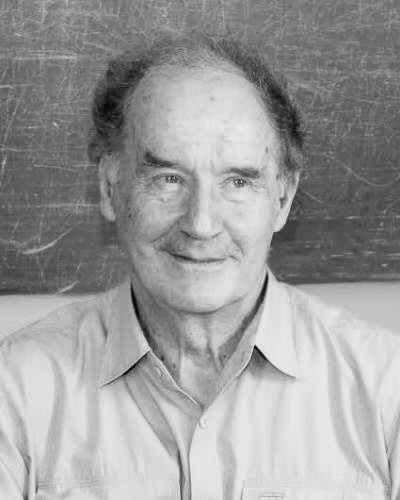}{Ritus}{0}
        \physicistPlot{2009}{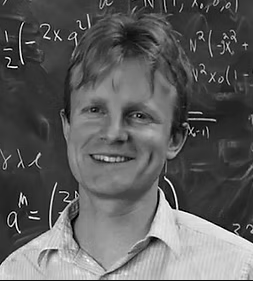}{Gralla}{1}
    \end{tikzpicture}
    } 
    \caption[Chronology of theory of radiation reaction]{Chronology of the theory of radiation reaction and the pivotal physicists involved in our understanding of radiation reaction. (Left to right) Lorentz, Abraham, Schott, Volkov, Dirac, Landau (bottom), Furry (top), Ritus, Gralla.}
    
    \label{fig:timeline}
\end{figure}

At the turn of the twentieth century, CED stood as a triumph of theoretical physics. Yet, this pristine framework harbored two fundamental pathologies that would silently dictate the trajectory of the field for the next hundred years (see Fig.~\ref{fig:timeline}). The first was a deep theoretical crisis regarding the fundamental nature of the charged particle itself. The second was an applied pathology wherein physicists systematically traded physical realism for apparent mathematical and conceptual simplicity when describing the electromagnetic environments in which those particles interacted. The history of the radiation reaction is the history of these two intertwined pathologies --- one famously debated, the other implicitly accepted --- until both ultimately collided at the experimental frontier of the modern inverse Compton source.

The theoretical crisis was precipitated by J. J. Thomson’s discovery of the electron in 1897 and reinforced by the demands of Lorentz invariance in development of relativistic electrodynamics that immediately followed~\cite{Thomson:1897}. If the electron was treated as a point charge, then its self-energy, and consequently its electromagnetic mass, would diverge to infinity. To circumvent this foundational issue, Hendrik Lorentz in 1892, and later Max Abraham in 1903, formulated the first classical models of the radiation reaction by abandoning the point-particle paradigm. By modeling the electron as a rigid, extended sphere of continuous charge, they calculated the back-reaction of the emitted field upon the particle, yielding the Abraham-Lorentz force~\cite{Lorentz:1892,Abraham:1903}.

However, this rigid model was intrinsically flawed. It was incompatible with the emerging tenets of special relativity, and in 1912, G. A. Schott exposed its incompatibility with the contemporary view of electromagnetism. By identifying a non-positive-definite energy term in the derivation of the electron self-force, Schott demonstrated the impossibility of decomposing the total electromagnetic field into its constituent parts: the electron's Coulomb field and its radiated field. This seemed to contradict the assumed law of superposition~\cite{Schott:1912}.

While the theoretical crisis of the self-force was now glaringly obvious, it was almost immediately eclipsed. In 1900, Max Planck’s quantum hypothesis caused the focus of theoretical physicists to shift. As figures like Wolfgang Pauli helped propel the quantum revolution throughout the 1920s, theoretical physics became focused on the immediate, observable mysteries of atomic spectra and wavefunction mechanics. Because the energetic scales of early quantum interactions were low enough that the radiation reaction remained negligible, the mysterious breakdown of the classical self-force was relegated to an afterthought --- there it stayed until the modern high-energy, strong-field regimes were penetrated where it could no longer be ignored.

It was during this time that the second, applied pathology took root in QED. In 1935, D. M. Volkov derived the exact solutions for the Dirac equation of an electron propagating through a fixed electromagnetic background. Volkov's demand for an analytic solution to this differential equation severely constrained the form of the field.\footnote{It should be noted that the same problem is present in CED. If one demands an analytic solution to the classical equation of motion for a charged particle traversing an electromagnetic field, then the field is similarly constrained.} So, he was forced to model the background field as a plane-wave~\cite{Volkov:1935}. This background field method was hauled over from quantum mechanics, where the absence of field operators necessitates background field solutions.

Three years later, the theoretical pathology resurfaced. In his seminal 1938 paper, Paul Dirac systematically returned the electron to a dimensionless point charge. By rigorously enforcing relativistic covariance and utilizing mass renormalization to absorb the infinite self-energy, Dirac derived the Abraham-Lorentz-Dirac (LAD) equation~\cite{Dirac:1938}. Dirac's relativistic formulation exposed the theory's deepest physical contradictions, famously admitting acausal pre-acceleration and runaway solutions wherein an electron exponentially self-accelerates to the speed of light without an external driving force.

By 1951, Lev Landau and Evgeny Lifshitz had introduced an order-reduction of the LAD equation to eliminate the theoretical pathology, cementing the resulting Landau-Lifshitz equation as a more reasonable description of the radiation reaction~\cite{Landau:1951}. Meanwhile, as the theoretical pathology was on the verge of being resolved, the applied one was becoming pedagogy. W. H. Furry canonized Volkov's background states by creating a background field interaction picture, later coined the Furry picture~\cite{Furry:1951}.

In the latter half of the twentieth century, theorists such as V. I. Ritus leveraged this Volkov-Furry paradigm to found the modern theory of strong-field QED~\cite{Ritus:1975,Ritus:1985}. Ritus’s monumental work established the standard methodologies for calculating quantum processes in intense fields, but it remained structurally bound to the idealized plane wave description of radiation fields. It was only at this point that it became clear that the background field picture in QED was not as conceptually or mathematically simple as it initially seemed.

In the 1990s, the rigidity of this applied pathology was formally mapped by the extensive mathematical treatment of V. G. Bagrov. In his exhaustive classification of the exact solutions to the Dirac equation in a background field, Bagrov applied Hamilton-Jacobi theory to demonstrate that the Dirac equation is solvable via separation of variables precisely when the analogous classical Lorentz force law in that same field is also separable.
This is true provided that the background field has at least three continuous symmetries such as a plane-wave, Coulomb field, or constant crossed-field~\cite{Bagrov:1990}. Thus, if one demands analytic solutions, then one may only work with idealized-background fields, while if a realistic description is necessary, one is forced to solve for Volkov states numerically. This may not be expensive on its own, however, calculating the expectation values of observables entails evaluating highly oscillatory integrals involving these Volkov states. Combined, this approach is computationally impractical.

It was not until 2009 that the first pathology was finally resolved~\cite{Gralla:2009}. S. E. Gralla, A. I. Harte, and R. M. Wald definitively proved that the Landau-Lifshitz equation is the correct solution to Maxwell's equations coupled to the conservation of current and conservation of total stress-energy of the laser pulse-electron system, i.e. the correct solution in the classical interaction picture under a mathematically precise point particle limit~\cite{Gralla:2009}.

In recent years, the true bottleneck impeding the advancement of next-generation inverse Compton sources has come to the foreground. Standard strong-field QED models, built entirely upon the plane-wave approximations of Volkov, Furry, and Ritus, are fundamentally incompatible with realistic descriptions of laser pulses. Volkov states simply cannot accommodate them. 

This background field methodology has placed significant structural limitations on strong-field QED. The background field picture was hauled over to QED from quantum mechanics. By accepting that macroscopic classical fields must be relegated to an unquantized background, the standard pedagogy implicitly teaches that QED should be reserved strictly for high-energy particle-particle interactions. This has cultivated a generation of physicists who believe that QED is only necessary over standard quantum mechanics when relativity is introduced.

This is a fundamental misunderstanding of the theory's architecture. QED is the beneficiary of many technical upgrades. It is not solely Lorentz covariance that makes QED inherently more powerful than quantum mechanics; another crucial distinction is the existence of dynamical field operators. Standard quantum mechanics contains no field operators. So, it is permanently bound to the background field methodology, capable only of describing interactions within fixed, idealized ambient fields that are systematically replenished exactly as they are depleted, such as the textbook example of the hydrogen-like Coulomb field. QED, on the other hand, possesses these dynamical field operators and is thus capable of describing particle-field and field-field interactions exactly, without relegating unquantized fields to the background with an effective action. The Volkov-Furry paradigm severely limits the expressibility of these interactions, imposing the same constraints that are manifest in quantum mechanics.

Therefore, the ultimate mandate of modern electrodynamics is to abandon this restrictive paradigm. While the immediate focus of this dissertation and the research beyond is the accurate spectral modeling of charged particles traversing intense laser pulses, the coherent state framework developed in the subsequent chapters dissolves this century-old mathematical limitation. By treating the macroscopic field not as a fixed classical background, but as a dynamic quantum coherent state, we construct a framework that can be applied to any atom, any molecule, and any charged matter interacting with an arbitrary electromagnetic field.

This approach is not merely a theoretical novelty. It is analytically and computationally more feasible than the background field approach. It is conceptually and mathematically simpler. It extends naturally into the nonperturbative regime, bypassing the infinite plane-wave approximations of Volkov and returning the full predictive power of the quantized electromagnetic field to the macroscopic world. Utilizing coherent states of a classical electromagnetic field allows the reconciliation of the kinematics of the radiating electron with the reality of its electromagnetic environment. It is therefore a resolution to the oldest problem in electrodynamics.
\chapter{Particles and Fields in Classical Electrodynamics}

Classical electrodynamics admits two conceptually distinct formulations, depending on how charged matter is described and incorporated into the theory. In the first, matter appears
as particles or continuous charge distributions interacting with the electromagnetic field
--- the Maxwell--Lorentzian framework that constitutes what is conventionally called classical
electrodynamics (CED) in the literature. In the second, matter is described by a Dirac spinor field --- a continuous field carrying the spinor degrees of freedom and whose excitations, upon quantization, are interpreted as charged particles. Although this description remains entirely classical, it is the natural precursor to quantum electrodynamics (QED): the passage to QED is effected by promoting the Dirac field to an operator on Fock space, with the classical field theory providing the action, the symmetries, and the canonical structure that the quantum theory inherits. For this reason, we refer to the Dirac spinor-field formulation
as \textit{pre-quantum electrodynamics} (pre-QED).

The presentation in this chapter departs from the standard literature, in that it treats both
formulations within a unified discourse and emphasizes the conceptual parallelism between
classical and quantum field theory rather than developing either in isolation.

\section{Action, Lagrangians and Dynamical Equations of Motion}

Consider a closed, isolated electrodynamic system --- a physical system consisting of electromagnetic fields and charged matter interacting within a region of spacetime, in a fixed gravitational field.\footnote{It is more natural to include dynamically curved spacetime than to artificially place it in the background. However, since the focus of this dissertation is electrodynamics, we leave spacetime in the background from the beginning~\cite{Gralla:2009}.} The appropriate starting point for any modern physical theory is the action functional $S=\int d^{4}x\,\sqrt{-g}\,\mathcal{L}$, defined in terms of the Lagrangian density $\mathcal{L}$ and the determinant of the metric tensor $\sqrt{-g}$. We assume that the action of a given isolated electrodynamic system may be decomposed into electromagnetic and charged matter terms as
\begin{align}
    S[A,g,\Gamma]&=S_{\text{EM}}[A,g]+S_{\text{M}}[A,g,\Gamma],
    \label{GeneralAction}
\end{align}
where $A$ is the electromagnetic field of the entire system, $g$ is the non-dynamic metric
tensor, and $\Gamma$ describes the charged matter content, not necessarily a field. The
matter term $S_{\text{M}}$ includes the interaction between the electromagnetic field and
charged matter.

The electromagnetic term in Eq.~\eqref{GeneralAction} in Heaviside-Lorentz units is
\begin{align}
    S_{\text{EM}}[A,g]
    =\int d^{4}x\,\sqrt{-g}\,\mathcal{L}_{\text{EM}}[A]
    \equiv -\frac{1}{4}\int d^{4}x\,\sqrt{-g}\,F^{\mu\nu}F_{\mu\nu},
    \label{ElectromagneticAction}
\end{align}
where $F_{\mu\nu}\equiv \nabla_{\mu}A_{\nu}-\nabla_{\nu}A_{\mu}$ is the electromagnetic
field strength tensor. The matter term, on the other hand, is left unspecified for now:
it could describe particles, continua, or fields. The primary constraint is that the matter
sector be gauge- and diffeomorphism-invariant, so the natural form for the interaction
between the electromagnetic field and charged matter is
$\int d^{4}x\,\sqrt{-g}\,J^{\mu}A_{\mu}$, where $J$ is the electromagnetic current.

The decomposition in Eq.~\eqref{GeneralAction} reflects a fundamental asymmetry of
electrodynamics: isolated systems containing electromagnetic fields but devoid of charged
matter are physically valid, whereas it is impossible to separate charged matter from its
ambient electromagnetic field.\footnote{It is nevertheless very
useful to describe charged matter as if it were free from its ambient electromagnetic
field, particularly when that field has a negligible effect on the dynamics of the system
under consideration.}

Classicality is enforced by demanding the action be stationary:
\begin{align}
    \delta S[F,g,\Gamma]=0.
    \label{GeneralClassicalEOM}
\end{align}
All of CED and pre-QED follows from
Eq.~\eqref{GeneralClassicalEOM}. These two theories differ in the way charged matter is
described within them. In the former, we admit any description of charged matter,
macroscopic or microscopic, so the primary objects are generically the stress-energy tensor
$T$ and the electromagnetic current $J$. In the latter, charged matter is described by the
Dirac spinor field $\Psi$, whose localized excitations are interpreted as charged particles upon quantization, while the stress-energy-momentum and charge density emerge as derived
quantities.\footnote{It should be noted that the Dirac spinor field is not a measurable
quantity. This means that, while pre-QED is a perfectly valid classical
field theory, it is not a direct description of physical reality. It is, however, a crucial
object in QED as an operator, despite remaining unobservable.}

The remainder of this section elaborates on the dynamical similarities and differences that the two theories derive from this common variational origin.

\subsection{The Equations of Motion of Classical Electrodynamics}

Recall that the action Eq.~\eqref{GeneralAction} depends on three objects: the
electromagnetic field $A$, the metric tensor $g$, and the matter model $\Gamma$. Of these,
only $A$ is dynamic; $g$ is fixed in the background and $\Gamma$ is left unspecified. The
natural variational objects are therefore $A$ and $g$, with respect to which we define
the electromagnetic current
\begin{align}
    J^{\mu}\equiv-\frac{1}{\sqrt{-g}}\frac{\delta S_{\text{M}}}{\delta A_{\mu}},
    \label{ChargeCurrentDensity}
\end{align}
and the matter stress-energy tensor
\begin{align}
    T_{\text{M}}^{\mu\nu}\equiv\frac{2}{\sqrt{-g}}\frac{\delta S_{\text{M}}}{\delta g_{\mu\nu}}.
    \label{MatterStressEnergyTensor}
\end{align}
However, the action is solely stationary under variation of $A$. The Euler-Lagrange equation of motion for CED --- Maxwell's first equation is then
\begin{align}
    \frac{\delta S}{\delta A_{\nu}}=\nabla_{\mu}F^{\mu\nu}-J^{\nu}=0.
    \label{MaxwellsEquation1}
\end{align}
Maxwell's second equation is not an equation of motion but a geometric identity --- the second Bianchi identity for the field-strength tensor:
\begin{align}
    \nabla_{\lambda}F_{\mu\nu}+\nabla_{\mu}F_{\nu\lambda}+\nabla_{\nu}F_{\lambda\mu}=0.
    \label{MaxwellsEquation2}
\end{align}

Gauge and diffeomorphism invariance impose, respectively, conservation of electromagnetic
current,
\begin{align}
    \nabla_{\mu}J^{\mu}=0,
    \label{CurrentConservation}
\end{align}
and conservation of the total Hilbert stress-energy,
\begin{align}
    \nabla_{\mu}T^{\mu\nu}=0,
    \label{StressEnergyConservation}
\end{align}
where $T^{\mu\nu}\equiv T_{\text{EM}}^{\mu\nu}+T_{\text{M}}^{\mu\nu}$ and
\begin{align}
    T_{\text{EM}}^{\mu\nu}
    \equiv\frac{2}{\sqrt{-g}}\frac{\delta S_{\text{EM}}}{\delta g_{\mu\nu}}
    =-F^{\mu\lambda}F^{\nu}_{\;\lambda}+\frac{1}{4}g^{\mu\nu}F_{\kappa\lambda}F^{\kappa\lambda},
    \label{ElectromagneticStressEnergy}
\end{align}
so that $\nabla_{\mu}T_{\text{EM}}^{\mu\nu}=J_{\mu}F^{\mu\nu}$.

Together Eqs~\eqref{MaxwellsEquation1}--\eqref{StressEnergyConservation} describe the interaction picture of CED~\cite{Gralla:2009}. It should be noted that they are coupled and, crucially, nonlinear in $A$. This nonlinearity has an important structural consequence: superposition of electromagnetic fields is not a general property of electrodynamics. More fundamentally, the electromagnetic field of an isolated system cannot in general be decomposed into the fields of its subsystems --- indeed, the notion of the field ``of a subsystem'' does not make sense without further qualification, since isolating subsystems
irreversibly discards electromagnetic information encoded in the interactions. In practice, however, we are often interested in the dynamics of charged matter alone, and it is possible to proceed in a controlled manner in which the discarded information has negligible influence on those dynamics. We will primarily be concerned with specific regimes in which this
approximation is reasonable, e.g. \emph{free theories}, in which we view an electromagnetic field as if it were independent from its source or charged matter as if it were independent from its ambient electromagnetic field.

The question of when a model of charged matter may legitimately be idealized as a massive charged point-particle requires care, and the treatment differs between CED and pre-QED, though the physical interpretation is the same in both: a point-particle is an isolated subsystem whose internal degrees of freedom are entirely inaccessible to the external system via interaction and
within the confines of the theory. We now make this precise and derive the corresponding dynamical equations.

\subsubsection{The Classical Dynamics of an Electron in an Electromagnetic Field}

One might expect that Eqs.~\eqref{MaxwellsEquation1}--\eqref{StressEnergyConservation} admit solutions sourced by a massive charged point-particle, i.e.\ by delta-function distributions. Upon closer inspection, however, this limit is singular on the worldline of the particle~\cite{Gralla:2009}: a massive charged point-particle responding to its own
electromagnetic field is unphysical. Maxwell's equations alone do admit solutions for point-particle sources, and the conventional treatment of the Lorentz force in an external field provides a compelling description of point-particle dynamics, even though this rests on the physically inaccurate description of a charged particle free from its own field. It is therefore essential to understand precisely how point-particle dynamics emerge from the full coupled system Eqs.~\eqref{MaxwellsEquation1}--\eqref{StressEnergyConservation}.

To this end, we define a massive charged point-particle as \emph{charged matter contracted to zero size, below which no radiation can resolve any internal structure}. Realizing such a limiting process requires that the matter model $\Gamma$ be continuously deformable to a point while remaining relativistically valid at each stage of the deformation. The complete procedure is carried out in Ref.~\cite{Gralla:2009}. Once the limit is taken, self-interaction is mitigated and what remains is an effective description of a massive charged particle --- such as an electron --- traversing an external electromagnetic field $A_{\text{cl}}$ in a background spacetime that can be assumed to be flat Minkowski space: $g^{\mu\nu}=\eta^{\mu\nu}$.

The resulting hierarchy of equations of motion, organized by order in the coupling $e<0$, is as follows. To order $e^{0}$, interaction vanishes and the motion is that of a free particle. To order $e^{1}$, one recovers the Lorentz force law in the external field:
\begin{align}
    \frac{dv^{\mu}}{ds}=\tilde{F}_{\text{cl}}^{\mu\nu}v_{\nu},
    \label{LorentzForce}
\end{align}
where $v$ is the four-velocity of the electron normalized so that $v^{2}=1$,
$\tilde{A}_{\text{cl}}=\frac{e}{mc}A_{\text{cl}}$ is the dimensionless external field, and $\tilde{F}_{\text{cl}}^{\mu\nu}\equiv\partial^{\mu}\tilde{A}_{\text{cl}}^{\nu}-\partial^{\nu}\tilde{A}_{\text{cl}}^{\mu}$ is the corresponding field-strength tensor with
dimensions of reciprocal length. At order $e^{2}$, radiation reaction enters and the equation of motion becomes the Lorentz--Abraham--Dirac force law:
\begin{align}
    \frac{dv^{\mu}}{ds}
    =\tilde{F}_{\text{cl}}^{\mu\nu}v_{\nu}
    +\frac{2}{3}r_{e}
    \left(\frac{d^{2}v^{\mu}}{ds^{2}}
    +\frac{dv^{\nu}}{ds}\frac{dv_{\nu}}{ds}v^{\mu}\right),
    \label{LADForce}
\end{align}
where $r_{e}=\frac{e^{2}}{4\pi mc^{2}}$ is the classical electron radius. For self-consistency, Eq.~\eqref{LorentzForce} should be substituted into the second term on the right, reducing the equation to second-order. The result is the Landau-Lifshitz force law:
\begin{align}
    \frac{dv^{\mu}}{ds}
    =\tilde{F}_{\text{cl}}^{\mu\nu}v_{\nu}
    +\frac{2}{3}r_{e}
    \left(
    \partial^{\lambda}\tilde{F}_{\text{cl}}^{\mu\nu}v_{\lambda}v_{\nu}
    +\tilde{F}_{\text{cl}}^{\mu\nu}\tilde{F}_{\text{cl},\nu\lambda}v^{\lambda}
    -\tilde{F}_{\text{cl}}^{\nu\kappa}\tilde{F}_{\text{cl},\kappa\lambda}
    v^{\mu}v_{\nu}v^{\lambda}
    \right).
    \label{LLForce}
\end{align}
Equations~\eqref{LorentzForce} and~\eqref{LLForce} both admit analytic solutions when the external field possesses three continuous symmetries --- equivalently, when it may be written as a function of a single variable~\cite{Breev:2016,Bagrov:1990,Bagrov:2014}. In the remainder of this work, the primary instance of this will be a plane-wave description of a laser pulse.

\subsection{The Equations of Motion of Pre-quantum Electrodynamics}

Before specifying the matter sector of pre-QED, it is instructive to contrast the role of point-particles in the two classical theories. In CED, point-particles emerge from a careful limiting procedure applied to extended charged matter, as described in the preceding subsection. In QED, by contrast, the point-particle picture has a different status: a
massive charged particle is an effective description of an electron relative to the energy scale of the probing photon, which can only resolve the electron's internal structure up to some ultraviolet cutoff. This is the Wilsonian perspective~\cite{Wilson:1982,Georgi:1993,Weinberg:1997}. From a fundamental standpoint, electrons and photons in QED are excitations of the Dirac and electromagnetic quantum fields, respectively, and the introduction of point-particles as primitives --- rather than as emergent descriptions --- is precisely what gives rise to the ultraviolet divergences that necessitate regularization and renormalization. Pre-QED, being entirely classical, sidesteps these issues; the relevant divergences will, however, reappear upon quantization, and it is useful to keep them in view.

In the pre-quantum context, the fundamental ontology is the fields themselves. We introduce a specific model of charged matter: the Dirac field $\Psi$, which is a dynamical four-component complex spinor field whose spinor degrees of freedom encode the spin-$\frac{1}{2}$ structure
of the electron. Fixing a flat gravitational background from the outset,\footnote{We preemptively fix a flat gravitational background to avoid the nuances associated with the pre-quantum theory in curved spacetime.} the matter sector of the action Eq.~\eqref{GeneralAction} is~\cite{Jauch:1976,Peskin:1995,Mandl:2010}
\begin{align}
    S_{\text{M}}[A,\eta,\Psi]
    \equiv\int d^{4}x\,\bar{\Psi}^{\beta}
    \left(i\left(\slashed{\partial}+ie\slashed{A}\right)^{\alpha}_{\beta}
    -m\delta^\alpha_\beta\right)\Psi_{\alpha},
    \label{PreQuantumMatterSector}
\end{align}
where we have introduced Feynman slash notation $\slashed{q}=\gamma^{\mu}q_{\mu}$ and the Dirac conjugate $\bar{U}=U^{\dagger}\gamma^{0}$ for column spinors $U$ and $\bar{V}=\gamma^{0}V^{\dagger}$ for row spinors $V$.\footnote{Hermitian conjugation exchanges column and row spinors and vectors, and therefore simultaneously raises and lowers both Lorentz and spinor indices.} The $\gamma^{\mu}$, for $\mu=0,\ldots,3$, are the Dirac
matrices --- $4\times 4$ complex matrices satisfying the Clifford algebra relation
$\{\gamma^\mu,\gamma^\nu\}=2\eta^{\mu\nu}$ --- given explicitly by
\begin{align}
\gamma^{0}&=\begin{pmatrix}
    1 & 0 & 0 & 0\\
    0 & 1 & 0 & 0\\
    0 & 0 & -1 & 0\\
    0 & 0 & 0 & -1
\end{pmatrix},
&
\gamma^{1}&=\begin{pmatrix}
    0 & 0 & 0 & 1\\
    0 & 0 & 1 & 0\\
    0 & -1 & 0 & 0\\
    -1 & 0 & 0 & 0
\end{pmatrix},
&
\gamma^{2}&=\begin{pmatrix}
    0 & 0 & 0 & -i\\
    0 & 0 & i & 0\\
    0 & i & 0 & 0\\
    -i & 0 & 0 & 0
\end{pmatrix},
\nonumber\\
\gamma^{3}&=\begin{pmatrix}
    0 & 0 & 1 & 0\\
    0 & 0 & 0 & -1\\
    -1 & 0 & 0 & 0\\
    0 & 1 & 0 & 0
\end{pmatrix}.
\end{align}

The matter sector of the action separates naturally into a free Dirac action and an interaction, $S_{\text{M}}=S_{\text{D}}+S_{\text{I}}$, where
\begin{align}
    S_{\text{D}}\equiv\int d^{4}x\,\bar{\Psi}^{\beta}
    \left(i\slashed{\partial}^{\alpha}_{\beta}-m\delta^{\alpha}_{\beta}\right)\Psi_{\alpha},
    \label{DiracAction}
\end{align}
describes free Dirac field propagation, and
\begin{align}
    S_{\text{I}}[A,\eta,\Psi]
    \equiv -e\int d^{4}x\,\bar{\Psi}^{\beta}\slashed{A}^{\alpha}_{\beta}\Psi_{\alpha},
    \label{Interaction}
\end{align}
encodes the minimal coupling between the Dirac field and the electromagnetic field. The separation $S_{\text{M}}=S_{\text{D}}+S_{\text{I}}$ treats the Dirac field as propagating independently of its ambient electromagnetic field --- the classical analogue of the bare electron in QED. Upon quantization, this idealization is responsible for the appearance of
infrared divergences, which must be handled by appropriately dressing the electron in its ambient electromagnetic field.

Since the total pre-QED action depends on two dynamical fields, $A$ and $\Psi$, the electromagnetic current and total stress-energy tensor are fully determined by the fields. Varying with respect to $A$ at $g=\eta$ gives
\begin{align}
    J^{\mu}
    \equiv -\frac{1}{\sqrt{-g}}\left.\frac{\delta S}{\delta A_{\mu}}\right|_{g=\eta}
    =e\bar{\Psi}\gamma^{\mu}\Psi,
    \label{ElectromagneticCurrentQED}
\end{align}
where spinor indices are suppressed and products are understood as matrix multiplication. Varying with respect to $g_{\mu\nu}$ at $g=\eta$ gives the total stress-energy tensor,
\begin{align}
    T^{\mu\nu}
    \equiv\left.\frac{2}{\sqrt{-g}}\frac{\delta S}{\delta g_{\mu\nu}}\right|_{g=\eta}
    =-F^{\mu\lambda}F^{\nu}_{\;\lambda}
    +\frac{1}{4}\eta^{\mu\nu}F^{\kappa\lambda}F_{\kappa\lambda}
    +\frac{i}{2}\bar{\Psi}\gamma^{(\mu}\overleftrightarrow{D}^{\nu)}\Psi,
    \label{StressEnergyQED}
\end{align}
where $D^{\mu}=\partial^{\mu}+ieA^{\mu}$ is the gauge covariant derivative,
$\overleftrightarrow{D}^{\mu}=\overrightarrow{D}^{\mu}-\overleftarrow{D}^{\mu}$ denotes the antisymmetric action of $D^\mu$ forwards and backwards, and parentheses on Lorentz indices denote symmetrization: $X^{(\mu}Y^{\nu)}=\frac{1}{2}(X^{\mu}Y^{\nu}+X^{\nu}Y^{\mu})$.

The Euler-Lagrange equation of motion for the electromagnetic field is Maxwell's first
equation in flat spacetime,
\begin{align}
    \left.\frac{\delta S}{\delta A_{\nu}}\right|_{g=\eta}
    =\partial_{\mu}F^{\mu\nu}-J^{\nu}=0,
    \label{MaxwellsEquation1QED}
\end{align}
with Maxwell's second equation again the Bianchi identity,
\begin{align}
    \partial_{\lambda}F_{\mu\nu}
    +\partial_{\mu}F_{\nu\lambda}
    +\partial_{\nu}F_{\lambda\mu}=0.
    \label{MaxwellsEquation2QED}
\end{align}

The Euler-Lagrange equation of motion for the Dirac field is the Dirac equation,
\begin{align}
    \left.\frac{\delta S}{\delta\bar{\Psi}^{\beta}}\right|_{g=\eta}
    =\left(i\slashed{\partial}^{\alpha}_{\beta}-m\delta^{\alpha}_{\beta}
    -e\slashed{A}^{\alpha}_{\beta}\right)\Psi_{\alpha}=0,
    \label{DiracEquation}
\end{align}
and an analogous equation holds for the conjugate field $\bar{\Psi}$.

As in CED, the electromagnetic current is conserved:
\begin{align}
    \nabla_{\mu}J^{\mu}=0,
    \label{CurrentConservationQED}
\end{align}
and so is the total Hilbert stress-energy,
\begin{align}
    \nabla_{\mu}T^{\mu\nu}=0.
    \label{StressEnergyConservationQED}
\end{align}
These relations impose additional constraints on the electromagnetic and Dirac fields.

Together, Eqs.~\eqref{MaxwellsEquation1QED}--\eqref{StressEnergyConservationQED} are the pre-QED counterparts of Eqs.~\eqref{MaxwellsEquation1}--\eqref{StressEnergyConservation}: the same Maxwell equations now sourced by a Dirac current, coupled self-consistently to the spinor field whose dynamics they govern.

\section{Fourier Theory}

The preceding section developed the dynamical equations governing closed, interacting electrodynamic systems. For the remainder of this dissertation, we will be especially concerned with \emph{free} systems --- those in which the electromagnetic field, or the Dirac field in pre-QED, has been isolated from its source. Free electromagnetic fields serve a dual purpose: in CED they provide the natural description of freely propagating laser pulses, and in pre-QED they furnish the mode expansions that are promoted to operator expansions upon quantization. Both roles are essential for the coherent-state QED framework developed in the next chapter.

Since the Feynman gauge is the natural setting for the plane-wave mode expansions that underpin this work, we work exclusively in the Feynman gauge for the remainder of this dissertation --- a mathematically convenient choice within the Lorenz gauge family. This is enforced in the following subsection by gauge-fixing the electromagnetic action Eq.~\eqref{ElectromagneticAction}.

\subsection{Free Electromagnetic Fields}

In a free electromagnetic system the total action reduces to Eq.~\eqref{ElectromagneticAction}. Gauge-fixing in the Feynman gauge adds a term that leaves the physical content unchanged, yielding the action in Minkowski spacetime
\begin{align}
    S_{\text{EM}}[A,\eta]
    =-\int d^{4}x\,\left(\frac{1}{4}F^{\mu\nu}F_{\mu\nu}
    +\frac{1}{2}\left(\partial_{\mu}A^{\mu}\right)^{2}\right).
    \label{FreeElectromagneticActionFeynmanGauge}
\end{align}
The Euler-Lagrange equation of motion is the gauge-fixed Maxwell equation in vacuum,
\begin{align}
    \frac{\delta S_{\text{EM}}}{\delta A^{\mu}}=\partial^{2}A_{\mu}=0.
    \label{MaxwellEquationFeynmanGauge}
\end{align}

We assume the field is contained within a finite transverse area $L^{2}$ perpendicular to the scattering axis,\footnote{This constraint ensures that the electromagnetic fields considered throughout this work are square-integrable. Our prototypical laser pulse is described as a plane-wave, which is square-integrable only in the transverse coordinates under this confinement.} so that the field satisfies periodic spatial boundary conditions of period $L$ in those coordinates. The solutions to Eq.~\eqref{MaxwellEquationFeynmanGauge} take the form
\begin{align*}
    \frac{1}{(2\pi)L^{2}}\,\varepsilon^{\kappa}_{\mu}(\bm{k})\,e^{-ikx},
\end{align*}
where $k=(k^{0},\bm{k})$ is on-shell, i.e.\ light-like: $k^{2}=k_{0}^{2}-\bm{k}^{2}=0$, or $\omega_{\bm{k}}\equiv k^{0}=|\bm{k}|$. The periodic boundary conditions discretize the transverse momenta: $k_{i}=\frac{2\pi}{L}n_{i}$ for $i=1,2$. The index $\kappa$ labels the polarization, and $\varepsilon^{\kappa}_{\mu}(\bm{k})$ is the corresponding polarization
four-vector.\footnote{For clarity, we do not mix spatiotemporal Lorentz and spin/polarization indices.} The polarization four-vectors form a tetrad spanning the tangent space $\mathbb{R}^{1+3}$ at every point in Minkowski spacetime and satisfy the orthogonality relations
\begin{align}
    \sum_{\mu}\varepsilon^{*}{}^{\kappa'}_{\mu}(\bm{k})\varepsilon^{\mu}_{\kappa}(\bm{k})
    =\eta^{\kappa'}_{\kappa},
    \label{LorentzOrthogonality}
\end{align}
and
\begin{align}
    \sum_{\kappa}\varepsilon^{*}{}^{\mu'}_{\kappa}(\bm{k})\varepsilon^{\kappa}_{\mu}(\bm{k})
    =\eta^{\mu'}_{\mu}.
    \label{PolarizationOrthogonality}
\end{align}
For linear polarizations, the polarization four-vectors may be chosen as
\begin{subequations}
\begin{align}
    \varepsilon^{\mu}_{\kappa}(\bm{k})=\begin{cases}
    (1,\bm{0}), & \kappa=0\\
    (0,\bm{\varepsilon}_{1}(\bm{k})), & \kappa=1\\
    (0,\bm{\varepsilon}_{2}(\bm{k})), & \kappa=2\\
    (0,\bm{\varepsilon}_{3}(\bm{k})), & \kappa=3
    \end{cases},
    \label{Polarizations}
\end{align}
where $\bm{\varepsilon}_{i}(\bm{k})$ for $i=1,2$ are unit vectors spanning the plane transverse to $\bm{k}$, satisfying $\bm{k}\cdot\bm{\varepsilon}_{i}(\bm{k})=0$ and $\bm{\varepsilon}_{i}(\bm{k})\cdot\bm{\varepsilon}_{j}(\bm{k})=\delta_{ij}$, so that $(k\varepsilon_{\kappa})=0$ for $\kappa=1,2$, and $\bm{\varepsilon}_{3}(\bm{k})=\pm\frac{\bm{k}}{|\bm{k}|}$. For $k$ not necessarily on-shell, the longitudinal polarization vector is\footnote{The physicality of scalar ($\kappa=0$) and longitudinal ($\kappa=3$) photons has been a longstanding concern in the literature. The standard remedies --- factoring out the Minkowskian signature~\cite{Mandl:2010} or adopting the indefinite metric approach~\cite{Jauch:1976} --- we do not adopt. Instead, we note that the difficulties arise because Hermitian conjugation maps uncontracted tensors to their duals, and there is no natural way to account for the covariant/contravariant distinction within
Dirac bra-ket notation. Our resolution is the following convention, maintained throughout: \emph{kets carry covariant indices and bras carry contravariant indices}. A consequence is that we \emph{do not adopt the Einstein summation convention in general}; sums over polarization or Lorentz indices are written explicitly in Fourier integrals. This makes manifest that Lorentz and spin/polarization indices are Pontryagin duals of one another, in precise analogy with position and momentum.}
\begin{align}
    \varepsilon^{\mu}_{3}(\bm{k})
    =\frac{k^{\mu}-(k\varepsilon_{0})\varepsilon^{\mu}_{0}}
    {\sqrt{k^{2}-(k\varepsilon_{0})^{2}}}.
    \label{Polarization3}
\end{align}
\end{subequations}

We use the notation $\bm{K}\equiv(\kappa,\bm{k})$ and $K\equiv(\kappa,k)$ when convenient, and note that no repeated covariant or contravariant indices are permitted in products of four-vector components.\footnote{This restriction holds in the free Maxwell theory. When paired with the free Dirac theory, contraction with Dirac matrices $\gamma^{\mu}$ is
required, e.g.\ $\slashed{v}=v_{\mu}\gamma^{\mu}$ and $\slashed{T}^{\lambda}=
T{}^{\lambda}_{\mu}\gamma^{\mu}$, and the Hermitian conjugate is described algebraically via $\slashed{T}^{\lambda\dagger}=\gamma^{0}\slashed{T}^{\lambda}\gamma^{0}=T^{*}{}^{\mu}_{\lambda}\gamma_{\mu}=\slashed{T}^{*}_{\lambda}$.}
We also define the following distributional measures:
\begin{subequations}
    \begin{align*}
        \int d^{3}\bm{k}
        &\Longleftrightarrow
        \sum_{\bm{k}_{T}}\int\Delta^{2}\bm{k}_{T}\,dk_{\parallel},\\[4pt]
        \int d^{3}\lambda_{\gamma}(\bm{k})
        &\Longleftrightarrow
        \sum_{\bm{k}_{T}}\int\frac{\Delta^{2}\bm{k}_{T}\,dk_{\parallel}}
        {(2\pi)^{3}\sqrt{2\omega_{\bm{k}}}},\\[4pt]
        \int d^{3}\mu_{\gamma}(\bm{k})
        &\Longleftrightarrow
        \sum_{\bm{k}_{T}}\int\frac{\Delta^{2}\bm{k}_{T}\,dk_{\parallel}}
        {(2\pi)^{3}(2\omega_{\bm{k}})},
    \end{align*}
\end{subequations}
where $\Delta^{2}\bm{k}_{T}=\frac{(2\pi)^{2}}{L^{2}}$ is the transverse momentum spacing. In the continuum limit $L\rightarrow\infty$, the left-hand notation remains unchanged. Furthermore, we absorb the polarization sum into the momentum integral by a deliberate abuse of notation, writing $d^{3}\bm{K}$, $d^{3}\lambda_{\gamma}(\bm{K})$, and $d^{3}\mu_{\gamma}(\bm{K})$ respectively.

The general solution to Eq.~\eqref{MaxwellEquationFeynmanGauge} is
\begin{subequations}
\begin{align}
    A_{\mu}(x)
    =\int d^{3}\lambda_{\gamma}(\bm{K})\left[
    a^{*}_{\gamma}(\bm{K})\varepsilon^{*}{}^{\kappa}_{\mu}(\bm{k})e^{ikx}
    +a_{\gamma}(\bm{K})\varepsilon^{\kappa}_{\mu}(\bm{k})e^{-ikx}
    \right],
    \label{ElectromagneticFieldFourierExpansion}
\end{align}
where $a_{\gamma}(\bm{K})\equiv a_{\kappa}(\bm{k})$ is understood to carry covariant
indices. The time derivative of Eq.~\eqref{ElectromagneticFieldFourierExpansion} is
\begin{align}
    \dot{A}_{\mu}(x)
    =i\int d^{3}\lambda_{\gamma}(\bm{K})\,\omega_{\bm{k}}\left[
    a^{*}_{\gamma}(\bm{K})\varepsilon^{*}{}^{\kappa}_{\mu}(\bm{k})e^{ikx}
    -a_{\gamma}(\bm{K})\varepsilon^{\kappa}_{\mu}(\bm{k})e^{-ikx}
    \right].
    \label{ClassicalElectromagneticFieldCanonicalMomentum}
\end{align}
\end{subequations}
These can be inverted to give the Fourier coefficients:
\begin{subequations}
\begin{align}
    a_{\kappa}(\bm{k})
    &=\sum_{\mu}\int d^{3}\bm{x}\,\sqrt{\frac{\omega_{\bm{k}}}{2}}
    \left[A_{\mu}(x)+\frac{i}{\omega_{\bm{k}}}\dot{A}_{\mu}(x)\right]
    \varepsilon^{*}{}^{\mu}_{\kappa}(\bm{k})\,e^{ikx},
    \label{ClassicalAbsorptionCoefficient}
\end{align}
\begin{align}
    a^{*}_{\kappa}(\bm{k})
    &=\sum_{\mu}\int d^{3}\bm{x}\,\sqrt{\frac{\omega_{\bm{k}}}{2}}
    \left[A_{\mu}(x)-\frac{i}{\omega_{\bm{k}}}\dot{A}_{\mu}(x)\right]
    \varepsilon^{\mu}_{\kappa}(\bm{k})\,e^{-ikx}.
    \label{ClassicalEmissionCoefficient}
\end{align}
\end{subequations}
It is also useful to decompose the field into its positive- and negative-frequency parts,
\begin{subequations}
\begin{align}
    A^{+}_{\mu}(x)
    &=\int d^{3}\lambda_{\gamma}(\bm{K})\,
    a_{\gamma}(\bm{K})\varepsilon^{\kappa}_{\mu}(\bm{k})e^{-ikx},
    \label{ClassicalElectromagneticFieldAbsorption}
\end{align}
\begin{align}
    A^{-}_{\mu}(x)
    &=\int d^{3}\lambda_{\gamma}(\bm{K})\,
    a^{*}_{\gamma}(\bm{K})\varepsilon^{*\kappa}_{\;\;\mu}(\bm{k})e^{ikx},
    \label{ClassicalElectromagneticFieldEmission}
\end{align}
\end{subequations}
so that $A_{\mu}=A^{+}_{\mu}+A^{-}_{\mu}$, with $A^{+}_{\mu}$ carrying the
absorption (negative-frequency) structure and $A^{-}_{\mu}$ the emission
(positive-frequency) structure.

\subsubsection{Electromagnetic Field of a Plane-wave}

A plane-wave model for a laser pulse is given by
\begin{align}
    A_{\text{cl},\mu}(x)
    \equiv A_{\text{cl},\parallel}(k_{L}x)\,
    \varepsilon^{\kappa_{L}}_{\mu}(\bm{k}_{L}),
    \label{EMFieldPlaneWave}
\end{align}
where $k_{L}=(\omega_{\bm{k}_{L}},\bm{k}_{L})$ is the wave-vector with $k_{L}^{2}=0$, $\hat{k}_{L}=(1,\hat{\bm{k}}_{L})=\frac{k_{L}}{\omega_{\bm{k}_{L}}}$, $k_{L}x=\omega_{\bm{k}_{L}}x^{0}-\bm{k}_{L}\cdot x$, and
$\varepsilon^{\kappa_{L}}(\bm{k}_{L})=(0,\bm{\varepsilon}^{\kappa_{L}}(\bm{k}_{L}))$ is the transverse polarization satisfying $\bm{k}_{L}\cdot\bm{\varepsilon}^{\kappa_{L}}(\bm{k}_{L})=0$.
The Fourier coefficients evaluate to
\begin{align}
    a_{\text{cl},\kappa}(\bm{k})
    =\delta^{(2)}(\bm{k}_{T};\bm{0}_{T})
    \sqrt{2(\hat{\bm{k}}_{L}{\cdot}\bm{k})}
    \int d(\hat{k}_{L}x)\,
    A_{\text{cl},\parallel}(k_{L}x)\,
    e^{i(\hat{\bm{k}}_{L}{\cdot}\bm{k})(\hat{k}_{L}x)}
    \Theta(\hat{\bm{k}}_{L}{\cdot}\bm{k})
    \delta^{\kappa_{L}}_{\kappa},
    \label{FourierCoefficientPlaneWave}
\end{align}
where the discrete delta function carries units $\delta^{(2)}(\bm{k}_{T};\bm{0}_{T})\sim[K]^{-2}\sim[L]^{2}$, and $\Theta$ is the Heaviside step-function. The longitudinal Fourier coefficient is then defined to be
\begin{align}
    a_{\text{cl},\parallel}(K_{\parallel})
    &\equiv\sqrt{2(\hat{\bm{k}}_{L}{\cdot}\bm{k})}
    \int d(\hat{k}_{L}x)\,
    A_{\text{cl},\parallel}(k_{L}x)\,
    e^{i(\hat{\bm{k}}_{L}{\cdot}\bm{k})(\hat{k}_{L}x)}
    \Theta(\hat{\bm{k}}_{L}{\cdot}\bm{k})
    \delta^{\kappa_{L}}_{\kappa}\nonumber\\
    &=i\sqrt{\frac{2}{(\hat{\bm{k}}_{L}{\cdot}\bm{k})}}
    \int d(\hat{k}_{L}x)\,
    \frac{d A_{\text{cl},\parallel}(k_{L}x)}{d(\hat{k}_{L}x)}\,
    e^{i(\hat{\bm{k}}_{L}{\cdot}\bm{k})(\hat{k}_{L}x)}
    \Theta(\hat{\bm{k}}_{L}{\cdot}\bm{k})
    \delta^{\kappa_{L}}_{\kappa},
    \label{LongitudinalFourierCoefficientPlaneWave}
\end{align}
where $\hat{\bm{k}}_{L}{\cdot}\bm{k}=\pm k_{\parallel}$ according to the propagation direction of the plane-wave relative to the orientation of the scattering axis.

The units of $a_{\text{cl}}$ are fixed via the Fourier transform of $A_{\text{cl}}$ in Heaviside--Lorentz natural units,\footnote{We retain the freedom to scale the Fourier transform in Heaviside--Lorentz or SI units by powers of $\hslash$, $c$, and $\varepsilon_{0}$ (SI only).} and it will
be shown in the next chapter that $|a_{\text{cl}}|^{2}\sim[K]^{-3}$ is the momentum density of the electromagnetic field. For later use it is also advantageous to convert $a_{\text{cl},\parallel}$ into units of inverse square-root energy. Under the conversion $A_{\text{cl},\parallel}^{\text{HL}}=\sqrt{\varepsilon_{0}}A_{\text{cl},\parallel}^{\text{SI}}$, demanding $a_{\text{cl}}^{\text{HL}}=\sqrt{\varepsilon_{0}}\,a_{\text{cl}}^{\text{SI}}$ leaves Eq.~\eqref{LongitudinalFourierCoefficientPlaneWave}
unchanged. Since $eA_{\text{cl}}$ and $ea_{\text{cl}}$ are purely kinematic --- carrying no electromagnetic units --- the dimensionless combination $\tilde{A}_{\text{cl}}:=\frac{e}{mc}A_{\text{cl}}$ is invariant of the chosen unit convention. We accordingly define the rescaled longitudinal coefficient, valid in both Heaviside-Lorentz and SI units,
\begin{align}
    \tilde{a}_{\text{cl},\parallel}(K_{\parallel})
    \equiv\frac{e}{mc}a_{\text{cl},\parallel}(K_{\parallel})
    =i\sqrt{\frac{2}{\hslash(\hat{\bm{k}}_{L}{\cdot}\bm{k})c}}
    \int d(\hat{k}_{L}x)\,
    \frac{d\tilde{A}_{\text{cl},\parallel}(k_{L}x)}
    {d(\hat{k}_{L}x)}\,
    e^{i(\hat{\bm{k}}_{L}{\cdot}\bm{k})(\hat{k}_{L}x)}
    \Theta(\hat{\bm{k}}_{L}{\cdot}\bm{k})
    \delta^{\kappa_{L}}_{\kappa},
    \label{LongitudinalFourierCoefficientPlaneWaveSI}
\end{align}
so that $\tilde{a}_{\text{cl},\parallel}\sim[E]^{-\frac{1}{2}}$.

In later chapters, we will ultimately specialize the general framework to a physically concrete setting: a laser pulse with a Gaussian envelope. Specifically, we will take the classical field to be a linearly polarized Gaussian plane-wave,
\begin{align}
    A_{\text{cl},\parallel}(k_{L}x)
    =A_{\text{cl},\text{max}}
    \exp\!\left(-\frac{x_{+}^{2}}{2\sigma_{x_{+}}^{2}}\right)
    \cos\!\left(\frac{2\pi}{\lambda_{L}}x_{+}\right),
    \label{GaussianPlaneWave}
\end{align}
where the forward lightfront time $x_{+}=x_{0}+x_{\parallel}=(\hat{k}_{L}x)
=\frac{(k_{L}x)}{|\bm{k}_{L}|}$, with $|\bm{k}_L| = \frac{2\pi}{\lambda_L}$ the wave number, and $A_{\text{cl},\text{max}}$ is the peak field strength. As stated in Chapter~1, in the language of inverse Compton scattering, the dimensionless peak field strength $a_{0}\equiv\frac{eA_{\text{cl},\text{max}}}{mc}$ is called the classical nonlinearity parameter.

The longitudinal Fourier coefficient
of Eq.~\eqref{LongitudinalFourierCoefficientPlaneWave} for this field has the form
\begin{align}
    a_{\text{cl},\parallel}(k_{\parallel})
    =A_{\text{cl},\text{max}}\sqrt{-\frac{\pi\hslash k_{\parallel}c}{\sigma_{E_{L}}^{2}}}
    \Bigl[
    g(\hslash k_{\parallel}c - E_{L})
    +g(\hslash k_{\parallel}c + E_{L})
    \Bigr]
    \Theta(-k_{\parallel}),
    \label{LongitudinalFourierCoefficientGaussianPlaneWave}
\end{align}
where
\begin{align}
    g(\hslash k_{\parallel}c \pm E_{L})
    \equiv
    \exp\!\left(-\frac{(\hslash k_{\parallel}c \pm E_{L})^{2}}{2\sigma_{E_{L}}^{2}}\right).
\end{align}
Here we have defined the laser carrier energy $E_{L}=\hslash|\bm{k}_{L}|c$ and the energy spread $\sigma_{E_{L}}=\frac{\hslash c}{\sigma_{x_{+}}}$, in Heaviside-Lorentz units, and casted all classical observables to energy. This will be seen to be the natural choice in later chapters where our goal will be to calculate the scattered electron energy spectrum.

\subsection{Free Dirac Fields}

The Fourier theory of the free Dirac field runs in parallel to the electromagnetic case. In a free system consisting solely of charged matter the total action reduces to Eq.~\eqref{DiracAction}, and the Euler-Lagrange equation of motion is the free Dirac equation,
\begin{align}
    \frac{\delta S_{\text{D}}}{\delta\bar{\Psi}^{\beta}}
    =\left(i\slashed{\partial}^{\alpha}_{\beta}-m\delta^{\alpha}_{\beta}\right)
    \Psi_{\alpha}=0.
    \label{FreeDiracEquation}
\end{align}
The field is again assumed to be confined to the same transverse area $L^{2}$, so that the free Dirac field satisfies the same periodic boundary conditions. Setting $\Psi=(i\slashed{\partial}+m)\Phi$, the equation reduces to the Klein-Gordon equation for $\Phi$, whose solutions $\Phi\propto e^{\pm ipx}$ generate the mode functions of the Dirac field. The solutions to Eq.~\eqref{FreeDiracEquation} take the form
\begin{align*}
    \frac{1}{(2\pi)L^{2}}\,u^{\pi}_{\alpha}(\bm{p})\,e^{-ipx},
    \qquad
    \frac{1}{(2\pi)L^{2}}\,\bar{v}^{\pi}_{\alpha}(\bm{p})\,e^{ipx},
\end{align*}
where $p=(p^{0},\bm{p})$ is on-shell: $p^{2}=p_{0}^{2}-\bm{p}^{2}=m^{2}$, or
$E_{\bm{p}}\equiv p^{0}=\sqrt{m^{2}+\bm{p}^{2}}$. The transverse momenta are again discretized: $p_{i}=\frac{2\pi}{L}n_{i}$ for $i=1,2$. The index $\pi$ labels the spin, and $u^{\pi}_{\alpha}(\bm{p}),\,\bar{v}^{\pi}_{\alpha}(\bm{p})\in\mathbb{C}^{4}$ are Dirac spinors --- complex four-component Euclidean vectors, as required for the action to be a scalar. They satisfy
$\sum_{\alpha}(\slashed{p}-m)^{\alpha}_{\alpha'}u^{\pi}_{\alpha}(\bm{p})
=\sum_{\alpha}(\slashed{p}+m)^{\alpha}_{\alpha'}v^{\pi}_{\alpha}(\bm{p})=0$
and the orthogonality relations\footnote{Recall that Hermitian conjugation exchanges Lorentz and spinor indices: $u^{\pi'}_{\alpha}(\bm{p})^{\dagger}=
u^{\dagger}{}^{\alpha}_{\pi'}(\bm{p})$. Spin and polarization indices are always Lorentz-type.}
\begin{align}
    \sum_{\alpha}u^{\dagger}{}^{\alpha}_{\pi'}(\bm{p})u^{\pi}_{\alpha}(\bm{p})
    =\sum_{\alpha}v^{\alpha}_{\pi'}(\bm{p})v^{\dagger}{}^{\pi}_{\alpha}(\bm{p})
    =2E_{\bm{p}}\delta^{\pi}_{\pi'},
    \label{DiracEuclideanOrthogonality}
\end{align}
\begin{align}
    \sum_{\alpha}\bar{u}^{\alpha}_{\pi'}(\bm{p})u^{\pi}_{\alpha}(\bm{p})
    =-\sum_{\alpha}v^{\alpha}_{\pi'}(\bm{p})\bar{v}^{\pi}_{\alpha}(\bm{p})=2m\delta^{\pi}_{\pi'},
    \quad
    \sum_{\alpha}v^{\alpha}_{\pi'}(\bm{p})u^{\pi}_{\alpha}(-\bm{p})
    =\sum_{\alpha}\bar{u}^{\alpha}_{\pi'}(\bm{p})\bar{v}^{\pi}_{\alpha}(-\bm{p})=0,
    \label{DiracLorentzOrthogonality}
\end{align}
and the spin completeness relations
\begin{align}
    \sum_{\pi}u^{\pi}_{\alpha}(\bm{p})\bar{u}^{\alpha'}_{\pi}(\bm{p})
    =\slashed{p}^{\alpha'}_{\alpha}+m\delta^{\alpha'}_{\alpha},
    \qquad
    \sum_{\pi}\bar{v}^{\pi}_{\alpha'}(\bm{p})v^{\alpha}_{\pi}(\bm{p})
    =\slashed{p}^{\alpha}_{\alpha'}-m\delta^{\alpha}_{\alpha'}.
    \label{DiracSpinOrthogonality}
\end{align}
An explicit representation is
\begin{align}
    u^{\pi}(\bm{p})
    =\frac{\slashed{p}+m}{\sqrt{E_{\bm{p}}+m}}
    \begin{pmatrix}\chi^{\pi}(\bm{p})\\\bm{0}\end{pmatrix},
    \qquad
    \bar{v}^{\pi'}(\bm{p})
    =\frac{-\slashed{p}+m}{\sqrt{E_{\bm{p}}+m}}
    \begin{pmatrix}\bm{0}\\\chi^{\pi'}(\bm{p})\end{pmatrix},
    \label{DiracSpinors}
\end{align}
where $\chi^{\pi}(\bm{p})$ for $\pi=1,2$ are two-component unit spinors satisfying
$\chi^{\dagger}_{\pi'}(\bm{p})\chi^{\pi}(\bm{p})=\delta^{\pi}_{\pi'}$.

In analogy with the electromagnetic case, we write $\bm{P}\equiv(\pi,\bm{p})$ and $P\equiv(\pi,p)$ when convenient, and define the Lorentz-invariant distributional measures

\begin{align*}
    \int d^{3}\bm{p}
    &\Longleftrightarrow
    \sum_{\bm{p}_{T}}\int\Delta^{2}\bm{p}_{T}\,dp_{\parallel},\\[4pt]
    \int d^{3}\lambda_{e^{\pm}}(\bm{p})
    &\Longleftrightarrow
    \sum_{\bm{p}_{T}}\int\frac{\Delta^{2}\bm{p}_{T}\,dp_{\parallel}}
    {(2\pi)^{3}\sqrt{2E_{\bm{p}}}},\\[4pt]
    \int d^{3}\mu_{e^{\pm}}(\bm{p})
    &\Longleftrightarrow
    \sum_{\bm{p}_{T}}\int\frac{\Delta^{2}\bm{p}_{T}\,dp_{\parallel}}
    {(2\pi)^{3}(2E_{\bm{p}})},
\end{align*}
with $\Delta^{2}\bm{p}_{T}=\frac{(2\pi)^{2}}{L^{2}}$, and with the same continuum-limit and spin-absorption conventions as for the photonic measures. The general solution to
Eq.~\eqref{FreeDiracEquation} is then
\begin{subequations}
\begin{align}
    \Psi_{\alpha}(x)
    =\int d^{3}\lambda_{e^{\pm}}(\bm{P})\left[
    a^{*}_{e^{+}}(\bm{P})\bar{v}^{\pi}_{\alpha}(\bm{p})\,e^{ipx}
    +a_{e^{-}}(\bm{P})u^{\pi}_{\alpha}(\bm{p})\,e^{-ipx}
    \right],
    \label{ClassicalDiracField}
\end{align}
with time derivative
\begin{align}
    \dot{\Psi}_{\alpha}(x)
    =i\int d^{3}\lambda_{e^{\pm}}(\bm{P})\,E_{\bm{p}}\left[
    a^{*}_{e^{+}}(\bm{P})\bar{v}^{\pi}_{\alpha}(\bm{p})\,e^{ipx}
    -a_{e^{-}}(\bm{P})u^{\pi}_{\alpha}(\bm{p})\,e^{-ipx}
    \right].
    \label{ClassicalDiracCanonicalMomentum}
\end{align}
\end{subequations}
Inverting these expressions yields the Fourier coefficients:
\begin{subequations}
\begin{align}
    a_{e^{-}}(\bm{P})
    &=\frac{1}{2m}\sum_{\alpha}\int d^{3}\bm{x}\,\sqrt{\frac{E_{\bm{p}}}{2}}
    \left[\Psi_{\alpha}(x)+\frac{i}{E_{\bm{p}}}\dot{\Psi}_{\alpha}(x)\right]
    \bar{u}^{\alpha}_{\pi}(\bm{p})\,e^{ipx},
    \label{ClassicalElectronAbsorptionCoefficient}
\end{align}
\begin{align}
    a^{*}_{e^{+}}(\bm{P})
    &=-\frac{1}{2m}\sum_{\alpha}\int d^{3}\bm{x}\,\sqrt{\frac{E_{\bm{p}}}{2}}
    \left[\Psi_{\alpha}(x)-\frac{i}{E_{\bm{p}}}\dot{\Psi}_{\alpha}(x)\right]
    v^{\alpha}_{\pi}(\bm{p})\,e^{-ipx}.
    \label{ClassicalPositronEmissionCoefficient}
\end{align}
\end{subequations}
As with the electromagnetic field, it is convenient to define the positive- and negative-frequency parts of the Dirac field separately:
\begin{subequations}
\begin{align}
    \Psi^{+}_{\alpha}(x)
    &=\int d^{3}\lambda_{e^{\pm}}(\bm{P})\,
    a_{e^{-}}(\bm{P})\,u^{\pi}_{\alpha}(\bm{p})\,e^{-ipx},
    \label{ClassicalDiracAbsorption}
\end{align}
\begin{align}
    \Psi^{-}_{\alpha}(x)
    &=\int d^{3}\lambda_{e^{\pm}}(\bm{P})\,
    a^{*}_{e^{+}}(\bm{P})\,\bar{v}^{\pi}_{\alpha}(\bm{p})\,e^{ipx},
    \label{ClassicalDiracEmission}
\end{align}
\end{subequations}
so that $\Psi_\alpha = \Psi^+_\alpha + \Psi^-_\alpha$, in analogy with the electromagnetic decomposition $A_\mu = A^+_\mu + A^-_\mu$.

\section*{Summary}

This chapter has developed the classical foundations of electrodynamics within a unified variational framework, treating the Maxwell--Lorentzian and pre-quantum formulations as two realizations of the same underlying action principle, distinguished solely by their description of charged matter. In the former, matter enters as an unspecified current $J^\mu$ and stress-energy $T^{\mu\nu}$, with particle dynamics emerging from a careful limiting procedure that yields the Lorentz force law and, at order $e^2$, the Landau-Lifshitz equation describing radiation reaction. In the latter, matter is described by the Dirac spinor field $\Psi$, whose dynamics are governed by the Dirac equation coupled to Maxwell's equations through the minimal interaction $S_{\text{I}}$. The two formulations share the same Maxwell equations, the same gauge structure, and the same Fourier mode decomposition into positive- and negative-frequency components --- a parallelism that is not accidental but reflects their common variational origin.

The Fourier theory developed in the final section occupies a pivotal position in the dissertation. The mode expansions of the free electromagnetic and Dirac fields --- and the distributional measures defined alongside them --- are the objects that will be promoted to operators in the next chapter, where the passage from classical to quantum electrodynamics is effected by imposing canonical commutation and anticommutation relations on the Fourier coefficients $a_\kappa(\bm{k})$ and $a_{e^\pm}(\bm{P})$. The coherent photonic states that lie at the heart of the scattering model developed in later chapters are then defined directly in terms of these operators. In this sense, the present chapter is not merely preparatory: it establishes the classical substrate from which the quantum theory is constructed, and it is the pre-QED formulation --- not the Maxwell--Lorentzian one --- that provides the natural bridge.
\chapter{Quantization, Operators, and States in Quantum Electrodynamics}

Canonical \emph{quantization} in QED is the process of promoting the pre-QED electromagnetic and Dirac fields to operator-valued fields acting on Hilbert spaces, subject to commutation and anticommutation relations~\cite{Bjorken:1965,Berestetskii:1982,Greiner:2009,Jauch:1976,Kallen:1972,Mandl:2010,Pauli:1980,Peskin:1995,Thaller:1992,Weinberg:1995}. The coupled field equations Eqs.~\eqref{MaxwellsEquation1QED}--\eqref{StressEnergyConservationQED}, when promoted to
operator equations, describe the QED interaction picture: the dynamics of quantum field operators under electrodynamic interaction. Solutions to these coupled operator equations give rise to interacting states which are Hilbert space representations of interacting QED systems. The power of the operator-Hilbert space formalism is that for processes such as scattering, one need not solve the interacting equations of motion explicitly. Quantum theory does not require knowledge of intermediate processes to make predictions about asymptotic outcomes. For asymptotic free states --- Hilbert space representations of free systems in the distant past and future, long before
and after any interaction --- an evolution operator known as the S-matrix encodes the full effect of the interaction as a transition amplitude between states~\cite{Bjorken:1965,Berestetskii:1982,Greiner:2009,Jauch:1976,Kallen:1972,Mandl:2010,Peskin:1995,Pauli:1980,Thaller:1992,Weinberg:1995}.

The preceding chapter established the classical substrate on which this construction rests. The Fourier mode expansions of the free electromagnetic and Dirac fields, and the distributional measures accompanying them, were developed in deliberate preparation for what follows: the passage to quantum field theory is effected by promoting the classical Fourier coefficients $a_\kappa(\bm{k})$ and $a_{e^\pm}(\bm{P})$ to operators on a Hilbert
space, subject to canonical commutation and anticommutation relations respectively. Everything that distinguishes QED from its classical precursors --- the particle interpretation, the Hilbert space structure, the uncertainty relations, the statistics of identical particles --- is a consequence of this quantization.

The chapter divides naturally into two parallel developments, one for each field. For the electromagnetic field, canonical quantization yields creation and annihilation operators whose algebra generates the photonic Fock space --- the direct sum of all multi-photon Hilbert spaces. Within this Fock space, we introduce the central object of the inverse Compton scattering model developed in subsequent chapters: the \emph{photonic coherent state}, a free quantum macroscopic state parameterized by a classical field configuration that serves as the realization of the corresponding classical free electromagnetic field within quantum electrodynamics. Coherent states are the appropriate quantum description of a laser pulse~\cite{Glauber:1963Theory,Glauber:1963States,Sudarshan:1963,Mandel:1995}, and it is their special algebraic properties that make the coherent-state QED framework both computationally tractable and physically well-motivated. For the Dirac field, canonical anticommutation relations yield fermionic creation and annihilation operators for electrons and positrons, and the resulting Fock space accommodates the multi-particle states required for a general description of charged matter. We introduce the electronic \emph{spectral states} that encode the statistical description of a relativistic electron beam, which, together with the photonic coherent state, constitute the asymptotic initial states of the scattering calculations in later chapters.

The chapter concludes by assembling these free-field states and operators into an interacting theory via the S-matrix and perturbation theory. This approach sidesteps the interacting equations of motion entirely: rather than solving for the dynamics of interacting fields, one expands the S-matrix in powers of the coupling $e$, generating a perturbative series whose terms are organized by the Feynman rules of QED. Each order describes processes involving a fixed number of photon absorptions and emissions, and the free states constructed earlier in the chapter serve as precisely the asymptotic states between which these matrix elements are computed. This perturbative structure is the calculational engine of the scattering framework developed in subsequent chapters.

Throughout, the presentation maintains the conventions and notation established in
Chapter~2 (also see Notation and Conventions in the frontmatter).

\section{Quantization of the Electromagnetic field}

The first step toward canonical quantization is to promote the electromagnetic field to an operator-valued distribution. In the context of QED, we refer to $A_{\mu}(x)$ as the electromagnetic field operator; $A_{\gamma}^{+}(x)$ and $a_{\gamma}(\bm{K})$ are the
photon absorption operators, and $A_{\gamma}^{-}(x)$ and $a_{\gamma}(\bm{K})^{\dagger}$ are the photon emission operators, where the subscript $\gamma$ indicates a photonic quantity throughout.

Under quantization, Eq.~\eqref{ElectromagneticFieldFourierExpansion} becomes
\begin{align}
    A_{\mu}(x)
    =\int d^{3}\lambda_{\gamma}(\bm{K})\left[
    a^{\dagger}_{\gamma}(\bm{K})\varepsilon^{*}{}^{\kappa}_{\mu}(\bm{k})e^{ikx}
    +a_{\gamma}(\bm{K})\varepsilon^{\kappa}_{\mu}(\bm{k})e^{-ikx}
    \right],
    \label{QuantizedElectromagneticField}
\end{align}
and the operator versions of $\dot{A}_{\mu}(x)$, $A^{+}_{\mu}(x)$, $A^{-}_{\mu}(x)$, $a_{\gamma}(\bm{K})$, and $a_{\gamma}(\bm{K})^{\dagger}$ follow directly from Eqs.~\eqref{ClassicalElectromagneticFieldCanonicalMomentum}--\eqref{ClassicalElectromagneticFieldEmission} by the same replacement. Note that $(A_{\mu}(x))^{\dagger}=A^{\dagger\mu}(x)=A^{\mu}(x)$: the four-vector $A$ is Hermitian as an operator, but a chosen representation of its covariant components $A_{\mu}$ need not be. This is not surprising --- Hermitian conjugation exchanges covariant and contravariant indices, just as it exchanges column and row vectors, and does not affect the reality of the eigenvalues, which remain a real
Minkowski four-vector.

These operators satisfy the equal-time commutation relations
\begin{subequations}
\begin{align}
    \bigl[A_{\mu'}(x^{0},\bm{x}'),A^{\mu}(x^{0},\bm{x})\bigr]
    =\bigl[\dot{A}_{\mu'}(x^{0},\bm{x}'),\dot{A}^{\mu}(x^{0},\bm{x})\bigr]
    =0,
\end{align}
\begin{align}
    \bigl[A_{\mu'}(x^{0},\bm{x}'),\dot{A}^{\mu}(x^{0},\bm{x})\bigr]
    =i\delta^{\mu}_{\mu'}\delta^{(3)}(\bm{x}'-\bm{x}),
    \qquad
    \bm{x}_{T}\in\bigl[-\tfrac{L}{2},\tfrac{L}{2}\bigr]^{2}.
\end{align}
\end{subequations}
The photon emission and absorption operators satisfy
\begin{subequations}
\begin{align}
    \bigl[a_{\gamma}(\bm{K}'),a_{\gamma}(\bm{K})\bigr]
    =\bigl[a_{\gamma}(\bm{K}')^{\dagger},a_{\gamma}(\bm{K})^{\dagger}\bigr]
    =0.
\end{align}
For the mixed commutator, we introduce the photonic delta-function
\begin{align}
    \delta^{(3)}_{\gamma}(\bm{K}';\bm{K})
    \equiv (2\pi)^{3}(2\omega_{\bm{k}})\,
    \delta^{\kappa}_{\kappa'}\,
    \delta(\bm{k}'_{T};\bm{k}_{T})\,
    \delta(k'_{\parallel}-k_{\parallel}),
    \label{PhotonicDeltaFunction}
\end{align}
defined so that $f(\bm{K})=\int d^{3}\mu_{\gamma}(\bm{K}')\,
\delta^{(3)}_{\gamma}(\bm{K}';\bm{K})\,f(\bm{K}')$. In terms of this,
\begin{align}
    \bigl[a_{\gamma}(\bm{K}'),a_{\gamma}(\bm{K})^{\dagger}\bigr]
    =\frac{\delta^{(3)}_{\gamma}(\bm{K}';\bm{K})}{2\omega_{\bm{k}}}.
    \label{PhotonCommutator}
\end{align}
It is sometimes more natural to work with the rescaled operators
$c_{\gamma}(\bm{K})\equiv\sqrt{2\omega_{\bm{k}}}\,a_{\gamma}(\bm{K})$, for which
\begin{align}
    \bigl[c_{\gamma}(\bm{K}'),c_{\gamma}(\bm{K})^{\dagger}\bigr]
    =\delta^{(3)}_{\gamma}(\bm{K}';\bm{K}).
    \label{RescaledPhotonCommutator}
\end{align}
\end{subequations}
Throughout, delta-functions are defined to carry the reciprocal units of their arguments. The discrete transverse delta-function satisfies $(2\pi)^{2}\delta^{(2)}(\bm{k}'_{T};\bm{k}_{T})=L^{2}\delta^{(2)}(\bm{n}'_{T};\bm{n}_{T})$, consistent with $\bm{k}_{T}=\frac{2\pi}{L}\bm{n}_{T}$.

\subsection{Photonic states}

The emission operator creates a single-photon state from the vacuum,\footnote{We maintain the convention that kets carry contravariant polarization indices and bras carry covariant polarization indices. In our notation, $c_{\gamma}(\bm{K})=c_{\kappa}(\bm{k})$ and
$c_{\gamma}(\bm{K})^{\dagger}=c^{\dagger\kappa}(\bm{k})$.}
\begin{align}
    c_{\gamma}(\bm{K})^{\dagger}\ket{0}=\ket{\bm{K}},
    \label{SinglePhotonicState}
\end{align}
while the absorption operator annihilates the vacuum,
\begin{align}
    c_{\gamma}(\bm{K})\ket{0}=0.
    \label{AnnihilationOfVacuum}
\end{align}
Single-photon states satisfy the orthogonality relation
\begin{align}
    \braket{\bm{K}'|\bm{K}}
    =\bra{0}c_{\gamma}(\bm{K}')c_{\gamma}(\bm{K})^{\dagger}\ket{0}
    =\bra{0}\bigl[c_{\gamma}(\bm{K}'),c_{\gamma}(\bm{K})^{\dagger}\bigr]\ket{0}
    =\delta_{\gamma}^{(3)}(\bm{K}';\bm{K}),
    \label{SinglePhotonicStateOrthogonality}
\end{align}
where the second equality uses the fact that $c_\gamma(\bm{K})\ket{0}=0$.

Multi-photon states are built by acting with successive emission operators. Using the notation $\bm{K}^{(m)}=(\bm{K}_{0},\ldots,\bm{K}_{m-1})$ for an ordered $m$-tuple of mode labels, the ket and bra are defined as
\begin{align}
    \ket{\bm{K}^{(m)}}
    =\prod_{j=m-1}^{0}c_{\gamma}(\bm{K}_{j})^{\dagger}\ket{0},
    \label{MultiPhotonicStateKet}
\end{align}
\begin{align}
    \bra{\bm{K}'^{(m')}}
    =\bra{0}\prod_{j'=0}^{m'-1}c_{\gamma}(\bm{K}'_{j'}),
    \label{MultiPhotonicStateBra}
\end{align}
where the ordering of operators in Eq.~\eqref{MultiPhotonicStateKet} is significant: operators with smaller index act first on the vacuum, so the state is an ordered product. Since the photon operators commute, the physical state is symmetric under permutation of the labels, but the ordering convention fixes a canonical representative.

The orthogonality relation for multi-photon states is
\begin{align}
    \braket{\bm{K}'^{(m')}|\bm{K}^{(m)}}
    &=\delta_{m'm}
    \sum_{\sigma\in S_{m}}
    \prod_{j'=0}^{m'-1}
    \delta^{(3)}_{\gamma}(\bm{K}'_{j'};\bm{K}_{\sigma(j')})\nonumber\\
    &=\frac{\delta_{m'm}}{m'!}
    \sum_{\sigma'\in S_{m'}}
    \sum_{\sigma\in S_{m}}
    \delta^{(3m')}_{\gamma}
    \bigl(\sigma'{\cdot}\bm{K}'^{(m')};\sigma{\cdot}\bm{K}^{(m)}\bigr),
    \label{MultiPhotonicStateOrthogonality}
\end{align}
where $S_{m}$ denotes the symmetric group on $m$ letters, $\sigma\cdot\bm{K}^{(m)}$ denotes a group action, and the sums run over all permutations of the respective index sets. The content of this relation is transparent: the inner product is non-zero if and only if $m'=m$ and the bra labels are a permutation of the ket labels. Equivalently, there exists a permutation $\sigma\in S_{m}$ such that
$\bm{K}'_{j'}=\bm{K}_{\sigma(j')}$ for each $j'$. This is precisely the statement that photons are indistinguishable bosons: two multi-photon states are orthogonal unless they differ only in the labeling of identical particles.

\subsection{Photonic Fock space}

Let $\mathcal{H}_{\gamma}$ be the Hilbert space of single-photon states. The corresponding photonic Fock space is
\begin{align}
    \mathcal{F}(\mathcal{H}_{\gamma})
    =\bigoplus_{m=0}^{\infty}\Sym^{m}(\mathcal{H}_{\gamma}),
\end{align}
where $\Sym^{m}(\mathcal{H}_{\gamma})\subset\mathcal{H}_{\gamma}^{\otimes m}$ is the subspace of symmetric $m$-photon states, reflecting the bosonic nature of photons. The identity operator on this space is
\begin{align}
    \mathbf{1}_{\gamma}
    =\sum_{n}\frac{1}{n!}
    \int d^{3n}\mu_{\gamma}(\bm{L}^{(n)})\,
    \ket{\bm{L}^{(n)}}\bra{\bm{L}^{(n)}}.
    \label{PhotonicFockSpaceIdentity}
\end{align}

It is also useful to record the inner product between the single-photon state $\ket{\bm{K}}=c_{\gamma}(\bm{K})^{\dagger}\ket{0}$ and the field-generated state $\ket{X}\equiv A^{\mu}(x)\ket{0}$, where $X\equiv(\mu,x)$. A direct calculation gives
\begin{align}
    \braket{X|\bm{K}}=\varepsilon^{\kappa}_{\mu}(\bm{k})\,e^{-ikx},
    \label{PhotonicOverlap}
\end{align}
from which $\braket{\bm{K}|X}=\varepsilon^{*}{}^{\mu}_{\kappa}(\bm{k})\,e^{ikx}$ follows
by Hermitian conjugation.

The polarized photonic number density operator is defined as
\begin{align}
    \mathcal{N}_{\gamma}(\bm{K})
    \equiv a_{\gamma}(\bm{K})^{\dagger}a_{\gamma}(\bm{K})
    =a^{\dagger}{}^{\kappa}(\bm{k})\,a_{\kappa}(\bm{k}),
    \label{PhotonicNumberDensity}
\end{align}
where, as throughout, the Einstein summation convention is not assumed. Summing over polarization indices yields the unpolarized number density operator. In general, the expectation value of $\mathcal{N}_{\gamma}(\bm{K})$ in a given photonic state is the expected number density of photons with polarization and momentum $\bm{K}$ in that state.
Inserting the identity Eq.~\eqref{PhotonicFockSpaceIdentity} between the emission and absorption operators and applying the multi-photon orthogonality relation Eq.~\eqref{MultiPhotonicStateOrthogonality}, one finds that multi-photon states are eigenstates of the number density operator:
\begin{align}
    \mathcal{N}_{\gamma}(\bm{K})\ket{\bm{K}^{(m)}}
    =\frac{1}{2\omega_{\bm{k}}}
    \sum_{i=0}^{m-1}
    \delta_{\gamma}^{(3)}(\bm{K}_{i};\bm{K})\,
    \ket{\bm{K}^{(m)}}.
    \label{PhotonicNumberDensityMultiPhotonicState}
\end{align}

\paragraph{Generic photonic Fock states.}
A general photonic Fock state is constructed from an arbitrary sequence of wavefunctions. Let $\Lambda=\frac{(2\pi)^{2}}{L^{2}}\mathbb{Z}^{2}\times\mathbb{R}$ be the momentum space, with $N_{\gamma}$ a random photon number and momenta $\bm{K}_{\gamma,0},\bm{K}_{\gamma,1},\ldots$ drawn from some free electromagnetic system. For each $m\in\mathbb{N}$, let $\phi_{\gamma}(m,\cdot):(\mathbb{Z}_{4}\times\Lambda)^{m}\rightarrow\mathbb{C}$ be a square-integrable wavefunction such that
$|\phi_{\gamma}(m,\bm{K}^{(m)})|^{2}=f_{N_{\gamma}}(m)\,
f_{\bm{K}_{\gamma}^{(m)}|N_{\gamma}}(\bm{K}^{(m)}|m)$ is the probability of selecting $m$ photons with mode labels $\bm{K}^{(m)}$ from the
system. The corresponding generic photonic state is
\begin{align}
    \ket{\Phi_{\gamma}}
    =\sum_{m}\frac{1}{\sqrt{m!}}
    \int d^{3m}\lambda_{\gamma}(\bm{K}^{(m)})\,
    \phi_{\gamma}(m,\bm{K}^{(m)})\,
    \ket{\bm{K}^{(m)}},
    \label{PhotonicState}
\end{align}
which is normalized, $\braket{\Phi_{\gamma}|\Phi_{\gamma}}=1$, under the standard momentum measure $\frac{d^{3}\bm{k}}{(2\pi)^{3}}$. The expected photonic number density in this state is
\begin{align}
    \bra{\Phi_{\gamma}}\mathcal{N}_{\gamma}(\bm{K})\ket{\Phi_{\gamma}}
    =\sum_{m=1}^{\infty}f_{N_{\gamma}}(m)
    \sum_{i=0}^{m-1}f_{\bm{K}_{\gamma,i}|N_{\gamma}}(\bm{K}|m),
    \label{PhotonicNumberDensityPhotonicState}
\end{align}
where, given $m$, $f_{\bm{K}_{\gamma,i}|N_{\gamma}}$ is the marginal obtained by integrating out all momenta except $\bm{K}_{i}$:
\begin{align*}
    f_{\bm{K}_{\gamma,i}|N_{\gamma}}(\bm{K}|m)
    =\prod_{\substack{j=0\\j\neq i}}^{m-1}
    \int\frac{d^{3}\bm{K}_{j}}{(2\pi)^{3}}\,
    f_{\bm{K}^{(m)}_{\gamma}|N_{\gamma}}(\bm{K}^{(m)}|m).
\end{align*}

\paragraph{Photonic spectral states.}
When the momenta $\bm{K}_{i}$ are independent and identically distributed, the joint conditional distribution factorizes:
\begin{align}
    \left|\phi_{\gamma}(m,\bm{K}^{(m)})\right|^{2}
    &=f_{N_{\gamma}}(m)\prod_{j=0}^{m-1}f_{\bm{K}_{\gamma}}(\bm{K}_{j}).
    \label{PhotonicSpectralState}
\end{align}
The corresponding state $\ket{\Phi_{\gamma}}$ is called a photonic \emph{spectral state}, and its expected number density simplifies to
\begin{align}
    \bra{\Phi_{\gamma}}\mathcal{N}_{\gamma}(\bm{K})\ket{\Phi_{\gamma}}
    =\langle m\rangle\,f_{\bm{K}_{\gamma}}(\bm{K}),
    \label{PhotonicNumberDensityPhotonicSpectralState}
\end{align}
where $\langle m\rangle$ is the expected photon number in the Fock state. The spectral state need not describe the entire physical system: it can be constructed from any repeated random sample drawn from the momentum spectrum of the system.

\paragraph{Photonic coherent states.}
A photonic Fock state need not contain a definite number of photons~\cite{Glauber:1963Theory,Glauber:1963States,Sudarshan:1963,Mandel:1995}. When the underlying physical state admits a classical description --- that is, when it satisfies Maxwell's equations --- its representation as a Fock state is a superposition over all photon numbers. Such a state is a photonic \emph{coherent state}. Given a classical electromagnetic field $A_{\text{cl},\mu}(x)$ satisfying Eqs.~\eqref{ClassicalElectromagneticFieldAbsorption}--\eqref{ClassicalElectromagneticFieldEmission}, with Fourier coefficients $a_{\text{cl}}(\bm{K})$ given by Eqs.~\eqref{ClassicalAbsorptionCoefficient}--\eqref{ClassicalEmissionCoefficient}, the corresponding photonic coherent state is
\begin{align}
    \ket{A_{\text{cl}}}
    &\equiv c_{\gamma}(A_{\text{cl}})^{\dagger}\ket{0}\nonumber\\
    &=e^{\int\frac{d^{3}\bm{K}}{(2\pi)^{3}}
    \left(a_{\gamma}(\bm{K})^{\dagger}a_{\text{cl}}(\bm{K})
    -a_{\text{cl}}(\bm{K})^{\dagger}a_{\gamma}(\bm{K})\right)}\ket{0}\nonumber\\
    &=C\,e^{\int\frac{d^{3}\bm{K}}{(2\pi)^{3}}
    a_{\gamma}(\bm{K})^{\dagger}a_{\text{cl}}(\bm{K})}\ket{0}\nonumber\\
    &=\sum_{m}\frac{C}{m!}
    \int d^{3m}\lambda_{\gamma}(\bm{K}^{(m)})\,
    \prod_{j=0}^{m-1}a_{\text{cl}}(\bm{K}_{j})\,
    \ket{\bm{K}^{(m)}},
    \label{PhotonicCoherentState}
\end{align}
where the third line follows from the Baker-Campbell-Hausdorff identity, the \emph{displacement operator} is
\begin{align}
    c_{\gamma}(A_{\text{cl}})^{\dagger}
    =\exp\!\left(
    \int\frac{d^{3}\bm{K}}{(2\pi)^{3}}
    \Bigl(a_{\gamma}(\bm{K})^{\dagger}a_{\text{cl}}(\bm{K})
    -a_{\text{cl}}(\bm{K})^{\dagger}a_{\gamma}(\bm{K})\Bigr)
    \right),
    \label{PhotonicCoherentStateCreation}
\end{align}
and the normalization constant is
\begin{align}
    C=\exp\!\left(
    -\frac{1}{2}\int\frac{d^{3}\bm{K}}{(2\pi)^{3}}\,
    \left|a_{\text{cl}}(\bm{K})\right|^{2}
    \right).
    \label{CoherentStateNorm}
\end{align}
Introducing the total spectral weight
\begin{align}
    \mathcal{Z}
    \equiv\int\frac{d^{3}\bm{K}}{(2\pi)^{3}}\,
    \left|a_{\text{cl}}(\bm{K})\right|^{2},
    \label{TotalSpectralWeight}
\end{align}
the wavefunction of the coherent state can be written as
\begin{align}
    \phi_{\gamma}(m,\bm{K}^{(m)})
    =C\sqrt{\frac{\mathcal{Z}^{m}}{m!}}
    \prod_{j=0}^{m-1}
    \frac{a_{\text{cl}}(\bm{K}_{j})}{\sqrt{\mathcal{Z}}},
    \label{CoherentStateWavefunction}
\end{align}
confirming that $\sum_{m}\int\frac{d^{3m}\bm{K}^{(m)}}{(2\pi)^{3m}}
|\phi_{\gamma}(m,\bm{K}^{(m)})|^{2}=1$. The photon number distribution is Poissonian with mean $\mathcal{Z}$: the probability of finding $m$ photons is $f_{N_\gamma}(m)=\frac{e^{-\mathcal{Z}}\mathcal{Z}^m}{m!}$, and the marginal momentum distribution is $f_{\bm{K}_\gamma}(\bm{K})=\frac{|a_{\text{cl}}(\bm{K})|^{2}}{\mathcal{Z}}$. Thus a photonic coherent state is a special case of a photonic spectral state.

The expected photon number deserves comment. Strictly speaking, there are no photons in classical electrodynamics --- a classical electromagnetic field is not a collection of photons, and classical field amplitudes are not photon number densities. Here, however, we are describing a classical state from the vantage point of QED, as an element of Fock space. In this context, the expected photon number $\langle m\rangle$ in $\ket{A_{\text{cl}}}$ is a well-defined quantum observable, given simply by
\begin{align}
    \langle m\rangle
    =\mathcal{Z}
    =\int\frac{d^{3}\bm{K}}{(2\pi)^{3}}\,
    \left|a_{\text{cl}}(\bm{K})\right|^{2}.
    \label{MeanPhotonNumber}
\end{align}
The corresponding photonic number density, obtained from
Eq.~\eqref{PhotonicNumberDensityPhotonicSpectralState}, is
\begin{align}
    \bra{A_{\text{cl}}}\mathcal{N}_{\gamma}(\bm{K})\ket{A_{\text{cl}}}
    =\left|a_{\text{cl}}(\bm{K})\right|^{2}.
    \label{PhotonicCoherentStateNumberDensity}
\end{align}
This is the precise sense in which $|a_{\text{cl}}(\bm{K})|^{2}$ is the momentum density of the classical electromagnetic field, as claimed in Chapter~2.

Photonic coherent states have two further properties that are essential for the scattering calculations in later chapters. They are eigenstates of the photon absorption operator,
\begin{align}
    a_{\gamma}(\bm{K})\ket{A_{\text{cl}}}
    =a_{\text{cl}}(\bm{K})\ket{A_{\text{cl}}},
    \label{PhotonicCoherentStateAbsorption}
\end{align}
and consequently eigenstates of the positive-frequency field operator,
\begin{align}
    A^{+}_{\mu}(x)\ket{A_{\text{cl}}}
    =A^{+}_{\text{cl},\mu}(x)\ket{A_{\text{cl}}},
    \label{PhotonicCoherentStateFieldAbsorption}
\end{align}
with classical field amplitudes as eigenvalues. These two relations encode the defining property of a coherent state: acting with the absorption operator returns the classical Fourier coefficient, so quantum matrix elements involving coherent states reduce to classical field integrals. It is this reduction that makes the coherent-state QED framework computationally tractable.

\section{Quantization of the Dirac field}

Quantization of the Dirac field proceeds in parallel to the electromagnetic case, but with one essential difference: the fermionic nature of electrons and positrons requires anticommutation relations in place of commutation relations. The Dirac field operator $\Psi_{\alpha}(x)$ and the electron/positron absorption operators $\Psi_{\alpha}^{+}(x)$ and $a_{e^{\pm}}(\bm{P})$ are the operator-valued counterparts of their classical analogues, as are the emission operators $\Psi_{\alpha}^{-}(x)$ and $a_{e^{\pm}}(\bm{P})^{\dagger}$.

Under quantization, Eq.~\eqref{ClassicalDiracField} becomes
\begin{align}
    \Psi_{\alpha}(x)
    =\int d^{3}\lambda_{e^{\pm}}(\bm{P})\left[
    a^{\dagger}_{e^{+}}(\bm{P})\bar{v}^{\pi}_{\alpha}(\bm{p})\,e^{ipx}
    +a_{e^{-}}(\bm{P})\,u^{\pi}_{\alpha}(\bm{p})\,e^{-ipx}
    \right],
    \label{QuantizedDiracField}
\end{align}
and the operator versions of $\dot{\Psi}_{\alpha}(x)$, $\Psi^{+}_{\alpha}(x)$, $\Psi^{-}_{\alpha}(x)$, $a_{e^{\pm}}(\bm{P})$, and $a_{e^{\pm}}(\bm{P})^{\dagger}$ follow directly from Eqs.~\eqref{ClassicalDiracCanonicalMomentum}--\eqref{ClassicalDiracEmission} by the same promotion. Note that
$(\Psi_{\alpha}(x))^{\dagger}=\Psi^{\dagger\alpha}(x)$.

These operators satisfy the equal-time canonical anticommutation relations. Absorption and emission operators anticommute with themselves:
\begin{subequations}
\begin{align}
    \bigl\{a_{e^{\pm}}(\bm{P}'),a_{e^{\pm}}(\bm{P}))\bigr\}
    =\bigl\{a_{e^{\pm}}(\bm{P}')^{\dagger},a_{e^{\pm}}(\bm{P})^{\dagger}\bigr\}
    =0,
\end{align}
while cross-species anticommutators vanish identically. For the mixed anticommutator, we introduce the fermionic delta-function
\begin{align}
    \delta^{(3)}_{e^{\pm}}(\bm{P}';\bm{P})
    \equiv(2\pi)^{3}(2E_{\bm{p}})\,
    \delta^{\pi}_{\pi'}\,
    \delta^{(2)}(\bm{p}'_{T};\bm{p}_{T})\,
    \delta(p'_{\parallel}-p_{\parallel}),
    \label{FermionicDeltaFunction}
\end{align}
defined so that $f(\bm{P})=\int d^{3}\mu_{e^{\pm}}(\bm{P}')\,
\delta^{(3)}_{e^{\pm}}(\bm{P}';\bm{P})\,f(\bm{P}')$. In terms of this,
\begin{align}
    \bigl\{a_{e^{\pm}}(\bm{P}'),a_{e^{\pm}}(\bm{P})^{\dagger}\bigr\}
    =\frac{\delta^{(3)}_{e^{\pm}}(\bm{P}';\bm{P})}{2E_{\bm{p}}}.
    \label{FermionAnticommutator}
\end{align}
As in the electromagnetic case, it is natural to introduce the rescaled operators
$c_{e^{\pm}}(\bm{P})\equiv\sqrt{2E_{\bm{p}}}\,a_{e^{\pm}}(\bm{P})$, for which
\begin{align}
    \bigl\{c_{e^{\pm}}(\bm{P}'),c_{e^{\pm}}(\bm{P})^{\dagger}\bigr\}
    =\delta^{(3)}_{e^{\pm}}(\bm{P}';\bm{P}).
    \label{RescaledFermionAnticommutator}
\end{align}
\end{subequations}
Again, the delta-functions are defined to carry the reciprocal units of their arguments throughout, so that in particular $(2\pi)^{2}\delta^{(2)}(\bm{p}'_{T};\bm{p}_{T})=L^{2}\delta^{(2)}(\bm{n}'_{T};\bm{n}_{T})$,
consistent with $\bm{p}_{T}=\frac{2\pi}{L}\bm{n}_{T}$.

\subsection{Electronic and Positronic states}

The construction of fermionic states proceeds as in the photonic case, with one essential modification: the anticommutation relations introduce sign factors that reflect the antisymmetry of fermionic states under particle exchange. The emission operator creates a single electron or positron state from the vacuum (with kets carrying a covariant spin index, as established in the preceding section),
\begin{align}
    c_{e^{\pm}}(\bm{P})^{\dagger}\ket{0}=\ket{\bm{P}},
    \label{SingleFermionicState}
\end{align}
while the absorption operator annihilates the vacuum,
\begin{align}
    c_{e^{\pm}}(\bm{P})\ket{0}=0.
    \label{FermionicAnnihilationOfVacuum}
\end{align}
Single-particle states satisfy the orthogonality relation
\begin{align}
    \braket{\bm{P}'|\bm{P}}
    =\bra{0}c_{e^{\pm}}(\bm{P}')c_{e^{\pm}}(\bm{P})^{\dagger}\ket{0}
    =\bra{0}\bigl\{c_{e^{\pm}}(\bm{P}'),c_{e^{\pm}}(\bm{P})^{\dagger}\bigr\}\ket{0}
    =\delta_{e^{\pm}}^{(3)}(\bm{P}';\bm{P}),
    \label{SingleFermionicStateOrthogonality}
\end{align}
where the second equality uses $c_{e^\pm}(\bm{P})\ket{0}=0$.

Multi-particle states are built by successive application of emission operators. Using the notation $\bm{P}^{(m)}=(\bm{P}_{0},\ldots,\bm{P}_{m-1})$, the ket and bra are
\begin{align}
    \ket{\bm{P}^{(m)}}
    =\prod_{j=m-1}^{0}c_{e^{\pm}}(\bm{P}_{j})^{\dagger}\ket{0},
    \label{MultiFermionicStateKet}
\end{align}
\begin{align}
    \bra{\bm{P}'^{(m')}}
    =\bra{0}\prod_{j'=0}^{m'-1}c_{e^{\pm}}(\bm{P}'_{j'}).
    \label{MultiFermionicStateBra}
\end{align}
Unlike the bosonic case, the ordering in Eq.~\eqref{MultiFermionicStateKet} is essential: because the operators anticommute, exchanging any two adjacent emission operators
introduces a sign change, so different orderings yield physically distinct states related by a sign. The canonical ordering here fixes the sign convention.

The multi-particle orthogonality relation is
\begin{align}
    \braket{\bm{P}'^{(m')}|\bm{P}^{(m)}}
    &=\delta_{m'm}
    \sum_{\sigma\in S_{m}}
    \sgn(\sigma)
    \prod_{j'=0}^{m'-1}
    \delta_{e^{\pm}}^{(3)}(\bm{P}'_{j'};\bm{P}_{\sigma(j')})\nonumber\\
    &=\frac{\delta_{m'm}}{m'!}
    \sum_{\sigma'\in S_{m'}}
    \sum_{\sigma\in S_{m}}
    \sgn(\sigma')\sgn(\sigma)\,
    \delta_{e^{\pm}}^{(3m')}
    \bigl(\sigma'{\cdot}\bm{P}'^{(m')};\sigma{\cdot}\bm{P}^{(m)}\bigr).
    \label{MultiFermionicStateOrthogonality}
\end{align}
The structure here mirrors Eq.~\eqref{MultiPhotonicStateOrthogonality} for bosons, with one key difference: each term in the permutation sum carries a factor of $\sgn(\sigma)$, the sign of the permutation. The inner product is therefore non-zero if and only if $m'=m$ and the bra labels are a permutation of the ket labels, with the value determined
by the parity of that permutation. This is precisely the statement that electrons and positrons are indistinguishable fermions: multi-particle states are antisymmetric under the exchange of any two particle labels, so an odd permutation of the bra relative to the ket flips the sign of the inner product.

\subsection{Electronic and Positronic Fock space}

Let $\mathcal{H}_{e^{\pm}}$ be the Hilbert space of single-electron and single-positron states. The corresponding fermionic Fock space is
\begin{align}
    \mathcal{F}(\mathcal{H}_{e^{\pm}})
    =\bigoplus_{m=0}^{\infty}{\bigwedge}^{m}(\mathcal{H}_{e^{\pm}}),
\end{align}
where ${\bigwedge}^{m}(\mathcal{H}_{e^{\pm}})\subset\mathcal{H}_{e^{\pm}}^{\otimes m}$ is the subspace of antisymmetric $m$-particle states, reflecting the fermionic statistics established by the anticommutation relations of the preceding section. The identity operator on this space is
\begin{align}
    \mathbf{1}_{e^{\pm}}
    =\sum_{n}\frac{1}{n!}
    \int d^{3n}\mu_{e^{\pm}}(\bm{Q}^{(n)})\,
    \ket{\bm{Q}^{(n)}}\bra{\bm{Q}^{(n)}}.
    \label{FermionicFockSpaceIdentity}
\end{align}

The inner product between the single-electron state $\ket{\bm{P}}=c_{e^{-}}(\bm{P})^{\dagger}\ket{0}$ and the field-generated state $\ket{X}\equiv\bar{\Psi}^{\alpha}(x)\ket{0}$, where $X=(\alpha,x)$, is
\begin{align}
    \braket{\bm{P}|X}=\bar{u}^{\alpha}_{\pi}(\bm{p})\,e^{ipx},
    \label{FermionicOverlap}
\end{align}
from which $\braket{X|\bm{P}}=u^{\pi}_{\alpha}(\bm{p})\,e^{-ipx}$ follows by Dirac conjugation. Analogous relations hold for positron states.

The polarized number density operator for electrons and positrons is
\begin{align}
    \mathcal{N}_{e^{\pm}}(\bm{P})
    \equiv a_{e^{\pm}}(\bm{P})^{\dagger}a_{e^{\pm}}(\bm{P}),
    \label{FermionicNumberDensity}
\end{align}
with the Einstein summation convention not assumed throughout. Summing over spin indices yields the unpolarized number density. In general, the expectation value of $\mathcal{N}_{e^{\pm}}(\bm{P})$ in a given state is the expected number of electrons or positrons with spin and momentum $\bm{P}$ in that state. Inserting the identity Eq.~\eqref{FermionicFockSpaceIdentity} between the emission and absorption operators and applying the multi-particle orthogonality relation Eq.~\eqref{MultiFermionicStateOrthogonality}, one finds that multi-particle states are eigenstates of the number density operator:
\begin{align}
    \mathcal{N}_{e^{\pm}}(\bm{P})\ket{\bm{P}^{(m)}}
    =\frac{1}{2E_{\bm{p}}}
    \sum_{i=0}^{m-1}
    \delta_{e^{\pm}}^{(3)}(\bm{P}_{i};\bm{P})\,
    \ket{\bm{P}^{(m)}}.
    \label{FermionicNumberDensityMultiParticleState}
\end{align}

\paragraph{Generic electronic and positronic Fock states.}
A general electronic or positronic Fock state is constructed in direct analogy with the photonic case. Let $\Lambda=\frac{(2\pi)^{2}}{L^{2}}\mathbb{Z}^{2}\times\mathbb{R}$ be the momentum space, with $N_{e^{\pm}}$ a random particle number and momenta $\bm{P}_{e^{\pm},0},\bm{P}_{e^{\pm},1},\ldots$ drawn from some free charged matter system. For each $m\in\mathbb{N}$, let $\phi_{e^{\pm}}(m,\cdot):(\mathbb{Z}_{2}\times\Lambda)^{m}\rightarrow\mathbb{C}$ satisfy the conditional probability law $|\phi_{e^{\pm}}(m,\bm{P}^{(m)})|^{2}=f_{N_{e^{\pm}}}(m)\,f_{\bm{P}_{e^{\pm}}^{(m)}|N_{e^{\pm}}}(\bm{P}^{(m)}|m)$. The generic fermionic Fock state is then
\begin{align}
    \ket{\Phi_{e^{\pm}}}
    =\sum_{m}\frac{1}{\sqrt{m!}}
    \int d^{3m}\lambda_{e^{\pm}}(\bm{P}^{(m)})\,
    \phi_{e^{\pm}}(m,\bm{P}^{(m)})\,
    \ket{\bm{P}^{(m)}},
    \label{FermionicState}
\end{align}
normalized so that $\braket{\Phi_{e^{\pm}}|\Phi_{e^{\pm}}}=1$. The expected number density in this state is
\begin{align}
    \bra{\Phi_{e^{\pm}}}\mathcal{N}_{e^{\pm}}(\bm{P})\ket{\Phi_{e^{\pm}}}
    =\sum_{m=1}^{\infty}f_{N_{e^{\pm}}}(m)
    \sum_{i=0}^{m-1}
    f_{\bm{P}_{e^{\pm},i}|N_{e^{\pm}}}(\bm{P}|m),
    \label{FermionicNumberDensityFermionicState}
\end{align}
where the marginal $f_{\bm{P}_{e^{\pm},i}|N_{e^{\pm}}}$ is obtained by integrating out all momenta except $\bm{P}_{i}$:
\begin{align*}
    f_{\bm{P}_{e^{\pm},i}|N_{e^{\pm}}}(\bm{P}|m)
    =\prod_{\substack{j=0\\j\neq i}}^{m-1}
    \int\frac{d^{3}\bm{P}_{j}}{(2\pi)^{3}}\,
    f_{\bm{P}_{e^{\pm}}^{(m)}|N_{e^{\pm}}}(\bm{P}^{(m)}|m).
\end{align*}

\paragraph{Electronic and positronic spectral states.}
When the momenta $\bm{P}_{i}$ are independent and identically distributed, the joint conditional distribution takes the form
\begin{align}
    \left|\phi_{e^{\pm}}(m,\bm{P}^{(m)})\right|^{2}
    &=f_{N_{e^{\pm}}}(m)\,
    |\varepsilon(\bm{P}^{(m)})|^{2}
    \prod_{j=0}^{m-1}f_{\bm{P}_{e^{\pm}}}(\bm{P}_{j}),
    \label{FermionicSpectralState}
\end{align}
where $\varepsilon(\bm{P}^{(m)})$ is the \emph{Pauli function}: a function of the variable labels rather than their values, which encodes the Pauli exclusion principle. It is antisymmetric under the exchange of any two labels --- so that exchanging two particle momenta introduces a sign change --- and vanishes whenever any two momenta coincide. The corresponding state $\ket{\Phi_{e^{\pm}}}$ is an electronic or positronic
\emph{spectral state}, with expected number density
\begin{align}
    \bra{\Phi_{e^{\pm}}}\mathcal{N}_{e^{\pm}}(\bm{P})\ket{\Phi_{e^{\pm}}}
    =\langle m\rangle\,f_{\bm{P}_{e^{\pm}}}(\bm{P}),
    \label{ElectronicNumberDensityFermionicSpectralState}
\end{align}
where $\langle m\rangle$ is the expected particle number in the Fock state. As in the photonic case, the spectral state need not describe the entire physical system; it may be constructed from any repeated random sample drawn from the momentum spectrum.

\paragraph{Electronic/positronic versus photonic ontology.}
The electronic/positronic and photonic Fock spaces differ in a fundamental way that reflects the underlying physics. Since electrons and positrons are massive particles, their Fock states must contain a definite particle number. Moreover, unlike the photonic case, the Dirac field has no classical interpretation in the sense of a directly observable field configuration. Classical descriptions of electron and positron beams as closed systems free from their ambient electromagnetic fields are therefore purely statistical: individual particles can be sampled from a macroscopic classical system and their momenta measured.\footnote{The experiments of Thomson and Millikan provide direct
verification of this~\cite{Thomson:1897,Millikan:1913}.} The spectral state is precisely this statistical description, encoding the beam as a random sample from a momentum distribution.

The Pauli exclusion principle, encoded in $\varepsilon(\bm{P}^{(m)})$, further distinguishes the electronic/positronic case: because no two electrons or positrons may occupy the same momentum state, electronic and positronic coherent states in the photonic sense cannot be constructed directly.\footnote{One could formally introduce such states via Grassmann variables, but the resulting state loses its statistical interpretation and is therefore no longer a spectral state.} Remarkably, it is precisely this exclusion that ensures the same statistical description is valid both microscopically and macroscopically: an electron beam can be described as a collection of discrete electrons at any scale, whereas a laser pulse admits a particle description --- as a collection of photons --- only from the quantum viewpoint. This is the deepest structural difference between the two Fock spaces.

\paragraph{Electronic spectral states for inverse Compton scattering.}
Since the primary application is to model inverse Compton scattering of a laser beam by an electron beam, we record a few definitions specific to electronic spectral states. For a single electron, we write $\psi_{\text{cl}}(\bm{P})\equiv\phi_{e^{-}}(1,\bm{P})$ and define its Fourier transform as the positive-frequency part of the classical Dirac
spinor:
\begin{align}
    \Psi^{+}_{\text{cl},\alpha}(x)
    \equiv\int d^{3}\lambda_{e^{-}}(\bm{P})\,
    \psi_{\text{cl}}(\bm{P})\,u^{\pi}_{\alpha}(\bm{p})\,e^{-ipx}.
    \label{ClassicalDiracSpectrum}
\end{align}
For a fixed number $m$ of electrons, the corresponding spectral state is created by
\begin{align}
    \ket{\Psi_{\text{cl}}^{(m)}}
    \equiv c_{e^{-}}(\Psi_{\text{cl}}^{(m)})^{\dagger}\ket{0}
    =\frac{1}{\sqrt{m!}}
    \int d^{3m}\lambda_{e^{-}}(\bm{P}^{(m)})\,
    \varepsilon(\bm{P}^{(m)})
    \prod_{j=0}^{m-1}\psi_{\text{cl}}(\bm{P}_{j})\,
    c_{e^{-}}(\bm{P}_{j})^{\dagger}\ket{0}.
    \label{ElectronicSpectralStateCreation}
\end{align}
Finally, for $m=1$ the positive-frequency Dirac field operator acts on the single-electron spectral state to return the classical spinor amplitude:
\begin{align}
    \Psi^{+}_{e^{-}}(x)\ket{\Psi_{\text{cl}}}
    =\Psi^{+}_{e^{-}}(x)\,c_{e^{-}}(\Psi_{\text{cl}})^{\dagger}\ket{0}
    =\Psi^{+}_{\text{cl}}(x)\ket{\Psi_{\text{cl}}},
    \label{ElectronicSpectralStateFieldAbsorptionOperator}
\end{align}
the electronic analogue of the coherent state eigenvalue property
Eq.~\eqref{PhotonicCoherentStateFieldAbsorption}.

\section{Interactions and Scattering}

Computing expectation values of operators in scattered states requires specifying the Hilbert space in which those states live. We adopt the standard asymptotic assumption: long before and long after a scattering event, the system is non-interacting and may be
faithfully represented as an element of the tensor product of the three particle Fock
spaces,
\begin{align}
    \ket{\Phi_{i/f}}
    \in\mathcal{F}(\mathcal{H}_{\gamma})
    \otimes\mathcal{F}(\mathcal{H}_{e^{-}})
    \otimes\mathcal{F}(\mathcal{H}_{e^{+}}).
\end{align}
This is the setting in which the S-matrix and perturbative calculations are defined.

Before proceeding, we consolidate the notation from the individual Fock spaces into a unified scheme for the full QED Fock space. This is purely a matter of bookkeeping, but the calculations in the remainder of this section involve all three sectors simultaneously and the consolidated notation renders them considerably more compact. The following conventions hold \emph{within this section}:
\begin{enumerate}
    \item The symbols $\gamma$, $e^{-}$, and $e^{+}$ are used as subscripts to
    identify the particle species a given quantity corresponds to.

    \item The letters $P$ and $Q$ are reserved for incident and scattered momenta
    respectively. For instance, $\bm{P}_{\gamma}$ is the momentum of an incident
    photon and $\bm{Q}_{e^{-}}$ is the momentum of a scattered electron.

    \item The letters $m$ and $n$ are reserved for the number of incident and
    scattered particles respectively. For instance, $m_{e^{-}}$ denotes the number
    of incident electrons.

    \item Superscripts in parentheses denote tuples. For instance, an $m_\gamma$-tuple
    of incident photon momenta is
    $\bm{P}^{(m_{\gamma})}_{\gamma}=(\bm{P}_{\gamma,0},\ldots,\bm{P}_{\gamma,m_{\gamma}-1})$.

    \item The letters $\sigma$ and $\tau$ are reserved for permutations of incident
    and scattered momentum tuples respectively: $\sigma_{e^{-}}\in S_{m_{e^{-}}}$
    and $\tau_{\gamma}\in S_{n_{\gamma}}$. The group action is
    $\tau_{\gamma}{\cdot}\bm{Q}_{\gamma}^{(n_{\gamma})}
    =(\bm{Q}_{\gamma,\tau_{\gamma}(0)},\ldots,\bm{Q}_{\gamma,\tau_{\gamma}(n_{\gamma}-1)})$.

    \item Momentum tuples may be sliced:
    $\bm{P}_{e^{-}}{}_{(2)}^{(8)}=(\bm{P}_{e^{-},2},\ldots,\bm{P}_{e^{-},7})$.
    Slicing a permuted tuple is defined by
    $\bm{P}_{e^{-}}{}_{(\sigma_{e^{-}}(2))}^{(\sigma_{e^{-}}(8))}
    =(\bm{P}_{e^{-},\sigma_{e^{-}}(2)},\ldots,\bm{P}_{e^{-},\sigma_{e^{-}}(7)})$,
    with the understanding that $\sigma_{e^{-}}(2),\ldots,\sigma_{e^{-}}(7)$ need
    not be consecutive.

    \item Unless stated otherwise, reserved letters without a subscript denote triplets of the corresponding subscripted quantities: 
    $n=(n_{\gamma};n_{e^{-}};n_{e^{+}})$.

    \item Tupling distributes over triplets:
    $\bm{P}^{(m)}=(\bm{P}_{\gamma}^{(m_{\gamma})};\bm{P}_{e^{-}}^{(m_{e^{-}})};\bm{P}_{e^{+}}^{(m_{e^{+}})})$.

    \item Permutation and integer triplets compose in a component-wise manner:
    $\tau(n)=(\tau_{\gamma}(n_{\gamma});\tau_{e^{-}}(n_{e^{-}});\tau_{e^{+}}(n_{e^{+}}))$.

    \item Tuple slicing extends to triplets in a component-wise component-wise manner:
    $\bm{Q}_{(\tau(0))}^{(\tau(n))}
    =(\bm{Q}_{\gamma}{}_{(\tau_{\gamma}(0))}^{(\tau_{\gamma}(n_{\gamma}))};
    \bm{Q}_{e^{-}}{}_{(\tau_{e^{-}}(0))}^{(\tau_{e^{-}}(n_{e^{-}}))};
    \bm{Q}_{e^{+}}{}_{(\tau_{e^{+}}(0))}^{(\tau_{e^{+}}(n_{e^{+}}))})$.

    \item Arithmetic on integer and momentum triplets is component-wise:
    $m-m^{+}=(m_{\gamma}-m_{\gamma}^{+};m_{e^{-}}-m_{e^{-}}^{+};m_{e^{+}}-m_{e^{+}}^{+})$.

    \item Arithmetic between a scalar and an integer triplet is likewise
    component-wise: $m-1=(m_{\gamma}-1;m_{e^{-}}-1;m_{e^{+}}-1)$.

    \item Scalar multiplication is component-wise: $2n=(2n_{\gamma};2n_{e^{-}};2n_{e^{+}})$.

    \item The factorial of an integer triplet is the product of the component
    factorials: $n!=n_{\gamma}!\,n_{e^{-}}!\,n_{e^{+}}!$.

    \item The combined energy factor is defined as
    \begin{align*}
        E_{\bm{p}_{(0)}^{(m)}}
        =\prod_{j_{\gamma}=0}^{m_{\gamma}-1}\omega_{\bm{p}_{\gamma,j_\gamma}}
        \prod_{j_{e^{-}}=0}^{m_{e^{-}}-1}E_{\bm{p}_{e^{-},j_{e^-}}}
        \prod_{j_{e^{+}}=0}^{m_{e^{+}}-1}E_{\bm{p}_{e^{+},j_{e^+}}}.
    \end{align*}

    \item The combined measure is defined component-wise:
    \begin{align*}
        \int d^{3m}\lambda(\bm{P}^{(m)})
        \equiv
        \int d^{3m_{\gamma}}\lambda_{\gamma}(\bm{P}_{\gamma}^{(m_{\gamma})})
        \int d^{3m_{e^{-}}}\lambda_{e^{-}}(\bm{P}_{e^{-}}^{(m_{e^{-}})})
        \int d^{3m_{e^{+}}}\lambda_{e^{+}}(\bm{P}_{e^{+}}^{(m_{e^{+}})}),
    \end{align*}
    and similarly for $d^{3m}\mu(\bm{P}^{(m)})$.

    \item Delta-functions factorize over species:
    \begin{align*}
        \delta_{m',m}
        \equiv\delta_{m'_{\gamma},m_{\gamma}}\,
        \delta_{m'_{e^{-}},m_{e^{-}}}\,
        \delta_{m'_{e^{+}},m_{e^{+}}},
    \end{align*}
    \begin{align*}
        \delta(\bm{P}'{}^{(m)},\bm{P}^{(m)})
        \equiv
        \delta_{\gamma}(\bm{P}_{\gamma}'{}^{(m_{\gamma})},\bm{P}_{\gamma}^{(m_{\gamma})})\,
        \delta_{e^{-}}(\bm{P}_{e^{-}}'{}^{(m_{e^{-}})},\bm{P}_{e^{-}}^{(m_{e^{-}})})\,
        \delta_{e^{+}}(\bm{P}_{e^{+}}'{}^{(m_{e^{+}})},\bm{P}_{e^{+}}^{(m_{e^{+}})}).
    \end{align*}
\end{enumerate}

In this consolidated notation, the state of a non-interacting quantum electrodynamical
system, $\ket{\Phi}\equiv\ket{\Phi_{\gamma};\Phi_{e^{-}};\Phi_{e^{+}}}
\equiv\ket{\Phi_{\gamma}}\otimes\ket{\Phi_{e^{-}}}\otimes\ket{\Phi_{e^{+}}}$,
takes the compact form
\begin{align}
    \ket{\Phi}
    =\sum_{m}\frac{1}{\sqrt{m!}}
    \int d^{3m}\lambda(\bm{P}^{(m)})\,
    \phi(m,\bm{P}^{(m)})\,
    \ket{\bm{P}^{(m)}},
    \label{QEDFockState}
\end{align}
and the identity on the full Fock space is
\begin{align}
    \mathbf{1}
    \equiv\mathbf{1}_{\gamma}\otimes\mathbf{1}_{e^{-}}\otimes\mathbf{1}_{e^{+}}
    =\sum_{n}\frac{1}{n!}
    \int d^{3n}\mu(\bm{Q}^{(n)})\,
    \ket{\bm{Q}^{(n)}}\bra{\bm{Q}^{(n)}}.
    \label{QEDIdentity}
\end{align}
Operators from any individual sector extend naturally to the full Fock space by
tensoring with the appropriate identities from
Eqs.~\eqref{PhotonicFockSpaceIdentity} and~\eqref{FermionicFockSpaceIdentity}. For
example, the electronic number density operator extends as
\begin{align*}
    \mathcal{N}_{e^{-}}(\bm{Q}_{e^{-}})
    \mapsto
    \mathbf{1}_{\gamma}
    \otimes\mathcal{N}_{e^{-}}(\bm{Q}_{e^{-}})
    \otimes\mathbf{1}_{e^{+}}.
\end{align*}
In what follows, we use the left-hand notation for all such extensions without further
comment.

\subsection{The S-matrix}

The scattering matrix, or S-matrix, is the unitary operator that maps asymptotic incident
free states to asymptotic scattered free states:
\begin{align}
    \mathcal{S}\ket{\Phi_{i}}=\ket{\Phi_{f}}.
    \label{SmatrixProperty}
\end{align}
In perturbative QED, the S-matrix is expanded in a Dyson series~\cite{Dyson:1949}. The interacting
Lagrangian density is $\mathcal{L}_{\text{I}}=-e\bar{\Psi}(x)\slashed{A}(x)\Psi(x)$,
and the expansion reads
\begin{align}
    \mathcal{S}
    &=\mathcal{T}\exp\!\left(i\int d^{4}x\,\mathcal{L}_{\text{I}}\right)\nonumber\\
    &=\sum_{l}\frac{(-ie)^{l}}{l!}
    \int\cdots\int d^{4}x_{l-1}\cdots d^{4}x_{0}\,
    \mathcal{T}\!\left[
    :\!\bar{\Psi}(x_{l-1})\slashed{A}(x_{l-1})\Psi(x_{l-1})\!:\cdots
    :\!\bar{\Psi}(x_{0})\slashed{A}(x_{0})\Psi(x_{0})\!:
    \right]\nonumber\\
    &=\sum_{l}\frac{(-ie)^{l}}{l!}\mathcal{S}^{(l)}\nonumber\\
    &=\sum_{l}\sum_{F_{l}}\frac{(-ie)^{l}}{l!}N(l,F_{l})\,\mathcal{S}^{(l)}_{F_{l}},
    \label{SmatrixGeneral}
\end{align}
where $\mathcal{S}^{(0)}=1$, colons denote normal ordering, and $\mathcal{T}$ denotes
time ordering. Each term is evaluated by decomposing the fields into their absorptive
and emissive parts: $A_{\gamma}=A_{\gamma}^{+}+A_{\gamma}^{-}$ and
$\Psi=\Psi_{e^{-}}^{+}+\Psi_{e^{+}}^{-}$. For a given order $l$, the inner sum runs
over classes of Feynman diagrams $F_{l}$, and $N(l,F_{l})$ is the degeneracy of the
class --- the number of distinct field contractions that yield the same diagram topology,
up to a symmetry factor. The index $l$ in $F_l$ indicates that the diagram topology
depends on the perturbative order, not that $l$ enumerates diagrams.

For a given Feynman diagram $F_l$ at order $l$, suppose the process involves
$c_{\text{EM}}$ electromagnetic field contractions, $m^{+}_{\gamma}$ photon absorptions,
$n^{-}_{\gamma}$ photon emissions, $c_{\text{D}}$ Dirac field contractions,
$m^{+}_{e^{-}}$ electron absorptions, $n^{-}_{e^{-}}$ electron emissions,
$m^{+}_{e^{+}}$ positron absorptions, and $n^{-}_{e^{+}}$ positron emissions.
These quantities satisfy\footnote{Specifying the order $l$ and the absorption and
emission numbers determines both $c_{\text{EM}}$ and $c_{\text{D}}$ uniquely.}
\begin{align}
    2l &= 2c_{\text{D}}+m^{+}_{e^{-}}+n^{-}_{e^{-}}+m^{+}_{e^{+}}+n^{-}_{e^{+}},
    \nonumber\\
    l &= 2c_{\text{EM}}+m^{+}_{\gamma}+n^{-}_{\gamma}.
    \label{FeynmanDiagramClassification}
\end{align}

We now consider the S-matrix element for this process when the incident state contains
$m_{\gamma}\geq m^{+}_{\gamma}$ photons, $m_{e^{-}}\geq m^{+}_{e^{-}}$ electrons, and
$m_{e^{+}}\geq m^{+}_{e^{+}}$ positrons, and the scattered state contains
$n_{\gamma}\geq n^{-}_{\gamma}$ photons, $n_{e^{-}}\geq n^{-}_{e^{-}}$ electrons, and
$n_{e^{+}}\geq n^{-}_{e^{+}}$ positrons. Not all particles need participate in the
interaction. The element vanishes unless the spectator counts match for each species:
$m_{\gamma}-m^{+}_{\gamma}=n_{\gamma}-n^{-}_{\gamma}$,
$m_{e^{-}}-m^{+}_{e^{-}}=n_{e^{-}}-n^{-}_{e^{-}}$, and
$m_{e^{+}}-m^{+}_{e^{+}}=n_{e^{+}}-n^{-}_{e^{+}}$.
Writing $m^{+}=(m_{\gamma}^{+};m_{e^{-}}^{+};m_{e^{+}}^{+})$ and
$n^{-}=(n_{\gamma}^{-};n_{e^{-}}^{-};n_{e^{+}}^{-})$, the general S-matrix element is
\begin{align}
    \bra{\bm{Q}^{(n)}}\mathcal{S}^{(l)}_{F_{l}}\ket{\bm{P}^{(m)}}
    &=\frac{\delta_{m-m^{+},\,n-n^{-}}}{(m-m^{+})!}
    \sum_{\sigma}\sum_{\tau}
    \sgn(\sigma_{e^{\pm}})\sgn(\tau_{e^{\pm}})\nonumber\\
    &\quad\times\delta\!\left(\bm{Q}_{(\tau(0))}^{(\tau(n-n^{-}))},
    \bm{P}_{(\sigma(0))}^{(\sigma(m-m^{+}))}\right)\nonumber\\
    &\quad\times
    \langle\bm{Q}_{(\tau(n-n^{-}))}^{(\tau(n))}
    \parallel\mathcal{S}^{(l)}_{F_{l}}\parallel
    \bm{P}_{(\sigma(m-m^{+}))}^{(\sigma(m))}\rangle,
    \label{SmatrixElementGeneral}
\end{align}
where $\sigma_{e^{\pm}}=(\sigma_{e^{-}},\sigma_{e^{+}})$ selects the fermionic
permutations, whose signs enforce the antisymmetry of the fermionic Fock space; the
double-bar notation $\langle\cdots\parallel\cdots\parallel\cdots\rangle$ indicates
the reduced matrix element in which all permutations have been accounted for and only
the external contractions remain. The delta-functions eliminate the momenta of spectator
particles --- those not participating in the interaction described by $F_l$ --- and
Eq.~\eqref{SmatrixElementGeneral} is therefore a direct generalization of the
multi-particle orthogonality relations
Eqs.~\eqref{MultiPhotonicStateOrthogonality} and~\eqref{MultiFermionicStateOrthogonality}.

In vacuum QED, the reduced matrix element always contains a factor enforcing total
four-momentum conservation. It is decomposed as
\begin{align}
    \langle\bm{Q}_{(\tau(n-n^{-}))}^{(\tau(n))}
    \parallel\mathcal{S}^{(l)}_{F_{l}}\parallel
    \bm{P}_{(\sigma(m-m^{+}))}^{(\sigma(m))}\rangle
    &=i(2\pi)^{4}
    \delta^{(4)}\!\left(
    \sum_{a}\sum_{j_{a}}q_{a,j_{a}}
    -\sum_{a}\sum_{i_{a}}p_{a,i_{a}}
    \right)\nonumber\\
    &\quad\times
    \mathcal{M}^{(l)}_{F_{l}}\!\left(
    \bm{P}_{(\sigma(m-m^{+}))}^{(\sigma(m))}
    \,\Big|\,
    \bm{Q}_{(\tau(n-n^{-}))}^{(\tau(n))}
    \right),
    \label{ScatteringAmplitudeGeneral}
\end{align}
where the sums inside the delta-function run over particle species $a$ and the
corresponding incident and scattered particle indices.

\subsection{Unitarity of the S-matrix}

The S-matrix is unitary, $\mathcal{S}^{\dagger}\mathcal{S}=\mathbf{1}$. Expanding in
powers of $e$ and applying the Cauchy product formula yields a hierarchy of order-by-order
constraints. At each order $m\neq 0$,
\begin{align}
    \mathcal{S}^{\dagger}\mathcal{S}
    =\sum_{m}\frac{(ie)^{m}}{m!}
    \sum_{l=0}^{m}(-1)^{l}\binom{m}{l}
    \mathcal{S}^{(m-l)\dagger}\mathcal{S}^{(l)}
    =\mathbf{1}
    \quad\implies\quad
    \sum_{l=0}^{m}(-1)^{l}\binom{m}{l}
    \mathcal{S}^{(m-l)\dagger}\mathcal{S}^{(l)}=0.
    \label{Unitarity1}
\end{align}
Isolating the $l=0$ and $l=m$ terms gives
\begin{align}
    \mathcal{S}^{(m)\dagger}+(-1)^{m}\mathcal{S}^{(m)}
    =-\sum_{l=1}^{m-1}(-1)^{l}\binom{m}{l}
    \mathcal{S}^{(m-l)\dagger}\mathcal{S}^{(l)}.
    \label{Unitarity2}
\end{align}

This identity is useful in the subsequent calculation of forward scattering amplitudes in Chapter~5. There, we use it along with Cutkosky cut rules~\cite{Cutkosky:1960} to avoid regularization and renormalization procedures.

\subsection{Spin and polarization sums}

In many experimental settings, the spin states of the electrons are not resolved ---
either the incident beam is unpolarized, or the detector does not distinguish final spin
states, or both. In such cases the physically relevant quantity is the
unpolarized squared amplitude, and the spin degrees of freedom can be marginalized in a way that
simplifies the calculation considerably.

To see this, consider a Feynman amplitude with a single fermion line,
\begin{align*}
    \mathcal{M}_{F}
    =\bar{u}_{\phi}(\bm{q})\,\Gamma\,u^{\pi}(\bm{p})
    =\bigl(\bar{u}_{\pi}(\bm{p})\,\widetilde{\Gamma}\,u^{\phi}(\bm{q})\bigr)^{\dagger},
\end{align*}
where $\widetilde{\Gamma}=\gamma^{0}\Gamma^{\dagger}\gamma^{0}$. The squared amplitude
is proportional to the transition probability for the process described by $F$.
Averaging over initial spins and summing over final spins\footnote{This marginalization
is handled seamlessly via the spectral state framework.} gives
\begin{align*}
    \frac{1}{2}\sum_{\phi}\sum_{\pi}|\mathcal{M}_{F}|^{2}
    =\frac{1}{2}\operatorname{Tr}\bigl[(\slashed{p}+m)\,
    \widetilde{\Gamma}\,(\slashed{q}+m)\,\Gamma\bigr],
\end{align*}
where the spin completeness relations Eq.~\eqref{DiracSpinOrthogonality} reduce the
double spin sum to a single trace.

Photon polarization sums require a more careful treatment. Consider two Feynman
amplitudes $\mathcal{M}_F$ and $\mathcal{M}_{F'}$, each with a single fermion line and
at least one photon of momentum $k$ whose polarization is to be marginalized:
\begin{align*}
    \mathcal{M}_{F}
    =\bar{A}\,\slashed{\epsilon}^{\kappa}(\bm{k})\,B
    =\varepsilon^{\kappa}_{\mu}(\bm{k})\,\mathcal{M}_{F}^{\mu},
    \qquad
    \mathcal{M}_{F'}
    =\bar{A}'\,\slashed{\epsilon}^{\kappa}(\bm{k})\,B'
    =\varepsilon^{\kappa}_{\mu'}(\bm{k})\,\mathcal{M}_{F'}^{\mu'},
\end{align*}
where $A$, $A'$, $B$, $B'$ are spinors and
$\mathcal{M}_{F}^{\mu}=\bar{A}\gamma^{\mu}B$,
$\mathcal{M}_{F'}^{\mu'}=\bar{A}'\gamma^{\mu'}B'$. The Ward identity implies that
replacing the polarization vector by the photon momentum gives zero:
$k_{\mu}\mathcal{M}_{F}^{\mu}=k_{\mu'}\mathcal{M}_{F'}^{\mu'}=0$.

The unpolarized overlap for these diagrams is
\begin{align*}
    \sum_{\kappa=1,2}\mathcal{M}_{F'}^{\dagger}\mathcal{M}_{F}
    =\sum_{\kappa=1,2}
    \varepsilon^{\kappa}_{\mu'}(\bm{k})\,\varepsilon^{\kappa}_{\mu}(\bm{k})\,
    \mathcal{M}_{F'}^{\dagger\mu'}\mathcal{M}_{F}^{\mu},
\end{align*}
where the sum is restricted to transverse polarizations $\kappa=1,2$ since only physical
polarizations are observed. As noted in Chapter~2, the orthogonality relations
Eq.~\eqref{PolarizationOrthogonality} do not apply directly here because the sum is
restricted and the index structure involves two contravariant labels. We therefore derive
the appropriate replacement. Beginning with the full polarization sum,
\begin{align*}
    \sum_{\kappa}
    \slashed{\epsilon}^{*}_{\kappa}(\bm{k})\,\Gamma\,\slashed{\epsilon}^{\kappa}(\bm{k})
    =\sum_{\kappa}
    \varepsilon^{*}{}^{\mu'}_{\kappa}(\bm{k})\,\varepsilon^{\kappa}_{\mu}(\bm{k})\,
    \gamma_{\mu'}\Gamma\gamma^{\mu}
    =\gamma_{\mu}\Gamma\gamma^{\mu}.
\end{align*}
Left-multiplying by $\gamma^{0}$ and setting $\Delta=\gamma^{0}\Gamma$,
\begin{align*}
    \sum_{\kappa}
    \slashed{\epsilon}^{\kappa}(\bm{k})\,\Delta\,\slashed{\epsilon}^{\kappa}(\bm{k})
    =\gamma^{\mu}\Delta\gamma_{\mu}.
\end{align*}
To isolate the transverse contribution, we subtract the longitudinal ($\kappa=3$) and
scalar ($\kappa=0$) terms. For on-shell momentum $k$ the longitudinal polarization is
\begin{align*}
    \varepsilon^{3}_{\mu}(\bm{k})
    =\frac{k_{\mu}-(k{\cdot}\varepsilon^{0}(\bm{k}))\varepsilon^{0}_{\mu}(\bm{k})}
    {(k{\cdot}\varepsilon^{0}(\bm{k}))},
\end{align*}
from which one derives the completeness decomposition
\begin{align*}
    \sum_{\kappa=1,2}
    \varepsilon^{\kappa}_{\mu'}(\bm{k})\,\varepsilon^{\kappa}_{\mu}(\bm{k})
    =-\eta_{\mu'\mu}
    +\frac{k_{\mu'}\varepsilon^{0}_{\mu}(\bm{k})
    +\varepsilon^{0}_{\mu'}(\bm{k})k_{\mu}}
    {(k{\cdot}\varepsilon^{0}(\bm{k}))}
    -\frac{k_{\mu'}k_{\mu}}
    {(k{\cdot}\varepsilon^{0}(\bm{k}))^{2}}.
\end{align*}
Setting $\Delta=A'\bar{A}$, contracting with $\bar{B}'$ on the left and $B$ on the right,
and applying the Ward-Takahashi identity $k_\mu \mathcal{M}_F^\mu = 0$ to eliminate the momentum
terms~\cite{Ward:1950,Takahashi:1957,Peskin:1995,Mandl:2010}, the transverse polarization sum reduces to
\begin{align*}
    \sum_{\kappa=1,2}\mathcal{M}_{F'}^{\dagger}\mathcal{M}_{F}
    =-\eta_{\mu'\mu}\,
    \mathcal{M}_{F'}^{\dagger\mu'}\mathcal{M}_{F}^{\mu}.
\end{align*}
The physical transverse polarization sum is therefore equivalent to replacing the
polarization vectors by $-\eta_{\mu'\mu}$, with the unphysical longitudinal and scalar
contributions cancelled by the Ward identity.

\section*{Summary}

This chapter has constructed the quantum electrodynamical framework from which the
scattering calculations of subsequent chapters proceed. The development followed two
parallel tracks --- one for each field --- before unifying them in the full interacting
theory.

For the electromagnetic field, canonical quantization promoted the classical Fourier
coefficients to photon creation and annihilation operators, generating the photonic Fock
space as a direct sum of symmetric multi-photon Hilbert spaces. The central object
introduced within this space is the photonic coherent state $\ket{A_{\text{cl}}}$: a
free quantum state parameterized by a classical field configuration and characterized by
the eigenvalue properties
Eqs.~\eqref{PhotonicCoherentStateAbsorption}--\eqref{PhotonicCoherentStateFieldAbsorption},
which identify the coherent state as the precise quantum realization of the classical
electromagnetic field within QED. The photon number distribution of a coherent state is
Poissonian, and its expected number density is $|a_{\text{cl}}(\bm{K})|^2$ --- the
momentum density of the classical field. For the Dirac field, anticommutation relations
generated the electronic and positronic Fock spaces as a direct sum of antisymmetric multi-particle
spaces. The corresponding object is the electronic spectral state
$\ket{\Psi_{\text{cl}}^{(m)}}$: a statistical description of a relativistic electron
beam, encoding the beam as a random sample from a momentum distribution. Unlike the
photonic coherent state, the spectral state contains a definite particle number ---
a structural consequence of the Pauli exclusion principle that simultaneously prevents
the construction of electronic coherent states and ensures the particle description
of an electron beam is valid at all scales.

The two Fock spaces were unified in the final section into the full QED tensor product
space $\mathcal{F}(\mathcal{H}_\gamma)\otimes\mathcal{F}(\mathcal{H}_{e^-})
\otimes\mathcal{F}(\mathcal{H}_{e^+})$, within which asymptotic free states are defined.
The S-matrix maps incident to scattered asymptotic states and is expanded perturbatively
in powers of $e$ via the Dyson series, with each term organized by Feynman diagrams.
The unitarity condition on this expansion yields a hierarchy of operator identities, of
which Eq.~\eqref{Unitarity2} will be applied directly to the interference term in the
scattering calculations in Chapter~5. The spin and polarization sum results ---
the trace formula for spin-summed squared amplitudes and the Ward-identity reduction
$\sum_{\kappa=1,2}\varepsilon^\kappa_{\mu'}\varepsilon^\kappa_\mu=-\eta_{\mu'\mu}$ for
physical polarizations --- complete the standard toolkit required for explicit amplitude
calculations.

With the states, operators, and perturbative machinery now in place, the following
chapter takes the specific initial state of the inverse Compton scattering experiment
--- a photonic coherent state tensored with an electronic spectral state --- and applies
the S-matrix framework to derive the scattered electron energy spectrum.
\chapter{Contemporary Models of Inverse Compton Scattering and Radiation Reaction}

\begin{figure}[h!]
    \centering
    \includegraphics[width=0.8\linewidth]{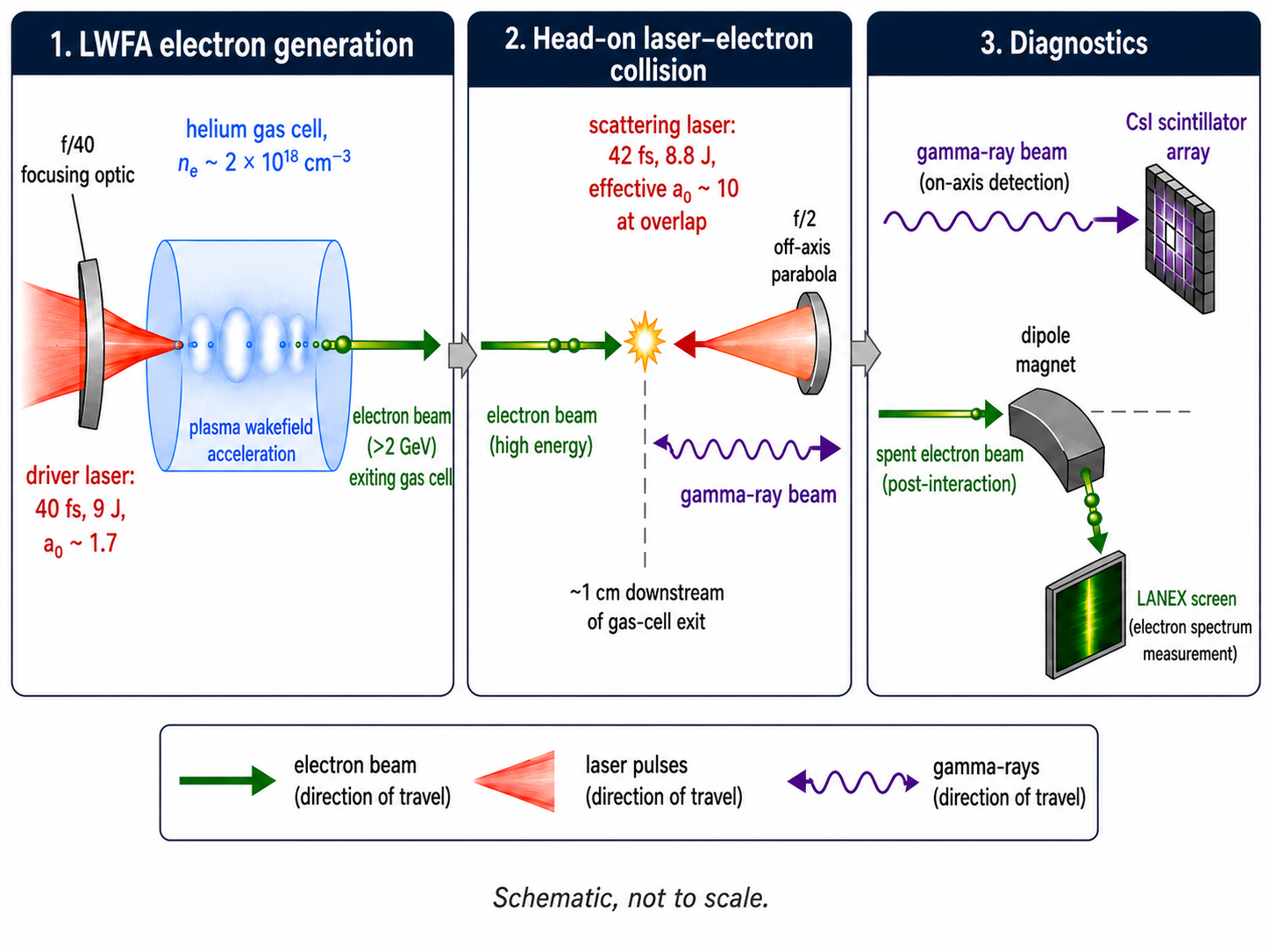}
    \caption[Experimental schematic for LWFA-ICS experiment in Ref.~\cite{Poder:2018}]{Experimental schematic for the all-optical inverse Compton scattering experiment. Figure reproduced from~\cite{Poder:2018}. An $f/40$-focused driver pulse, was incident on a helium gas cell with $n_{e}\simeq 2\times10^{18}~\text{cm}^{-3}$, generating a broadband multi-GeV electron beam. Approximately $1~\text{cm}$ downstream of the gas cell, the electrons collided head-on with an $f/2$-focused scattering pulse of duration $42\pm 3~\text{fs}$ and carrier energy $1.59~\text{eV}$, producing nonlinear Compton gamma rays and radiation-reaction energy loss. The post-interaction electrons were measured with a dipole-magnet/LANEX spectrometer, while the gamma-ray beam was recorded on axis with a CsI scintillator array.}
    \label{fig:poderexperiment}
\end{figure}

The interaction of a relativistic electron with an intense laser field is among the most richly structured problems in modern electrodynamics. At modest field strengths, the dynamics are well described by the Lorentz force alone. As the field intensity increases, however, the exchange of energy and momentum between the electron and the ambient electromagnetic field becomes non-negligible. This exchange modifies the field itself, which in turn alters the subsequent dynamics of the electron. This back-reaction --- the self-consistent coupling between the electron's motion and the field it disturbs --- is \emph{radiation reaction}, and its correct treatment is essential for modeling relativistic electrons in intense laser fields. One hundred and thirty years of theoretical effort has been spent on the underlying physical theory, and a persistent source of difficulty has been the temptation to characterize radiation reaction as a simple cumulative recoil force acting on the electron in a fixed background. This framing, while computationally convenient, obscures the fundamentally self-consistent nature of the interaction and has contributed to the slow progress of the field.

At sufficiently high intensities, quantum corrections to the classical radiation reaction picture become non-negligible. These are not standard non-perturbative quantum effects encountered in the literature such as pair production and QED cascades which lie well beyond the reach of any current model, including the one developed in this dissertation, and are negligible in the experimental regime considered here. The quantum corrections considered in this dissertation are a consequence of quantum radiation reaction. These are corrections to the emitted spectrum and the electron recoil that already appear at low orders in the coupling $e$ and are computable, to reasonable approximation, within the framework of perturbative QED. Understanding when and how these corrections become significant is the central theoretical question motivating the experiment of Poder et al.~\cite{Poder:2018} and the broader modeling programs this chapter reviews.

In this experiment, twin laser beams were delivered by the petawatt-class Astra Gemini laser system. The driver laser, a high-intensity ultrashort pulse, was focused into a gas cell target, where its intense electromagnetic field ionized the medium via tunneling ionization. The laser's ponderomotive force then displaced these free electrons, creating a plasma wake bubble that accelerated a secondary trailing electron bunch to relativistic energies over distances of millimeters to centimeters. The scattering laser was then directed head-on into the resulting relativistic electron beam, as shown in Fig.~\ref{fig:poderexperiment}. The highest-energy electrons were driven into a regime where signatures of the quantum radiation reaction are expected to be visible in the scattered electron energy spectrum.

\begin{figure}[h!]
    \centering
    \includegraphics[width=0.8\linewidth]{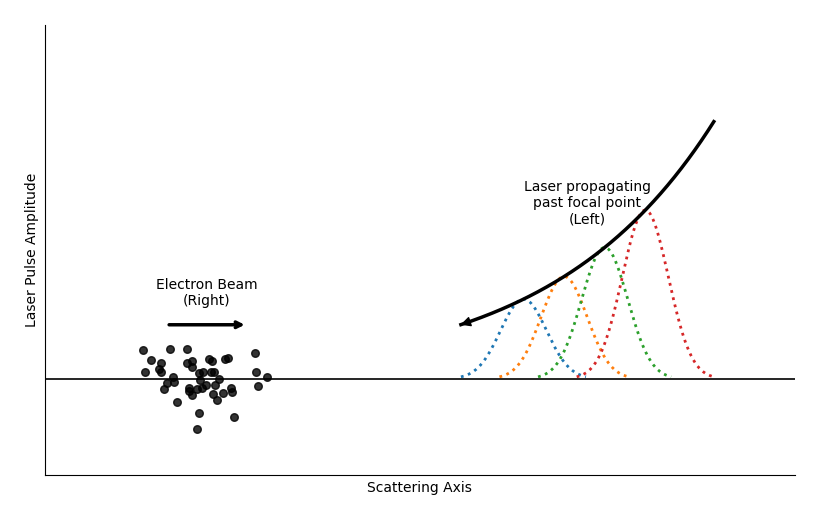}
    \caption[Visualization of an electron beam scattered by diffracting laser pulse]{An infographic depicting an electron beam moving to the right down the scattering axis towards a laser pulse propagating in the opposite direction. The laser pulse experiences field strength decay due to diffraction as it propagates past its focus. Idealized laser pulse models (e.g. plane-wave) cannot accommodate this.}
    \label{fig:LWFAICSCartoon}
\end{figure}

The electron energy spectra were compared against four models of increasing sophistication: the \emph{Lorentz-Larmor} model utilizing the Lorentz force Eq.~\eqref{LorentzForce} for electron dynamics and accounting for electron energy loss post-propagation via the relativistic Larmor formula in CED, the \emph{Landau-Lifshitz} model utilizing the Landau-Lifshitz force Eq.~\eqref{LLForce} to account for both electron dynamics and energy loss in CED, the \emph{Furry} model utilizing the locally constant field approximation (LCFA) in the Furry picture of QED, and the semiclassical \emph{Baier} model which utilizes the LCFA to rescale the Landau-Lifshitz correction to the Lorentz force. The Baier model achieved the best agreement with the data, though even this agreement is imperfect and the absolute quality of the fit in Ref.~\cite{Poder:2018} leaves room for interpretation. Poder et al.\ attributed part of the residual mismatch to a breakdown of the LCFA, which assumes photon formation times much shorter than the timescale of laser field variation --- an assumption of uncertain validity in this experiment. We note additionally that the laser pulse had propagated past its focal point before interacting with the electron beam, causing significant decay of the intensity profile throughout the interaction (see Fig.~\ref{fig:LWFAICSCartoon}); the precise range of field strengths experienced by the electrons is therefore not straightforwardly characterized by the nominal focal value, nor is it described by idealized plane-wave laser pulses --- a point we will return to when discussing model limitations in this chapter and Chapter~5.

The broader theoretical landscape is also unsatisfying. The Baier model --- the best-fitting model in the study --- is phenomenological rather than fundamental. It corrects a classical equation of motion by a quantum emission factor, and it inherits the LCFA whose validity it implicitly assumes. The Furry model, while more principled, is computationally intensive, LCFA-dependent, and underperformed the semiclassical correction in this experiment. Existing analytic QED approaches based on Volkov states or Furry-picture perturbation theory are either restricted to monochromatic fields or burdened by highly oscillatory integrals that resist analytical and numerical treatment. The field currently lacks a model that is simultaneously derived from first principles, analytically tractable for general laser pulse shapes, and free of the LCFA.

This chapter addresses both the experimental and theoretical situation. We begin by reviewing the existing modeling landscape systematically: the hierarchy of CED radiation reaction models, the background field QED (Furry picture) model, including the LCFA and its domain of validity and the stochastic QED algorithm used in the Poder et al.\ simulations, and finally the semiclassical model. In discussing CED models, we also present our own spectral pushforward model from the CED framework developed in Chapter~2 --- a closed-form analytic expression for the scattered electron energy spectrum of a single electron traversing a Gaussian laser pulse, derived from the Landau-Lifshitz equation. We end the chapter fitting our CED model to the data in Poder et al.\ and discuss this fit in the context of the simulations presented therein. The intent is not to position this single-particle Gaussian model as a direct competitor to the multi-particle simulations of Poder et al., which sample $10^7$ electrons across an energy-dependent divergence distribution and model the laser field more elaborately. Rather, the CED model presented here serves as a benchmark: a clean, analytic proof of concept demonstrating that the Landau-Lifshitz radiation reaction dynamics are correctly captured, and establishing the baseline against which the coherent-state QED framework of Chapter~5 is validated. Despite the single-particle Gaussian approximation, the fit compares favorably against the existing simulations. The comparison also suggests that the effective field strength experienced during the interaction was lower than the nominal peak value reported, consistent with the intensity decay expected from propagation past the focal point.

The successful classical comparison is not an endpoint but a foundation. As will be established in Chapter~5, the CED model developed in this chapter is the exact ultrarelativistic limit of a first-principles QED calculation based on the coherent-state framework: the two are identical to second order in perturbation theory because the ultrarelativistic limit forces the classical limit at this order. The benchmark established in this chapter therefore simultaneously grounds the spectral pushforward model empirically and provides experimental motivation for the QED extension that follows.

\section{CED Models}

This section reviews two CED models of radiation reaction. The first is the Lorentz-Larmor model based on the Lorentz force. It treats the electron trajectory and the emitted radiation separately: the trajectory is determined by the Lorentz force, and the energy radiated is accounted for post-propagation via the Larmor formula rather than as a dynamical influence on the motion itself. This approach has a long history in accelerator physics~\cite{Emma:1997,Borland:2000,Borland:2001,Demaria:2023,Li:2024}, where it provides adequate accuracy in regimes of weak emission~\cite{Heifets:2002,Poder:2018}, but it is fundamentally at odds with the self-consistent picture of radiation reaction as a continuous exchange of energy and momentum between the electron and its ambient electromagnetic field developed in Chapter~2. However, it serves as a useful baseline precisely because its failure mode is transparent: by decoupling the radiation from the dynamics, it overestimates the field strength the electron experiences and therefore overestimates the total energy loss. This manifests as an excessive downshift in the predicted scattered electron energy spectrum. This is precisely what the second model --- the Landau-Lifshitz model --- corrects. The Landau-Lifshitz model follows from the self-consistent classical equation of motion derived in Chapter~2. It is the appropriate classical description of radiation reaction~\cite{Landau:1951,Gralla:2009,DiPiazza:2008,Yaghjian2021,DiPiazza:2012,Poder:2018}. We present an analytic spectral pushforward approach to computing the scattered electron energy spectrum from the Landau-Lifshitz equation: for a linearly polarized Gaussian laser pulse and an incident electron beam described by an energy spectrum, the Landau-Lifshitz equation admits a closed-form solution whose action on the initial spectrum is a smooth deformation that can be evaluated without numerical integration of the equations of motion, stochastic sampling, or approximation of the laser pulse profile.

\subsection{Lorentz-Larmor}

The Lorentz force on an electron traversing a linearly polarized plane-wave, Eq.~\eqref{EMFieldPlaneWave}, is
\begin{align}
    \frac{dv}{ds}&=(\tilde{A}_{\text{cl}}'v)k_{L}-(k_{L}v)\tilde{A}_{\text{cl}}',
    \label{LorentzForcePlanewave}
\end{align}
where, as in Chapter~2, $\tilde{A}_{\text{cl}}=\frac{e}{mc}A_{\text{cl}}$ is the dimensionless electromagnetic field, primes denote differentiation with respect to $(\hat{k}_{L}x)$, and $v$, $k_{L}$, and $A_{\text{cl}}$ are Lorentz four-vectors. This is a separable first-order ODE whose solution, given incident four-velocity $u_{i}$, is
\begin{align}
    v=u_{i}-\tilde{A}_{\text{cl}}+\frac{2(\tilde{A}_{\text{cl}}u_{i})-\tilde{A}_{\text{cl}}^{2}}{2(k_{L}u_{i})}k_{L}.
    \label{ElectronVelocityPlanewaveLorentz}
\end{align}

The instantaneous power radiated by the electron while traversing the laser pulse is given by the Larmor formula:
\begin{align}
    P_{\text{rad}}
    =-\frac{e^{2}}{6\pi}\left(\frac{dv}{ds}\right)^{2},
    \label{LarmorRadiantPower}
\end{align}
whose textbook derivation via Liénard-Wiechert potentials is interesting in that it is entirely independent of the motion of the electron, although the final result crucially depends on it.

\paragraph{Single-particle simulation.}
Simulation of inverse Compton scattering including radiation reaction under the Lorentz-Larmor model is the simplest of the various simulations discussed in this chapter, because the model is a simulation algorithm. It proceeds via perturbative tracking and post-hoc subtraction of the radiated energy.

The electron is first propagated through the laser pulse via the Lorentz force. For any physical laser pulse, $A_{\text{cl}}\rightarrow 0$ as $t\rightarrow\infty$. So, according to Eq.~\eqref{ElectronVelocityPlanewaveLorentz}, the scattered four-momentum $q_{f}\equiv\lim_{t\rightarrow\infty}mcv$ and the incident four-momentum $p_{i}\equiv mcu_{i}$ are equal. Therefore, no net momentum is exchanged between the electron and the electromagnetic field: $\Delta q\equiv q_{f}-p_{i}=0$. Radiation reaction is absent from the dynamics entirely, which is precisely the structural deficiency of the Lorent-Larmor model: the energy loss must be introduced by hand.

The radiated energy is accounted for by integrating the relativistic Larmor formula for the instantaneous radiated power along the Lorentz-force trajectory. Restricting to an on-axis electron ($\bm{p}_{i,T}=\bm{0}_{T}$) traversing a Gaussian pulse Eq.~\eqref{GaussianPlaneWave}, and substituting Eq.~\eqref{ElectronVelocityPlanewaveLorentz} into Eq.~\eqref{LorentzForcePlanewave} and then into Eq.~\eqref{LarmorRadiantPower}, the total radiated energy is obtained by integrating. Subtracting this from the incident electron energy and taking the ultrarelativistic limit, the Lorentz-Larmor model predicts a scattered electron energy of
\begin{align}
    E_{\bm{q}}\approx E_{\bm{p}}-\frac{2E_{\bm{p}}^{2}}{E_{\bm{q},\text{max}}},
    \label{ScatteredElectronEnergyLorentz}
\end{align}
for an incident electron of energy $E_{\bm{p}}$, where
\begin{align}
    E_{\bm{q},\text{max}}
    =\frac{6\sqrt{\pi}\,\sigma_{E}\,m^{2}}{e^{2}a_{0}^{2}E_{L}^{2}}.
    \label{MaxScatteredElectronEnergy}
\end{align}
The significance of the label $E_{\bm{q},\text{max}}$ will become clear in the
following subsection, where the Landau-Lifshitz model yields a scattered spectrum that is naturally bounded above by this quantity.

\subsection{Landau-Lifshitz}

From Chapter~2, the Landau-Lifshitz equation for an electron traversing a linearly polarized plane-wave as in Eq.~\eqref{EMFieldPlaneWave} is

\begin{align}
    \frac{dv}{ds}
    &=(\tilde{A}_{\text{cl}}'v)k_{L}-(k_{L}v)\tilde{A}_{\text{cl}}'
    +\frac{2}{3}r_{e}\Bigl[
    (\tilde{A}_{\text{cl}}''v)(k_{L}v)k_{L} \nonumber\\
    &\qquad\qquad\quad
    -(k_{L}v)^{2}\tilde{A}_{\text{cl}}''
    -\tilde{A}_{\text{cl}}'^{2}(k_{L}v)k_{L}
    +\tilde{A}_{\text{cl}}'^{2}(k_{L}v)^{2}v
    \Bigr],
    \label{LandauLifshitzForcePlanewave}
\end{align}
where $r_e$ is the classical electron radius and primes denote differentiation with respect to $(\hat{k}_Lx)$. Like the Lorentz force equation Eq.~\eqref{LorentzForcePlanewave}, this is a separable first-order ODE. To express its solution compactly, define the transverse update to the four-velocity to be
\begin{align}
    \Delta v_{T}
    \equiv 
    -\frac{2}{3}r_{e}(k_{L}u_{i})
    \frac{d\tilde{A}_{\text{cl}}(k_{L}x)}{d(\hat{k}_{L}x)}
    -\int_{-\infty}^{\hat{k}_{L}x}d(\hat{k}_{L}y)\,
    \frac{(k_{L}u_{i})}{(k_{L}v)}
    \frac{d\tilde{A}_{\text{cl}}(k_{L}y)}{d(\hat{k}_{L}y)}.
    \label{ElectronTransverseVelocityChangeLL}
\end{align}
The solution for the four-velocity $v$ given incident four-velocity $u_i$ is then
\begin{subequations}
\begin{align}
    (k_{L}v)
    &=\frac{(k_{L}u_{i})}
    {1-\frac{2}{3}r_{e}(k_{L}u_{i})
    \int_{-\infty}^{\hat{k}_{L}x}d(\hat{k}_{L}y)\,
    \left(\frac{d\tilde{A}_{\text{cl}}(k_{L}y)}{d(\hat{k}_{L}y)}\right)^{2}},
    \label{ElectronForwardLightconeVelocityPlanewaveLandauLifshitz}
\end{align}
\begin{align}
    \frac{(k_{L}u_{i})}{(k_{L}v)}v
    &=u_{i}+\Delta v_{T}
    +\frac{1}{2(k_{L}u_{i})}
    \left[
    \left(\frac{k_{L}u_{i}}{k_{L}v}\right)^{2}
    -(u_{i}+\Delta v_{T})^{2}
    \right]k_{L}.
    \label{ElectronVelocityPlanewaveLandauLifshitz}
\end{align}
\end{subequations}
The net electron momentum shift to order $e^{4}$, valid for an arbitrary pulse
envelope, is
\begin{align}
    \Delta q
    \approx
    -\frac{e^{2}}{6\pi m^{2}}
    (\hat{k}_{L}p_{i})
    \int d(\hat{k}_{L}x)\,
    \left|\frac{d\tilde{A}_{\text{cl},\parallel}(k_{L}x)}
    {d(\hat{k}_{L}x)}\right|^{2}
    \left(p_{i}-\frac{m^{2}}{(\hat{k}_{L}p_{i})}\hat{k}_{L}\right).
    \label{ElectronScatteredMomentumShiftLL}
\end{align}

\paragraph{Single-particle simulation.}
Simulation of inverse Compton scattering including radiation reaction under the Landau-Lifshitz model is direct. For a Gaussian pulse, the scattered electron momentum is $q_{f}=\lim_{t\rightarrow\infty}mcv$ using Eq.~\eqref{ElectronVelocityPlanewaveLandauLifshitz}. Evaluating this limit, one finds that the scattered electron energy is related to the sampled incident energy by the rational expression
\begin{align}
    E_{\bm{q}}
    &\approx \frac{E_{\bm{p}}}{1+\dfrac{E_{\bm{p}}}{E_{\bm{q},\text{max}}}}
    \approx E_{\bm{p}}-\frac{E_{\bm{p}}^{2}}{E_{\bm{q},\text{max}}}.
    \label{ScatteredElectronEnergyLL}
\end{align}
Comparing with Eq.~\eqref{ScatteredElectronEnergyLorentz}, the Landau-Lifshitz model predicts an energy loss that is approximately half that of the Lorentz-Larmor model at the same incident energy: the factor of $E_{\bm{p}}^2$ versus $2E_{\bm{p}}^2$ in the correction term makes explicit the overestimate inherent in the post-propagation subtraction approach.

\paragraph{Spectral pushforward model.}
The relation Eq.~\eqref{ScatteredElectronEnergyLL} establishes a diffeomorphism between the domains of incident and scattered electron energies:
$E_{\bm{p}}\in[m,\infty)$ if and only if $E_{\bm{q}}\in[E_{\bm{q},\text{min}},
E_{\bm{q},\text{max}})$, where $E_{\bm{q},\text{min}}\approx m$ in the ultrarelativistic limit, confirming that $E_{\bm{q},\text{max}}$ is indeed the asymptotic maximum of the scattered energy. Inverting Eq.~\eqref{ScatteredElectronEnergyLL} gives the incident energy as a function of the scattered energy:
\begin{align}
    E_{\bm{p}}
    \approx \frac{E_{\bm{q}}}{1-\dfrac{E_{\bm{q}}}{E_{\bm{q},\text{max}}}}.
    \label{IncidentElectronEnergyLL}
\end{align}
The scattered electron energy spectrum is then the pushforward of the incident spectrum under this map. By the standard change-of-variables formula for probability densities,
\begin{align}
    f_{E_{\bm{q}_{f}}}(E_{\bm{q}})
    =f_{E_{\bm{p}_{i}}}(E_{\bm{p}})\,
    \left|\frac{dE_{\bm{p}}}{dE_{\bm{q}}}\right|,
    \label{ScatteredElectronEnergySpectrumPushforward}
\end{align}
where $E_{\bm{p}}$ is understood as a function of $E_{\bm{q}}$ via
Eq.~\eqref{IncidentElectronEnergyLL}. Substituting gives the scattered spectrum for a given electron sampled from the incident beam:
\begin{align}
    f_{E_{\bm{q}_{f}}}(E_{\bm{q}})
    \approx
    f_{E_{\bm{p}_{i}}}\!\left(\frac{E_{\bm{q}}}{1-\dfrac{E_{\bm{q}}}
    {E_{\bm{q},\text{max}}}}\right)
    \frac{1}{\left(1-\dfrac{E_{\bm{q}}}{E_{\bm{q},\text{max}}}\right)^{2}}.
    \label{ElectronScatteredEnergySpectrumLLCED}
\end{align}

It should be noted that all other models presented in this chapter are models of individual electron dynamics, necessitating scattered electron energy spectra to be assembled via repeated simulation. The spectral pushforward model eliminates this need for simulation by deriving the predicted scattered spectrum directly for any incident electron energy distribution, given entirely in closed form with no numerical integration and no stochastic sampling.

\section{QED Model}

The classical models of the preceding section, however carefully formulated, share a common limitation: they describe the electron's dynamics and the emitted radiation through classical equations of motion rather than through the amplitude structure of QED. In the regime where the quantum nonlinearity parameter $\chi$ from Eq.~\eqref{QuantumNonlinearityParameter} becomes a non-negligible fraction of unity --- as in the Poder et al.\ experiment --- this is no longer adequate, and a treatment that accounts for the discrete, stochastic nature of photon emission within a quantum field-theoretic framework is required. The natural framework for this is QED, but the intense-field regime introduces a technical obstacle: the standard perturbative expansion of QED in powers of the coupling $e$ is organized around the free vacuum, and is not straightforwardly suited to the regime $a_{0} \gg 1$ where multi-photon processes from the background field are significant. This has motivated a different organizational principle.

The background field methodology, known in the QED literature as the Furry picture, provides this. Rather than expanding around the free vacuum, the Furry picture splits the electromagnetic field into a prescribed classical background and quantum fluctuations, and reorganizes perturbation theory so that the electron propagator is exact in the background field while the quantum fluctuations --- the emission and absorption of incoherent photons --- is treated perturbatively. The electron states in the background field are
known as Volkov states: exact solutions to the Dirac equation in an arbitrary plane-wave background, which automatically incorporate the multi-photon absorptions from the background field to all orders in $a_0$. Perturbation theory in the Furry picture is therefore an expansion in $\alpha$ at fixed $a_0$, and is designed for the regime $a_{0} \gg 1$, $\alpha \ll 1$ that characterizes modern intense-laser experiments.

This approach has provided a systematic theoretical framework, yielding formally exact results for Compton scattering in monochromatic plane-wave backgrounds and forming the bases for the LCFA models used in strong-field QED simulations. However, it is not without limitations. Volkov states are exact only for backgrounds that are true plane waves --- infinite in transverse extent and either monochromatic or of a specific pulse form --- and the extension to realistic laser pulses with finite transverse profiles and spectrally broad envelopes introduces approximations whose validity must be assessed. More significantly for practical computation, the transition amplitudes in the Furry picture involve highly oscillatory integrals over the laser phase that are analytically intractable for general pulse shapes and resist numerical treatment without approximation.

The locally constant field approximation (LCFA) is designed to resolve this oscillatory integral problem by exploiting a separation of scales available when $a_{0} \gg 1$: the formation length for photon emission is much shorter than the laser wavelength, so, within the LCFA, the electron effectively sees a static field during each emission event. The emission probability can then be computed in a constant crossed field and integrated over the classical trajectory, replacing the oscillatory phase integral with a local rate. This is the approximation underlying the Furry model of Poder et al. The LCFA is well-controlled when $a_{0} \gg 1$ and $\chi \ll 1$, but its validity becomes uncertain when either condition is marginal --- precisely the situation in the Poder et al.\ experiment, as the authors themselves note.

This section reviews the Furry picture and the LCFA, examining both their
theoretical basis. The stochastic QED framework of Poder et al.\ is discussed as the primary experimental application of these ideas. The review is intended to establish clearly what the existing QED models are, setting the stage for discussion of where their limitations lie later in this chapter and for the coherent-state framework of Chapter~5, which approaches the problem from a different direction: rather than fixing $a_{0}$ and incorporating multi-photon background-field absorptions via Volkov states, the coherent-state approach treats the laser field perturbatively in the coupling $e$ without this analyticity restriction, and is therefore applicable to general square-integrable pulse profiles and eliminates the need for the LCFA.

\subsection{Furry}

The Furry picture modifies the pre-QED and QED frameworks of Chapter~2 and Chapter~3 by splitting the electromagnetic field into a dynamic quantum fluctuation $A$ and a prescribed classical background $A_{\text{cl}}$, so that $A\mapsto A+A_{\text{cl}}$, with only the dynamic field ultimately quantized. The effect on the Dirac sector is to replace the free Dirac action Eq.~\eqref{DiracAction} with the background-field Dirac action
\begin{align}
    S_{\text{D}}\equiv\int d^{4}x\,\bar{\Psi}^{\beta}
    \left[\left(i\slashed{\partial}-e\slashed{A}_{\text{cl}}\right)^{\alpha}_{\beta}-m\delta^{\alpha}_{\beta}\right]\Psi_{\alpha},
    \label{FurryDiracAction}
\end{align}
and the corresponding equation of motion becomes
\begin{align}
    \frac{\delta S_{\text{D}}}{\delta\bar{\Psi}^{\beta}}
    =\left[\left(i\slashed{\partial}-e\slashed{A}_{\text{cl}}\right)^{\alpha}_{\beta}
    -m\delta^{\alpha}_{\beta}\right]\Psi_{\alpha}=0.
    \label{DiracEquationBackgroundField}
\end{align}
The solutions to Eq.~\eqref{DiracEquationBackgroundField} are the Volkov
states~\cite{Volkov:1935,Nikishov:1964,Ritus:1975,Berestetskii:1982,Ritus:1985,
Bagrov:1990,DiPiazza:2012,Bagrov:2014,Seipt:2017}. For a linearly polarized plane-wave background Eq.~\eqref{EMFieldPlaneWave}, they take the analytic form
\begin{align*}
    \psi_{p}(x)\,u^{\pi}_{\alpha}(\bm{p}),
    \qquad
    \bar{\psi}_{p}(x)\,\bar{v}^{\pi}_{\alpha}(\bm{p}),
\end{align*}
where the Ritus matrix is
\begin{align}
    \psi_{p}(x)=\left(1+\frac{e\slashed{k}_{L}\slashed{A}_{\text{cl}}}{2(k_{L}p)}\right)\exp\!\left[{-ipx-i\int_{-\infty}^{k_{L}x}d(k_{L}y)\,\frac{2e(pA_{\text{cl}})-e^{2}A_{\text{cl}}^{2}}{2(k_{L}p)}}\right],
    \label{VolkovState}
\end{align}
for a given spin-momentum $P=(\pi,p)$ with $u^{\pi}(\bm{p})$ the free Dirac spinor of Eq.~\eqref{DiracSpinors}. The leading factor encodes the Volkov dressing of the electron by the background field, while the exponential contains the classical Hamilton-Jacobi action of an electron in the plane wave. These are, more specifically, positive-energy (electron) Volkov in-states. Negative-energy (positron) solutions are constructed analogously, and Volkov out-states are obtained by flipping the sign of the phase integral and replacing the domain of integration with $[k_Lx,\infty)$. The quantized Dirac field in the Furry picture has the expansion
\begin{align}
    \Psi_{\alpha}(x)
    =\int d^{3}\lambda_{e^{\pm}}(\bm{P})\left[
    a^{\dagger}_{e^{+}}(\bm{P})\bar{v}^{\pi}_{\alpha}(\bm{p})\,\bar{\psi}_{p}(x)
    +a_{e^{-}}(\bm{P})\,u^{\pi}_{\alpha}(\bm{p})\,\psi_{p}(x)
    \right],
    \label{QuantizedDiracFieldFurryPicture}
\end{align}
with creation and annihilation operators satisfying the same anticommutation relations as in Chapter~3, now acting on Volkov-dressed rather than free particle states.

Interactions and scattering then proceed as in standard QED, with one critical
exception: the plane-wave background breaks full four-dimensional translation
invariance, so kinetic four-momentum is not conserved. Only the transverse  and the backward lightcone momenta are conserved; the forward lightcone momentum is exchanged with the background. This has major ramifications from an analytic and computational perspective. Nevertheless, the S-matrix elements have the standard form. For nonlinear Compton scattering, the S-matrix element is~\cite{Ritus:1985,Seipt:2017}
\begin{align}
    S_{\text{NLCS}}&=-ie\int d^{4}x\,\bar{u}_{\phi}(\bm{q})\,\bar{\psi}_{q}(x)\slashed{\varepsilon}^{*}_{\lambda}(\bm{l})e^{ilx}u^{\pi}(\bm{p})\psi_{p}(x),
    \label{SmatrixElementNLCS}
\end{align}
and the differential probability for single-photon emission is~\cite{Ritus:1985,Seipt:2017}
\begin{align}
    d\mathbb{P}
    =\frac{|S_{\text{NLCS}}|^{2}}{2(\hat{k}_{L}p)}\,d\Pi,
    \label{DifferentialSingleEmissionProbability}
\end{align}
where $d\Pi=d^{3}\mu_{\gamma}(\bm{L})\,d^{3}\mu_{e^{-}}(\bm{Q})$ is the Lorentz-invariant phase-space differential for the final-state photon and electron.

Because only three of the four momentum components are conserved, integrating out the final-state electron momentum via the three conserved delta functions leaves a residual integral over the laser phase rather than yielding a simple invariant amplitude. The probability for single emission at a given photon frequency and solid angle is accordingly~\cite{Seipt:2017}
\begin{subequations}
\begin{align}
    \frac{d^{3}\mathbb{P}}{d^{2}\omega_{\bm{l}}d\Omega_{l}}=\frac{\alpha\omega_{\bm{l}} \langle|\mathcal{M}_{\text{NLCS}}|^{2}\rangle}{16\pi^{2}(k_{L}p)(k_{L}q)},
    \label{SingleEmissionProbabilityGivenPhotonEnergy}
\end{align}
with transition amplitude
\begin{align}
    \mathcal{M}_{\text{NLCS}}&=\int d(k_{L}x)\bar{u}_{\phi}(\bm{q})\Gamma u^{\pi}(\bm{p})\exp{\left[i\int(k_{L}y)\frac{(lq_{\text{cl}})}{(k_{L}q)}\right]},
    \label{TransitionAmplitudeNLCS}
\end{align}
and dressed electron vertex function
\begin{align}
    \Gamma&=\slashed{\varepsilon}^{*}_{\lambda}+\frac{e\slashed{A}_{\text{cl}}\slashed{k}_{L}\slashed{\varepsilon}^{*}_{\lambda}}{2(k_{L}p)}-\frac{e\slashed{\varepsilon}^{*}_{\lambda}\slashed{k}_{L}\slashed{A}_{\text{cl}}}{2(k_{L}q)}.
    \label{DressedElectronVertexFunction}
\end{align}
\end{subequations}
Here $q_{\text{cl}}=mcv$, with $v$ the solution to the Lorentz force law
Eq.~\eqref{ElectronVelocityPlanewaveLorentz}, is the laser-dressed kinetic momentum of the electron, and $\langle\cdots\rangle$ denotes averaging over initial spins and summing over final spins and polarizations. The dressed vertex $\Gamma$ encodes three contributions: the bare photon emission vertex $\slashed{\varepsilon}^*_\lambda$, and two Volkov dressing terms from the incoming and outgoing electron lines respectively, representing background-field photon absorptions before and after the emission event. The phase integral in $\mathcal{M}_{\text{NLCS}}$ is the Volkov phase, and it is here
that the highly oscillatory integrals characteristic of the Furry picture appear: for a general pulse envelope, $|\mathcal{M}_{\text{NLCS}}|^2$ is a double phase integral that is analytically intractable without further approximation.

\paragraph{Locally constant field approximation.}
In the LCFA, the laser field is assumed to be approximately constant over the photon formation time $\tau_{\text{form}}\sim \frac{1}{a_{0}\omega_{\bm{k}_{L}}}$. Under this assumption, the Taylor expansion of $A_{\text{cl}}$ around the average phase point is valid to linear order, which collapses the double phase integral in Eq.~\eqref{SingleEmissionProbabilityGivenPhotonEnergy} via a stationary phase
argument. The residual integral over the separation in laser phase reduces to a cubic phase integral, which evaluates to modified Bessel functions. After also integrating over the emission solid angle using the ultrarelativistic collinear approximation, the differential probability to leading order takes the form~\cite{Ritus:1985,Blackburn:2020,Blackburn:2026}
\begin{align}
    \frac{d\mathbb{P}(\chi)}{df}\approx\int d\tau\;\underbrace{\frac{\alpha}{\sqrt{3}\pi\gamma\tau_{\text{C}}}\left[\left(1-f+\frac{1}{1-f}\right)K_{\frac{2}{3}}(\xi)-\int_{\xi}^{\infty}d\eta\,K_{\frac{1}{3}}(\eta)\right]}_{%
    \displaystyle\frac{d^{2}\mathbb{P}(\chi)}{df\,d\tau}},
    \label{LCFADifferentialProbability}
\end{align}
where $\tau_{\text{C}}=\frac{\hslash}{mc^{2}}$ is the Compton time, $f=\frac{\omega_{\bm{l}}}{E_{\bm{p}}}$ is the ratio of the emitted photon energy to the incident electron energy, and $\xi=\frac{2f}{3\chi(1-f)}$. The integrand $\frac{d^{2}\mathbb{P}(\chi)}{df\,d\tau}$ is the joint probabilistic rate of emitting a photon with energy fraction $f$ per unit proper time --- the local emission rate for photons of a given fraction of incident electron energy at the instantaneous value of $\chi$. This is Ritus's exact result in a constant crossed field~\cite{Ritus:1985}, promoted to a local rate by the LCFA. Marginalizing $f$ gives the local emission rate for photons of any energy:
\begin{align}
    \frac{d\mathbb{P}(\chi)}{d\tau}=\int_{0}^{1} df\;\frac{\alpha}{\sqrt{3}\pi\gamma\tau_{\text{C}}}\left[\left(1-f+\frac{1}{1-f}\right)K_{\frac{2}{3}}(\xi)-\int_{\xi}^{\infty}d\eta\,K_{\frac{1}{3}}(\eta)\right],
    \label{ProbabilisticEmissionRateLCFA}
\end{align}
which has no closed form in terms of elementary functions and must be tabulated numerically as a function of $\chi$ prior to simulation.

\paragraph{Single-particle simulation.}
Similar to the Lorentz-Larmor model, the simulation of the Furry model also implements perturbative tracking along with post-propagation subtraction~\cite{Elkina:2011,Duclous:2011,Ridgers:2014}: the electron trajectory is determined by the Lorentz force law, but energy loss is accounted for post-hoc at each timestep via a Monte Carlo procedure incorporating stochastic emissions. Prior to the simulation, Eq.~\eqref{ProbabilisticEmissionRateLCFA} is evaluated over a dense grid in $\chi$ and cached. At each timestep $\Delta\tau$, the local value of $\chi$ is computed from the electron's current momentum and the local field. Since $\chi$ is locally constant over timescales shorter than the formation time $\tau_{\text{form}}$, so is $\frac{d\mathbb{P}(\chi)}{d\tau}$, and the probability of emitting a photon during $\Delta\tau$ is simply $\mathbb{P}(\chi)=\frac{d\mathbb{P}(\chi)}{d\tau}\Delta\tau$, which requires $\Delta\tau\ll\tau_{\text{form}}$. For the Monte Carlo procedure, a pseudorandom number $r\in[0,1]$ is drawn; if $r<\mathbb{P}(\chi)$, a photon is emitted. If emission occurs, the energy fraction $f$ is sampled from the conditional distribution
\begin{align*}
    \left(\frac{d\mathbb{P}(\chi)}{d\tau}\right)^{-1}
    \frac{d^{2}\mathbb{P}(\chi)}{df\,d\tau}
\end{align*}
via inverse transform sampling. The photon is assumed to be emitted parallel to the electron velocity --- justified by the ultrarelativistic approximation --- and the electron momentum is updated by subtracting the emitted photon four-momentum. 

Some implementations use an optical depth formulation~\cite{Los:2026}: each electron is assigned a random target optical depth at the start, the cumulative emission probability $\int \frac{d\mathbb{P}(\chi)}{d\tau}d\tau$
is tracked across multiple time steps on which it is piecewise constant, and emission is triggered when the accumulation reaches the target optical depth, after which a new random target is assigned.

\section{Semiclassical Model}

Combining the classical Landau-Lifshitz and LCFA frameworks of the preceding sections yields semiclassical Baier model. It is commonly claimed that the motivation for the development of this model is that the Landau-Lifshitz model allows for emitted radiation with unbounded frequency, consequently allowing the electron to radiate more energy than it initially possesses~\cite{Poder:2018}. This claim is incorrect. As detailed in Chapter~2, the Landau-Lifshitz equation is derived as a self-consistent statement of momentum conservation between the electron and its ambient electromagnetic field --- the radiation reaction is precisely this momentum exchange.\footnote{In fact, any self-consistent, physically valid force law is a statement of momentum conservation. The claim is only valid for externally driven dynamical equations in which it is implicitly assumed that the driving agent draws from an energy reservoir that cannot be depleted by the internal dynamical system. The Lorentz force law is one such equation, as is seen from Eq.~\eqref{ElectronVelocityPlanewaveLorentz}.} Moreover, for plane-wave fields, the electron energy can be expressed in a form that is manifestly positive, as can be verified from Eq.~\eqref{ElectronVelocityPlanewaveLandauLifshitz}.

The true motivation for the development of the Baier model is to rescale the classical mean radiated power according to the mean radiated power according to the LCFA, thereby extending the validity of the Landau-Lifshitz model deeper into the quantum regime. The confusion arises because, as will be shown, the expression for the mean power superficially resembles the expected fraction of radiated energy, and under this misinterpretation it appears that this expected fraction is being rescaled.

\subsection{Baier}

As indicated, the Baier model incorporates the quantum correction to the mean radiated power as a continuous modification of the Landau-Lifshitz force. The key observation is that the LCFA emission rate Eq.~\eqref{LCFADifferentialProbability} defines a quantum mean radiated power by weighting each emission by its energy fraction $f$ and integrating:
\begin{align}
    P_{\text{q}}(\chi)
    =E_{\bm{p}}\int_{0}^{1}df\;f\;\frac{d^{2}\mathbb{P}(\chi)}{df\,d\tau}.
    \label{QuantumRadiatedPower}
\end{align}
The analogous classical mean radiated power has the same form~\cite{Blackburn:2020}
\begin{subequations}
    \begin{align}
        P_{\text{cl}}(\chi)
        =E_{\bm{p}}\int_{0}^{1}df\;f\;\frac{d^{2}\mathbb{P}_{\text{cl}}(\chi)}{df\,d\tau},
    \label{ClassicalRadiatedPower}
    \end{align}
    with
    \begin{align}
        \frac{d^{2}\mathbb{P}_{\text{cl}}(\chi)}{df\,d\tau}=\frac{\alpha}{\sqrt{3}\pi\gamma\tau_{\text{C}}}\left[2K_{\frac{2}{3}}(\zeta)-\int_{\zeta}^{\infty}d\eta\,K_{\frac{1}{3}}(\eta)\right]
    \end{align}
    and $\zeta=\frac{2f}{3\chi}$.
\end{subequations}

The ratio of the quantum radiated power to the classical radiated power
defines the Gaunt factor~\cite{Erber:1966,Baier:1998,Blackburn:2020}
\begin{align}
    g(\chi)\equiv\frac{P_{\text{q}}(\chi)}{P_{\text{cl}}(\chi)},
    \label{GauntFactor}
\end{align}
which satisfies $g(0)=1$ and decreases monotonically to zero as $\chi\rightarrow\infty$, encoding the suppression of radiated power due to the hard quantum cutoff on photon energies. Like the LCFA from which it is derived, $g(\chi)$ has no closed form and is approximated in practice by the interpolation formula~\cite{Erber:1966,Baier:1998,Blackburn:2020}
\begin{align}
    g(\chi)
    \approx\frac{1}{\left[1+4.8(1+\chi)\ln(1+1.7\chi)+2.44\chi^{2}\right]^{2/3}},
    \label{GauntFactorInterpolation}
\end{align}
accurate to better than $2\%$ for all $\chi$.

The Baier model then replaces the Landau-Lifshitz radiation reaction force with its $g(\chi)$-weighted counterpart, so that continuous classical emission is rescaled pointwise by the local quantum correction. The resulting equation of motion for the electron traversing a linearly polarized plane-wave laser pulse is
\begin{align}
    \frac{dv}{ds}
    &=(\tilde{A}_{\text{cl}}'v)k_{L}-(k_{L}v)\tilde{A}_{\text{cl}}'
    +\frac{2}{3}r_{e}g(\chi)\Bigl[
    (\tilde{A}_{\text{cl}}''v)(k_{L}v)k_{L} \nonumber\\
    &\qquad\qquad\qquad
    -(k_{L}v)^{2}\tilde{A}_{\text{cl}}''
    -\tilde{A}_{\text{cl}}'^{2}(k_{L}v)k_{L}
    +\tilde{A}_{\text{cl}}'^{2}(k_{L}v)^{2}v
    \Bigr].
    \label{SemiclassicalLandauLifshitzForcePlanewave}
\end{align}

\paragraph{Single-particle simulation.}
Similar to the Landau-Lifshitz model, on which it is based, simulation is direct. Here, the electron is propagated by Eq.~\eqref{SemiclassicalLandauLifshitzForcePlanewave}. However, unlike the Landau-Lifshitz case, there is no analytic solution, nor is there a simple analytic expression for the scattered electron energy. Therefore, the equation of motion must be solved numerically.

\section{Multi-Particle and Particle-in-Cell Simulations}

All four single-particle models reviewed in this chapter are deployed in Poder et al.\ within the same multi-particle simulation framework, in which $N = 10^7$ electrons are independently sampled and propagated. The particle-in-cell (PIC) simulation applies this same framework to the stochastic QED model with the addition of self-consistent electromagnetic field evolution, allowing collective effects to be assessed. This section describes the multi-particle framework common to all models and the PIC simulation used to validate the independent-particle assumption.

\subsection{Multi-Particle Simulation}

A beam of $N = 10^7$ electrons is generated by independently sampling each electron's energy from the experimentally measured incident energy spectrum and its two divergence angles from an energy-dependent zero-mean Gaussian with FWHM extracted from spectrometer data. The three-dimensional electron momentum is then reconstructed from the sampled energy and divergence angles. To account for the free propagation of electrons from the gas cell to the interaction point, the initial transverse spatial distribution of the bunch is obtained by assuming ballistic propagation over $1~\text{cm}$ from a pointlike source. The longitudinal distribution is taken to be Gaussian with a duration of $40~\text{fs}$ FWHM.

The laser field model is factorized into a transverse spatial part and a temporal envelope. The transverse profile is obtained by fitting the experimentally measured spatial intensity distribution with a linear superposition of two Gaussian distributions, each expanded to fifth order in the diffraction angle, and propagated according to Gaussian beam optics to model the temporal decay of the field. The temporal profile is taken to be Gaussian with a $42~\text{fs}$ FWHM. The interaction is timed so that the peak of the scattering laser meets the peak of the electron bunch $64~\text{fs}$ after the laser reaches its focal point, resulting in a reduction of the peak normalized field strength from $a_0\approx 22.5$ at focus to $a_0\approx 10$ at the interaction point, and a broadening of the transverse intensity FWHM from $2.5~\mu\text{m}$ to approximately $6.9~\mu\text{m}$.

The role of the assumed $40~\text{fs}$ bunch duration is to spread the effective field strength experienced by different electrons across the ensemble. Since the Gaussian beam profile continues to evolve after focus, an electron arriving at the interaction point at time $t$ after the bunch peak sees a transverse profile that has decayed further than one arriving earlier. The transverse position of each electron, determined by its divergence angle and its longitudinal position within the assumed Gaussian bunch, together with the Gaussian beam propagation formula, then assigns it an effective local
$a_0$. If the bunch duration were zero, all electrons would sample the same transverse profile and the longitudinal distribution would be irrelevant.

Each electron is assumed to be propagated through a Gaussian plane-wave whose effective peak field strength is determined by its transverse position and arrival time at the interaction point. This is simulated by the single-particle simulation algorithms developed within this chapter. The temporal decay of the field between successive electrons is therefore accounted for through the spread of effective field strengths across the ensemble, but the decay of the transverse profile during each individual electron's transit through the pulse is not. Incorporating this decay ocurring during the scattering process would require abandoning the plane-wave solutions entirely and performing full numerical integration of the equations of motion in a spatiotemporally varying field for each of the $10^7$ electrons --- a computationally formidable undertaking that, for the Furry model, would additionally require LCFA table evaluation and Monte Carlo sampling at
every timestep. The assumed $40~\text{fs}$ bunch duration therefore enters the simulation only through the ensemble spread of effective peak field strengths determined at the interaction point, not through any field variation experienced by individual electrons during their interaction with the pulse beyond the idealized plane-wave variation discussed in this chapter.

In all cases the scattered electron energy spectrum is assembled from the $N$ final electron energies. Since each electron is propagated independently, inter-particle electromagnetic interactions --- space charge forces, beam loading, and any collective response of the electron bunch to the laser field --- are absent by construction. As confirmed by comparison with the PIC simulation, these collective effects are negligible in the experimental conditions of Poder et al.

\subsection{Particle-in-Cell Simulation}

The PIC simulation was performed using the EPOCH code~\cite{Arber:2015}. Unlike the multi-particle codes presented in this chapter, a PIC simulation injects the laser profile at the boundary and self-consistently evolves the electromagnetic field on a spatial grid using the discretized Maxwell equations, and advances macroparticles --- each representing a large number of physical electrons --- through the Lorentz force using the field values interpolated from the grid. This allows collective electromagnetic effects to be captured, including space charge forces and the response of the electron bunch to its own radiation field.

The simulation domain extended $78.7\,\mu\text{m}$ along the laser propagation
direction, discretized over $1020$ cells, and $40\,\mu\text{m}$ in each transverse direction, discretized over $920$ cells. The electron bunch was represented by $1.5\times 10^7$ macroparticles using third-order particle weighting, and the laser and electron bunch parameters were identical to those used in the multi-particle simulation. Stochastic photon emission was incorporated at the single-macroparticle level using the same LCFA procedure: at each timestep, $\chi$ is computed for each macroparticle, the emission probability $\mathbb{P}(\chi) = (d\mathbb{P}/d\tau)\Delta\tau$ is evaluated from the precomputed LCFA tables, and photon emission and the corresponding
electron recoil are handled stochastically via Monte Carlo sampling.

The PIC and multi-particle simulation of the Furry model yield nearly identical scattered electron energy spectra~\cite{Poder:2018}. This agreement is expected: both implement the same LCFA stochastic emission model at the single-particle level, and the PIC simulation confirms that collective electromagnetic effects --- the only physical content it adds
--- are negligible under the experimental conditions. The agreement is therefore a validation of the collective-effects assumption rather than a test of the underlying emission model.

Both simulations share the same structural limitations identified in the preceding subsection. The temporal decay of the laser field as it propagates past its focal point modifies the electron dynamics in a way that cannot be correctly captured by either code: the multi-particle simulation treats each electron as independently propagating through a fixed field model, and the PIC simulation, while self-consistently evolving the field and capturing decay during transit, cannot recover the longitudinal energy-position correlation within the electron bunch that would be required to correctly assign each electron to the field strength it encounters. Both codes exclusively sample energies from the marginal incident electron energy spectrum, which is the only tractable approach for these simulations but leaves the dominant source of uncertainty unaddressed.

\section{Comparison and Motivation for Macroscopic QED}

\begin{figure}[h!]
    \centering
    \includegraphics[width=0.8\linewidth]{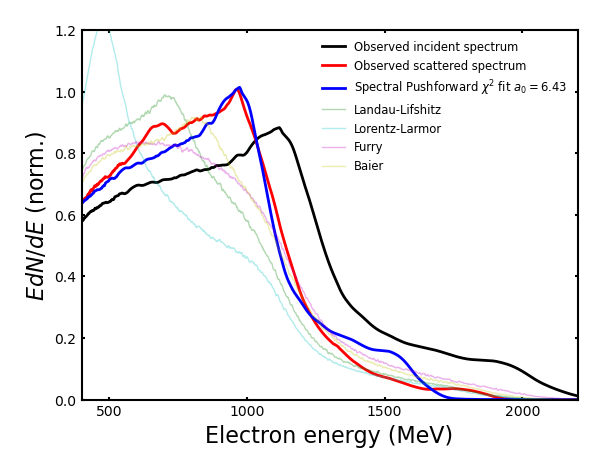}
    \caption[Chi-square fit of Landau-Lifshitz-based spectral pushforward model]{Chi-square fit of Landau-Lifshitz-based spectral pushforward model prediction of the scattered electron energy spectrum due to inverse Compton scattering of a LWFA generated electron beam (incident spectrum black) by a Gaussian-shaped laser pulse with optimal $a_{0}=6.43$ to the observed scattered spectrum (red) from the experiment in Ref.~\cite{Poder:2018} alongside simulations therein. }
    \label{fig:poderCEDfit}
\end{figure}

The four models reviewed in this chapter --- Lorentz-Larmor, Landau-Lifshitz, Baier, and Furry --- represent the current state of the art in modeling inverse Compton scattering with radiation reaction. A direct comparison of their predictions against the Poder experimental data, alongside the analytic spectral pushforward model developed in this dissertation, is shown in Fig.~\ref{fig:poderCEDfit}.

\subsection{Model Comparison}

The Lorentz-Larmor model overestimates the energy loss across the spectrum, as expected from its post-hoc accounting of radiated energy along an unmodified Lorentz force trajectory. The Landau-Lifshitz model substantially outperformed the Lorentz-Larmor model, correctly capturing the mean radiation reaction dynamics and producing a scattered spectrum that is in reasonable agreement with the data. The Baier model achieves the best statistical agreement ($R^2=96\%$), with the $g(\chi)$ correction rescaling the mean radiated power toward the quantum prediction. However, this improvement is likely an artifact of the $g(\chi)$ factor partially compensating for the overestimated field strength used in the simulation rather than correctly capturing quantum radiation reaction physics. The Furry model, despite being the most physically elaborate, performs only marginally better than the unmodified Landau-Lifshitz model ($R^2=92\%$), suggesting that the stochastic structure of individual photon emissions is not the dominant effect in this experimental regime. The spectral pushforward model, despite its single-particle Gaussian simplicity, compares favorably against all simulation results and --- uniquely among all models --- fits the effective field strength as a free parameter, pointing to a lower effective $a_0$ than the nominal value used in the simulations, consistent with the field strength decay analysis of the preceding section. The spectrum plotted in Fig.~\ref{fig:poderCEDfit} should be interpreted as the best fitting expected spectrum assuming a Gaussian laser pulse for a given electron sampled from the incident electron beam under the Landau-Lifshitz model.

\subsection{Fundamental Limitations}

Despite the breadth of approaches represented by these models, they share a set of
fundamental limitations that no amount of additional computational resources or
refinement of the existing frameworks can fully resolve. These limitations motivate not only new theoretical frameworks — the coherent-state QED model of Chapter 5 — but also new experimental geometries designed to provide a more uniform field interaction, which the coherent-state framework is uniquely positioned to describe. We now discuss these limitations in detail:

\paragraph{Single-particle dynamics versus spectral observables.}
Perhaps the most fundamental structural limitation shared by all existing models in the literature is that they are, without exception, models of single-particle dynamics, each describing the trajectory and energy loss of a single electron traversing a laser pulse. None of them directly produces a scattered electron energy spectrum --- the actual observable measured in experiment. The spectrum is instead constructed as a post-hoc statistical aggregate. For instance, in Poder et al.\ $10^7$ independent single-particle simulations are run, each producing a single scattered electron energy, and the histogram of these energies is taken as the predicted spectrum. The spectrum is therefore not a prediction of any of these models in the proper sense --- it is an empirical distribution assembled from an ensemble of single-particle predictions, and its quality depends entirely on the quality of the statistical sampling and the fidelity of the assumed beam distribution from which electrons are drawn.

This is a significant conceptual gap. A genuine spectral model takes a description of the incident beam as its input and produces the scattered spectrum as its output, without passing through the expensive intermediate step of simulating individual trajectories. The spectral pushforward model developed in this dissertation is the only existing model of this type for the Landau-Lifshitz equation: it takes the incident energy spectrum as input and produces the scattered spectrum in closed form via a change of variables, with no simulation, no sampling, and no statistical assembly. The coherent-state QED framework of Chapter~5 achieves the same for the full QED calculation: the initial state is a photonic coherent state tensored with an electronic spectral state, and the scattered electron energy spectrum follows directly from the S-matrix calculation as a closed-form expression. In neither case is it necessary to simulate individual electrons and aggregate their energies. The spectrum is a primary output of the theory, not a statistical construction built on top of it, and this is the only approach that treats the scattered spectrum as the fundamental observable it actually is.

\paragraph{The locally constant field approximation.}
The LCFA is the foundational approximation underlying both the stochastic QED model and the semiclassical model. It is valid when $a_0\gg 1$ and the photon formation length is much shorter than the scale of field variation, but in the Poder experiment $a_0\lesssim 10$ places both conditions in a regime of uncertain validity. As Poder et al.\ themselves note, the LCFA is expected to overestimate the energy loss when its validity is marginal, and this is likely a contributing factor to the mismatch between the stochastic QED model and the data. More fundamentally, the LCFA is an approximation to the exact Furry-picture transition probability that cannot be systematically improved without confronting the oscillatory integral problem it was designed to avoid. The coherent-state QED framework of Chapter~5 does not suffer from the LCFA.

\paragraph{Oscillatory integrals in background-field QED and realistic laser descriptions.}
The calculation of the probability of nonlinear Compton emission and other processes in the Furry picture involve highly oscillatory integrals, which are infamous in mathematics and physics for being resistent to analytic and numerical treatment. For a monochromatic background these integrals can be evaluated analytically via stationary phase. For a general plane-wave laser pulse, the integral is analytically intractable and resists direct numerical evaluation without approximation. For realistic laser descriptions, there is no analytic solution to the Dirac equation in the background laser field analogous to Volkov states. Such solutions only exist when the background field can be expressed as a function of a single variable, e.g. plane-waves or Coulomb fields. Thus, the consideration of realistic laser descriptions necessitates numerical solutions for the electron eigenstates, further complicating the oscillatory integral problem. Moreover, even in the context of plane-wave solutions, the LCFA is the standard resolution, but it introduces its own domain of validity and its own errors already discussed. Any model that remains within the Furry picture framework is therefore forced to either restrict to monochromatic fields, invoke the LCFA, or face the oscillatory integral problem directly, and it is not clear that the Furry picture is even viable for realistic pulse shapes. The coherent-state QED model of Chapter~5 circumvents this problem entirely.

\paragraph{Field strength decay of the laser pulse.}
None of the physical models correctly account for the continuous field strength decay of the laser field throughout the interaction, rather, with the exception of the PIC simulation, they relegate this effect to statistical tracking and modeling during simulation. The multi-particle simulations assign each electron an effective field strength based on its assumed longitudinal position within the bunch and the Gaussian beam propagation formula, treating each electron as traversing a fixed plane-wave with a constant effective $a_0$. This decay is the dominant source of field strength uncertainty in the experiment, as evidenced by the reduction from $a_0\approx 22.5$ at focus to $a_0\approx 10$ at the nominal interaction point over just $64~\text{fs}$ of propagation. The coherent-state QED framework of Chapter~5 addresses this directly: by treating the laser field as a general square-integrable classical field rather than restricting to a fixed plane-wave profile from the outset, it can accommodate a spatiotemporally decaying pulse without modification to the theoretical framework.

\paragraph{The electron ordering problem.}
The most prohibitive limitation, and the one most resistant to resolution within any existing framework, is the inability to account for the ordering of electrons through the decaying pulse during the experiment. In a LWFA-generated beam, higher-energy electrons tend to occupy the head of the bunch and encounter a stronger field than lower-energy electrons at the tail. An accurate model must account for this correlation between longitudinal position and energy. Both the multi-particle and PIC simulations exclusively sample energies from the marginal incident electron energy spectrum, which is the only tractable approach but leaves this dominant source of uncertainty unaddressed. 

It is instructive to consider what the simulation would achieve if the longitudinal position and energy were completely uncorrelated. In that case, every ordering of the $N$ electrons through the decaying pulse would be equally likely, and the simulation presented would be sampling uniformly from all $N!$ possible orderings. It, along with the spectral pushforward model, would be the most obvious digital reproductions of the experiment. But, with $N=10^7$ electrons, the probability of any single simulation reproducing the actual ordering of the physical experiment would be essentially zero, meaning that there would be no way to predict the scattered energy spectrum.

The LWFA beam is, however, not the uncorrelated case. The energy-longitudinal position correlation is real and structured: the ordering is not a uniform draw from all $N!$ permutations but is determined by the physics underlying LWFA. Thus, the simulations sample from the wrong distribution --- the uniform distribution over orderings --- when the true distribution is sharply peaked on a specific correlated structure. The simulation is therefore not merely imprecise but systematically biased in a way that depends on the energy-position correlation, whose direction and magnitude are neither measured nor modeled. This makes the situation strictly worse than the uncorrelated case: rather than achieving the best possible prediction under ignorance, the simulation is making a specific wrong assumption about the ordering that biases the result in an uncontrolled manner.

Correctly accounting for the ordering of electrons in the bunch would require knowledge of the joint longitudinal-energy distribution of the bunch, which is not directly measurable with current diagnostics, and whose resolution may be further complicated by the quantum energy-time uncertainty relation. In the absence of this information, a complete multi-particle treatment would require weighted averaging over all $N!$ possible orderings of the $N$ electrons through the decaying field --- a combinatorially intractable problem. Consequently, multi-particle and contemporary PIC simulations simply cannot predict scattered electron spectra in a LWFA-ICS without significantly stronger models of LWFA.

Crucially, however, the strong energy-position correlation that makes the current
simulations systematically biased is also the source of hope for dramatically better models. Precisely because the ordering is not random but is a physical consequence of the wakefield acceleration dynamics, it is in principle recoverable. A particle-in-cell simulation of the LWFA acceleration stage itself --- already a well-studied and increasingly well-characterized process --- could generate a realistic electron bunch with the correct joint longitudinal-energy distribution, providing a physically motivated input to the scattering simulation that replaces the assumed 40\,fs Gaussian with a beam whose ordering is determined by the beam physics rather than assumed away. This would transform the ordering problem from an intractable statistical obstacle into a well-defined modeling task, and would make the coherent-state QED framework of Chapter~5 --- which can accommodate a general momentum distribution without restriction to a factorized marginal spectrum --- the natural scattering model to pair with such
an input.

The coherent-state QED framework of Chapter~5 points toward a resolution: what is required is not a more elaborate sampling of individual electron trajectories but a genuinely macroscopic description of the electron beam as a continuous charged matter distribution, in the same sense that the laser field is described by a coherent state rather than as a collection of individual photons. The construction of such a macroscopic quantum state of charged matter, and the role of the Pauli exclusion principle in constraining it, is left as a direction for future work. However, the coherent-state framework in its current form is explicitly designed to be extendable to accommodate this consideration.

\subsection{Setting the Stage for Coherent-State QED}

The failures catalogued above share a common thread: they all arise from attempting to describe a fundamentally macroscopic quantum scattering experiment using models that are either classical with quantum corrections, or quantum at the single-particle level with classical statistics applied to the ensemble. The coherent-state QED framework developed in Chapter~5 takes a different approach from the ground up. The laser field is represented as a photonic coherent state --- the precise QED realization of a classical electromagnetic field --- and the electron beam is represented as an electronic spectral state encoding its momentum distribution as a quantum state. The S-matrix is then evaluated between these macroscopic quantum states, yielding a scattered electron energy spectrum that is derived from first principles, free of the LCFA, applicable to general square-integrable laser pulse profiles, and analytic in closed form for a Gaussian pulse. The oscillatory integrals of the Furry picture never appear because the approach avoids solving differential equations in background fields by design.
\chapter{A Macroscopic Model of Radiation Reaction via Quantum Electrodynamics}

Compton scattering --- the elastic scattering of a photon by a free electron --- is the archetypal electron-photon interaction in QED. In a realistic scattering experiment, we do not control individual particles but particle beams. As a continuation of Chapter~4, the goal of this chapter is to produce a QED model of one such experiment: the scattering of a laser beam by a relativistic electron beam, specifically the all-optical inverse Compton scattering experiment of~\cite{Poder:2018}. 

In the asymptotic past, prior to interaction, we assume that both beams admit classical descriptions which are known to us.\footnote{In practice, we have direct control over certain properties of the scattering laser --- e.g.\ the longitudinal profile at low field strengths --- and we can observe the spectrum of the electron beam.} Specifically, we take as given a classical field $A_{\text{cl}}$ for the incident laser and a spin-momentum distribution $f_{\bm{P}_{i}}$ for the incident electron beam. Since the laser field is in a high-occupation coherent state, the classical field description is justified in this regime. We further assume there are no positrons in the system.

The petawatt-class pulses delivered by Astra Gemini, while intense, have an important practical limitation for this experiment: the transverse intensity profile undergoes significant decay from its peak value, both spatially and temporally, as the pulse propagates past its focal point as shown in Fig.~1c of~\cite{Poder:2018} and discussed in previous chapters. Any adequate model of this experiment must therefore account for radiation reaction while accommodating a general pulse profile $A_{\text{cl}}$. A real electron beam may further consist of $m_{e^{-},i}$ electrons with a broad distribution of spins, energies, and incident angles, and the model should accommodate an arbitrary sample of such electrons. Beyond the free-beam case treated here, the framework should also be extendable to more general macroscopic descriptions of charged matter. Finally, in light of the computational limitations of alternative approaches discussed in the previous chapter, we require the model to be analytic and free of the highly oscillatory integrals that afflict those methods.\footnote{The classical Landau-Lifshitz model is not itself burdened by such integrals, provided one is only concerned with the scattered electron energy spectrum of a single electron. However, because the Landau-Lifshitz equation has no closed-form solution for a general field $A_{\text{cl}}$, it does not appear possible to replicate the spectral calculation in Eq.~\eqref{ElectronScatteredEnergySpectrumLLCED} numerically.} To summarize, we seek a model that is:
\begin{enumerate}
    \item capable of incorporating radiation reaction for a general, square-integrable laser field
          $A_{\text{cl}}$;
    \item applicable to an electron beam with an arbitrary spin-momentum distribution
          $f_{\bm{P}_i}$;
    \item extendable to general macroscopic descriptions of charged matter; and
    \item analytic, and free of highly oscillatory integrals.
\end{enumerate}

We present such a framework in this chapter. Notably, it will recover the Landau-Lifshitz dynamics as its classical limit --- providing not merely a computational alternative to Landau-Lifshitz dynamics, but a QED derivation of them, which establishes the regime of validity of the classical model from first principles.
 
To validate the framework against conventional models, we focus on two specific limiting cases: 
\begin{enumerate}
    \item the scattering of a single electron with a given initial momentum by a linearly polarized plane-wave laser pulse $A_{\text{cl}}$, yielding the classical Landau-Lifshitz result for the scattered momentum found in the literature and in Eq.~\eqref{ElectronScatteredMomentumShiftLL}; and
    \item the scattering of a single electron drawn from an initial energy spectrum $f_{E_{\bm{p}_i}}$\footnote{Spins and angles will be marginalized.} by a linearly polarized \emph{Gaussian} plane-wave laser pulse $A_{\text{cl}}$, yielding the Landau-Lifshitz result Eq.~\eqref{ElectronScatteredEnergySpectrumLLCED} from the preceding chapter.
\end{enumerate}
Together, these constitute a proof of concept for the far more general scenarios --- arbitrary beam distributions, realistic pulse profiles, electromagnetic interactions in macroscopic matter --- to which the framework will be extended in subsequent work.

\section{A Spectral State Scattered by a Coherent State}

The initial quantum electrodynamical state describing our system consists of a coherent photonic state tensored with a single-electron spectral state, $\ket{\Phi_{i}}=\ket{A_{\text{cl}};\Phi_{i,e^{-}};0}$, where the three entries denote the coherent laser field, the electron spectral state, and the positron vacuum, respectively. Such a state may be interpreted as the realization of a classical electrodynamical system within quantum electrodynamics. We assume the two beams collide head-on, and our objective is to predict the scattered electron energy spectrum.\footnote{We could alternatively calculate the scattered radiation energy spectrum.}

The contribution to the scattered electron momentum spectrum due to the $l^{\text{th}}$ and $l'^{\text{th}}$ order processes described by Feynman diagrams $F_{l}$ and $F'_{l'}$ is proportional to the quantity $\bra{\Phi_{i}}\mathcal{S}^{(l')\dagger}_{F'_{l'}}\mathcal{N}_{e^{-}}(\bm{Q}_{e^{-},f})\mathcal{S}^{(l)}_{F_{l}}\ket{\Phi_{i}}$. In principle, one would wish to compute the full S-matrix and allow the statistical properties of the incident beams to determine which processes dominate and which are suppressed; direct calculation of all diagrams is however infeasible, so we restrict attention to diagrams containing a single fermion line. For such diagrams, odd-order contributions vanish: a single fermion line with an odd number of external photon legs cannot simultaneously satisfy four-momentum conservation at each vertex given the on-shell conditions, and so $\mathcal{S}^{(1)}$ and $\mathcal{S}^{(3)}$ do not contribute. Accordingly, we truncate the S-matrix at fourth order in perturbation theory,
\begin{align}
    \mathcal{S}
    &\approx \sum_{l=0}^{4}\frac{(-ie)^{l}}{l!}\mathcal{S}^{(l)}\nonumber\\
    &\approx 1-\frac{e^{2}}{2!}\mathcal{S}^{(2)}+\frac{e^{4}}{4!}\mathcal{S}^{(4)}.
    \label{SmatrixExpansion}
\end{align}
The scattered electron momentum spectrum to order $e^{4}$ is then
\begin{align}
     f_{\bm{q}_{f}}(\bm{q})
    &=\bra{\Phi_{i}}\mathcal{S}^{\dagger}\mathcal{N}_{e^{-}}(\bm{q})\mathcal{S}\ket{\Phi_{i}}
    \nonumber\\
    &\approx f_{\bm{p}_{i}}(\bm{q})
    +\frac{e^{4}}{(2!)^{2}}\bra{\Phi_{i}}\mathcal{S}^{(2)\dagger}
    \mathcal{N}_{e^{-}}(\bm{q})\mathcal{S}^{(2)}\ket{\Phi_{i}}
    +\frac{2e^{4}}{4!}\,\text{Re}\bra{\Phi_{i}}\mathcal{S}^{(4)\dagger}
    \mathcal{N}_{e^{-}}(\bm{q})\ket{\Phi_{i}},
    \label{ScatteredElectronMomentumSpectrum}
\end{align}
where we have used Eq.~\eqref{ElectronicNumberDensityFermionicSpectralState} and Eq.~\eqref{Unitarity2}, and marginalized over final spins so that $\mathcal{N}_{e^{-}}(\bm{q})=\sum_{\phi}\mathcal{N}_{e^{-}}(\phi,\bm{q})$.
 
The zeroth-order term $f_{\bm{p}_{i}}(\bm{q})$ is simply the unscattered initial distribution. The order-$e^{2}$ interference term $\text{Re}\bra{\Phi_{i}}\mathcal{S}^{(2)\dagger}\mathcal{N}_{e^{-}}(\bm{q})\ket{\Phi_{i}}$ vanishes, as we demonstrate in the proceeding subsection; it is therefore omitted from Eq.~\eqref{ScatteredElectronMomentumSpectrum}. We consider the remaining $e^{4}$ contributions individually. The fourth-order interference term is further decomposed into two sub-processes:
single-photon exchange (one photon absorbed and one emitted) and two-photon exchange (two
photons absorbed and two emitted). Diagrams with no external photon legs do not contribute to the correction because they do not alter electron energy-momentum.

\subsection{The Interference Term}

\begin{figure}[htbp]
  \centering
  \begin{tikzpicture}[scale=0.72, every node/.style={font=\small}]


    \begin{feynman}
      \vertex (v1) at (0.0, 0.0);
      \vertex (v2) at (2.0, 0.0);
      \vertex (v3) at (4.0, 0.0);
      \vertex (v4) at (6.0, 0.0);

      \vertex [label={[label distance=3pt]225:$q$}] (ie) at (-1.5, -1.3);
      \vertex [label={[label distance=3pt]135:$k$}] (ig) at (-1.5,  1.3);
      \vertex [label={[label distance=3pt]315:$r$}] (oe) at ( 7.5, -1.3);
      \vertex [label={[label distance=3pt]45:$l$}]  (og) at ( 7.5,  1.3);

      \diagram*{
        (ie) -- [fermion] (v1) -- [fermion] (v2)
             -- [fermion] (v3) -- [fermion] (v4) -- [fermion] (oe),
        (ig) -- [photon] (v1),
        (v4) -- [photon] (og),
        (v2) -- [photon, half left, looseness=1.8] (v3),
      };
    \end{feynman}

    \node[font=\normalsize] at (3.3, -2.2) {(a)};


    \begin{feynman}
      \vertex (v5) at (10.5, 0.0);
      \vertex (v6) at (12.5, 0.0);
      \vertex (v7) at (14.5, 0.0);
      \vertex (v8) at (16.5, 0.0);

      \vertex [label={[label distance=3pt]225:$q$}]  (ie2) at ( 9.0, -1.3);
      \vertex [label={[label distance=3pt]135:$k$}]  (ig2) at ( 9.0,  1.3);
      \vertex [label={[label distance=3pt]315:$r$}]  (oe2) at (18.0, -1.3);
      \vertex [label={[label distance=3pt]45:$l$}]   (og2) at (18.0,  1.3);

      \vertex [label={[label distance=3pt]90:$k'$}]  (ig3) at (11.0,  1.3);
      \vertex [label={[label distance=3pt]90:$l'$}]  (og3) at (16.0,  1.3);

      \diagram*{
        (ie2) -- [fermion] (v5) -- [fermion] (v6)
              -- [fermion] (v7) -- [fermion] (v8) -- [fermion] (oe2),
        (ig2) -- [photon] (v5),
        (ig3) -- [photon] (v6),
        (v7)  -- [photon] (og3),
        (v8)  -- [photon] (og2),
      };
    \end{feynman}

    \node[font=\normalsize] at (13.6, -2.2) {(b)};

  \end{tikzpicture}
  \caption[Feynman diagrams involved in calculation of the interference term]{Fourth-order Feynman diagrams for Compton scattering
           $e^{-}(p)+\gamma(k)\to e^{-}(q)+\gamma(l)$.
           Diagram~(a): Single photon exchange: the electron absorbs $k$, undergoes a one-loop virtual photon correction, then emits $l$. There are $12$ distinct topologies for single photon exchange.
           Diagram~(b): Double photon exchange: the electron successively absorbs $k$
           and $k'$ at the first two vertices, then successively
           emits $l'$ and $l$ at the final two vertices. There are $6$ distinct topologies for double photon exchange. Note: These specific diagrams admit Cutkosky cuts, but this is not true of all topologies obtained by permuting vertices.}
  \label{fig:compton-4th-order}
\end{figure}
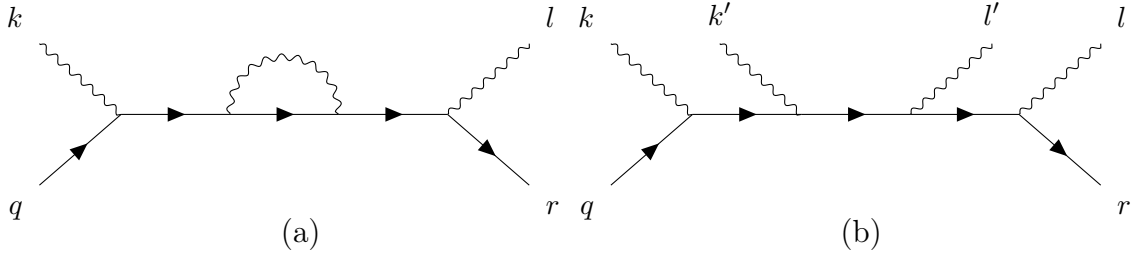

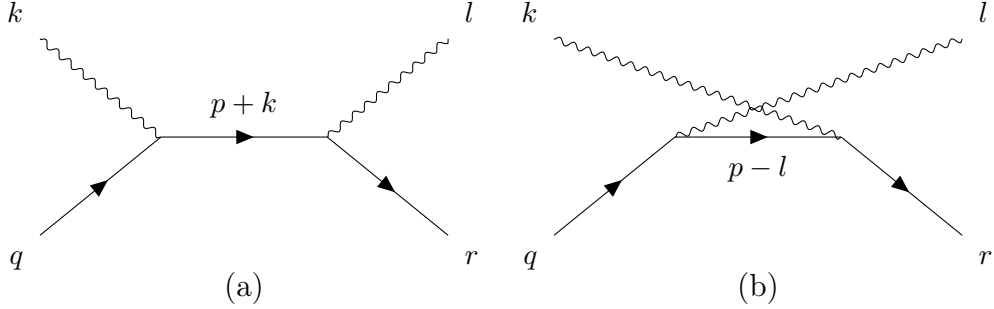
\begin{figure}[htbp]
  \centering
%
%

\begin{tikzpicture}[every node/.style={font=\small}]


  \begin{feynman}
    \vertex (a) at (0.0, 0.0);
    \vertex (b) at (2.2, 0.0);

    \vertex [label={[label distance=3pt]225:$q$}] (ie) at (-1.6, -1.3);
    \vertex [label={[label distance=3pt]135:$k$}] (ig) at (-1.6,  1.3);
    \vertex [label={[label distance=3pt]315:$r$}] (oe) at ( 3.8, -1.3);
    \vertex [label={[label distance=3pt]45:$l$}]  (og) at ( 3.8,  1.3);

    \diagram*{
      (ie) -- [fermion] (a) -- [fermion] (b) -- [fermion] (oe),
      (ig) -- [photon]  (a),
      (b)  -- [photon]  (og),
    };
  \end{feynman}

  \node[above, yshift=3pt] at (1.1, 0) {$p + k$};

  \node[font=\normalsize] at (1.1, -2.0) {(a)};


  \begin{feynman}
    \vertex (c) at ( 6.8, 0.0);
    \vertex (d) at ( 9.0, 0.0);

    \vertex [label={[label distance=3pt]225:$q$}] (ie2) at ( 5.2, -1.3);
    \vertex [label={[label distance=3pt]135:$k$}] (ig2) at ( 5.2,  1.3);
    \vertex [label={[label distance=3pt]315:$r$}] (oe2) at (10.6, -1.3);
    \vertex [label={[label distance=3pt]45:$l$}]  (og2) at (10.6,  1.3);

    \diagram*{
      (ie2) -- [fermion] (c) -- [fermion] (d) -- [fermion] (oe2),
      (ig2) -- [photon]  (d),
      (c)   -- [photon]  (og2),
    };
  \end{feynman}

  \node[below, yshift=-3pt] at (7.9, 0) {$p - l$};

  \node[font=\normalsize] at (7.9, -2.0) {(b)};

    \end{tikzpicture}
        \caption[Feynman diagrams involved in calculation of the direct term]{Second order Feynman diagrams for Compton scattering at tree-level. $e^-(p) + \gamma(k) \to e^-(q) + \gamma(l)$.
           Diagram~(a) is the $s$-channel contribution with internal momentum $q+k$;
           diagram~(b) is the $u$-channel contribution with internal momentum $q-l$.}
  \label{fig:compton-2nd-ord-tree}
\end{figure}

The structure of the interference terms indicates that it is natural to appeal to the optical theorem, which makes this calculation particularly
elegant: rather than computing forward amplitudes directly, the unitarity condition reduces the
problem to a sum over physical cut diagrams, each admitting a transparent particle-physics
interpretation. We calculate the interference terms at orders $l=2,\,4$ in
Eq.~\eqref{ScatteredElectronMomentumSpectrum}, obtaining the fourth-order contribution in closed form and establishing simultaneously that the second-order
contribution vanishes. As previously noted, we
restrict our attention to Feynman diagrams composed of a single fermion line with no positrons. At fourth order, two classes contribute:
single and double photon exchange as in Fig.~(\ref{fig:compton-4th-order}); at second
order, the relevant diagrams are precisely the tree-level Compton diagrams with a single photon
absorption and emission in Fig.~(\ref{fig:compton-2nd-ord-tree}).

We begin by expanding the spectral states in the interference matrix elements and applying the
number operator, which gives
\begin{align*}
    \bra{\Phi_{i}}\mathcal{S}^{(l)\dagger}\mathcal{N}_{e^{-}}(\bm{Q})\ket{\Phi_{i}}
    &=\frac{\psi_{\text{cl}}(\bm{Q})}{\sqrt{2E_{\bm{q}}}}
    \int d^{3}\lambda_{e^{-}}(\bm{P}')\,\psi_{\text{cl}}(\bm{P}')^{\dagger}
    \bra{A_{\text{cl}};\bm{P}';0}\mathcal{S}^{(l)\dagger}\ket{A_{\text{cl}};\bm{Q};0},
\end{align*}
and similarly for the Hermitian conjugate. Denoting the contribution to the scattered spectrum
at order $e^{l}$ due to interference by $f^{(l)}_{\bm{Q}_{f}}(\bm{Q})_{\text{I}}$,
we add the conjugate terms, integrate over the scattered momentum, and relabel
$\bm{P}'\leftrightarrow\bm{Q}$ in one copy:
\begin{align*}
    \int\frac{d^{3}\bm{Q}}{(2\pi)^{3}}
    f^{(l)}_{\bm{Q}_{f}}(\bm{Q})_{\text{I}}
    &=\int\frac{d^{3}\bm{Q}}{(2\pi)^{3}}
    \frac{e^{l}}{l!}
    \frac{\psi_{\text{cl}}(\bm{Q})^{\dagger}}{\sqrt{2E_{\bm{q}}}}
    \int d^{3}\lambda_{e^{-}}(\bm{P})\,\psi_{\text{cl}}(\bm{P})
    \bra{A_{\text{cl}};\bm{Q};0}
    \bigl(\mathcal{S}^{(l)\dagger}+\mathcal{S}^{(l)}\bigr)
    \ket{A_{\text{cl}};\bm{P};0}.
\end{align*}
The integrands on either side can differ at most by a term antisymmetric under
$\bm{P}\leftrightarrow\bm{Q}$. Since the right-hand side must be real, power-counting at the
vertices shows that this antisymmetric term vanishes identically at $l=2,4$: all real
contributions at these orders arise from diagrams whose central propagator can be placed on
shell and split into two physical sub-diagrams via the Cutkosky cutting
rules.\footnote{One could verify this diagram-by-diagram for each class listed above; our
approach handles it uniformly, with the Cutkosky interpretation emerging as a consequence
rather than an input.}

Applying the unitarity condition (Eq.~\eqref{Unitarity2}) gives
\begin{align*}
    f^{(l)}_{\bm{Q}_{f}}(\bm{Q})_{\text{I}}
    =-\frac{e^{l}}{l!}
    \frac{\psi_{\text{cl}}(\bm{Q})^{\dagger}}{\sqrt{2E_{\bm{q}}}}
    \sum_{k=1}^{l-1}(-1)^{k}\binom{l}{k}
    \int d^{3}\lambda_{e^{-}}(\bm{P})\,\psi_{\text{cl}}(\bm{P})
    \bra{A_{\text{cl}};\bm{Q};0}
    \mathcal{S}^{(l-k)\dagger}\mathcal{S}^{(k)}
    \ket{A_{\text{cl}};\bm{P};0}.
\end{align*}
At $l=2$, the interference term vanishes because $\mathcal{S}^{(1)}$ contains no physical
processes. At $l=4$, only the tree-level Compton diagrams from $\mathcal{S}^{(2)}$ survive,
selected by the valid Cutkosky cuts.

We now evaluate the non-vanishing $l=4$ interference term. Calculations involving plane-wave coherent
states proceed in two steps: first, the spectator photons are cleared from the coherent state
expansions; second, because the incident photonic state is sharply localized in transverse
momentum, a collinear delta-function identity collapses the kinematics and sets initial momenta
equal. We
insert the identity for the full photonic Fock space tensored with the single-electron Hilbert
space, defining
\begin{align}
    \mathcal{I}(\bm{P};\bm{Q};\bm{R})
    \equiv\sum_{n}\frac{1}{n!}
    \int d^{3n}\mu_{\gamma}(\bm{L}^{(n)})\,
    \bra{A_{\text{cl}};\bm{Q};0}\mathcal{S}^{(2)\dagger}_{\text{C}}
    \ket{\bm{L}^{(n)};\bm{R};0}
    \bra{\bm{L}^{(n)};\bm{R};0}\mathcal{S}^{(2)}_{\text{C}}
    \ket{A_{\text{cl}};\bm{P};0},
\end{align}
Then, noting that for the second-order Compton diagrams the symmetry factor is given by $N(2,\text{C})=2!$, the scattered electron momentum spectrum may be written in the form
\begin{align}
    f^{(4)}_{\bm{Q}_{f}}(\bm{Q})_{\text{I}}
    &=-e^{4}
    \frac{\psi_{\text{cl}}(\bm{Q})^{\dagger}}{\sqrt{2E_{\bm{q}}}}
    \int d^{3}\lambda_{e^{-}}(\bm{P})\,\psi_{\text{cl}}(\bm{P})
    \int d^{3}\mu_{e^{-}}(\bm{R})\;
    \mathcal{I}(\bm{P};\bm{Q};\bm{R}).
    \label{ElectronScatteredMomentumSpectrumInterferencePolarized}
\end{align}

Before proceeding, a remark on the structure of the calculation is in order. One might expect
that the QED momentum-space Feynman rules extend to coherent and spectral states via modified
external-leg vertex functions. This, however, fails for multi-particle macroscopic
states. As the calculation below makes explicit, an electron interacting with a coherent photon
field can absorb and re-emit a coherent photon, and such \emph{coherent subprocesses} arise
whenever spectator photons are cleared. The interference calculation via Cutkosky rules is
particularly illuminating here: placing the central propagator on shell forces the absorbed and
emitted coherent photon to carry the same energy, so the net electron momentum is unchanged by
this subprocess. We will see in the following subsection that these coherent contributions
cancel against a corresponding term from the direct calculation; whether this persists at
higher orders, where more elaborate coherent subprocesses appear, is a non-trivial question.

Using Eq.~\eqref{SmatrixGeneral} together with the absorption and emission identities in
Eqs.~\ref{PhotonicCoherentStateFieldAbsorption}--\ref{FermionicOverlap}, the S-matrix element
evaluates to
\begin{align*}
    \bra{\bm{L}^{(n)};\bm{R};0}
    \mathcal{S}^{(2)}_{\text{C}}
    \ket{A_{\text{cl}};\bm{P};0}&=iC\sum_{i=0}^{n-1}
    \Bigl(\prod_{\substack{j=0\\j\neq i}}^{n-1}
    \sqrt{2\omega_{\bm{l}_{j}}}\,a_{\text{cl}}(\bm{L}_{j})\Bigr)\\
    &\times\int d^{3}\lambda_{\gamma}(\bm{K})\,a_{\text{cl}}(\bm{K})\,
    (2\pi)^{4}\delta^{(4)}(l_{i}+r-k-p)\,
    \mathcal{M}^{(2)}_{\text{C}}(\bm{K};\bm{P}|\bm{L}_{i};\bm{R}),
\end{align*}

where all quantities have been cast into momentum space and the Compton amplitude is~\cite{Bjorken:1965,Kallen:1972,Jauch:1976,Pauli:1980,Berestetskii:1982,Thaller:1992,Weinberg:1995,Peskin:1995,Greiner:2009,Mandl:2010}
\begin{align}
    \mathcal{M}^{(2)}_{\text{C}}(\bm{K};\bm{P}|\bm{L};\bm{R})
    =\bar{u}_{\rho}(\bm{r})
    \left[
    \slashed{\epsilon}^{*}_{\lambda}(\bm{l})
    \frac{\slashed{k}+\slashed{p}+m}{2(kp)}
    \slashed{\epsilon}^{\kappa}(\bm{k})
    -\slashed{\epsilon}^{\kappa}(\bm{k})
    \frac{-\slashed{k}+\slashed{r}+m}{2(kr)}
    \slashed{\epsilon}^{*}_{\lambda}(\bm{l})
    \right]u^{\pi}(\bm{p}).
\end{align}

\paragraph{Step 1: Clearing spectator photons.}
Expanding the coherent states inside the product of S-matrix elements produces a double sum
over indices $i'$ and $i$. Splitting into diagonal ($i'=i$) and off-diagonal ($i'\neq i$)
contributions and performing a change of variables, the normalization constant and the sum
over $n$ cancel in each term. To write the result compactly, define
\begin{align}
    \mathcal{T}_{\text{C}}(\bm{K};\bm{P}|\bm{L};\bm{R})
    \equiv (2\pi)^{4}\delta^{(4)}(l+r-k-p)\,
    \mathcal{M}^{(2)}_{\text{C}}(\bm{K};\bm{P}|\bm{L};\bm{R}).
    \label{Tshorthand}
\end{align}

\begin{subequations}
The aforementioned diagonal and off-diagonal terms correspond to incoherent and coherent emission, respectively:
    \begin{align}
        \mathcal{I}(\bm{P};\bm{Q};\bm{R})&=\mathcal{I}(\bm{P};\bm{Q};\bm{R})_{\text{I}}+\mathcal{I}(\bm{P};\bm{Q};\bm{R})_{\text{C}}
    \end{align}
The term $\mathcal{I}(\bm{P};\bm{Q};\bm{R})_{\text{I}}$ corresponds to \emph{incoherent} forward scattering, arising
diagrammatically from cutting the central fermion propagator and photon loop in the
fourth-order single-photon exchange diagrams, as in Fig.~(\ref{fig:compton-4th-order}). It is given by
    \begin{align}
        \mathcal{I}(\bm{P};\bm{Q};\bm{R})_{\text{I}}=\int d^{3}\lambda_{\gamma}(\bm{K}')\int &d^{3}\lambda_{\gamma}(\bm{K})\,
    a_{\text{cl}}(\bm{K}')^{\dagger}a_{\text{cl}}(\bm{K})
    \int d^{3}\mu_{\gamma}(\bm{L})\nonumber\\
    &\times\mathcal{T}_{\text{C}}(\bm{K}';\bm{Q}|\bm{L};\bm{R})^{\dagger}
    \mathcal{T}_{\text{C}}(\bm{K};\bm{P}|\bm{L};\bm{R})
    \end{align}
The term $\mathcal{I}(\bm{P};\bm{Q};\bm{R})_{\text{C}}$ corresponds to
\emph{coherent} forward scattering, arising from cutting the central propagator in the
fourth-order double-photon exchange diagrams, as in Fig.~(\ref{fig:compton-4th-order}): a laser photon is
absorbed and re-emitted coherently with identical energy, so the net electron momentum is
unchanged. As noted above, the contribution to the scattered electron spectrum Eq.~\eqref{ElectronScatteredMomentumSpectrumInterferencePolarized} due to $\mathcal{I}(\bm{P};\bm{Q};\bm{R})_{\text{C}}$ will exactly cancel with a term from the direct
calculation. This term is given by
    \begin{align}
        \mathcal{I}(\bm{P};\bm{Q};\bm{R})_{\text{C}}=\int d^{3}\lambda_{\gamma}&(\bm{L}')\int d^{3}\lambda_{\gamma}(\bm{L})\,
        a_{\text{cl}}(\bm{L})^{\dagger}a_{\text{cl}}(\bm{L}')
        \int d^{3}\lambda_{\gamma}(\bm{K}')\int d^{3}\lambda_{\gamma}(\bm{K})\nonumber\\
        &\times a_{\text{cl}}(\bm{K}')^{\dagger}a_{\text{cl}}(\bm{K})
        \times\mathcal{T}_{\text{C}}(\bm{K}';\bm{Q}|\bm{L}';\bm{R})^{\dagger}
        \mathcal{T}_{\text{C}}(\bm{K};\bm{P}|\bm{L};\bm{R})
    \end{align}
\end{subequations}

\paragraph{Step 2: Collinear delta-function identity.}
Since the classical laser pulse envelope is a plane-wave, its Fourier coefficients are sharply localized
in transverse momentum, as in Eq.~\eqref{FourierCoefficientPlaneWave}:
\begin{align}
    a_{\text{cl}}(\bm{K})
    =(2\pi)^{2}\delta^{(2)}(\bm{k}_{T})\,
    a_{\text{cl},\parallel}(k_{\parallel})\,
    \delta^{\kappa_{L}}_{\kappa}.
\end{align}
Simultaneously, the transverse delta-functions satisfy 
\begin{align*}
    \delta^{(2)}(\bm{k}'_{T})\delta^{(2)}(\bm{k}_{T})
    =\delta^{(2)}(\bm{k}'_{T}-\bm{k}_{T})\delta^{(2)}(\bm{k}_{T})
\end{align*}
and since 
\begin{align*}
    \frac{\omega_{\bm{k}}E_{\bm{p}}}{|k_{\parallel}E_{\bm{p}}-\omega_{\bm{k}}p_{\parallel}|}
    =\frac{\omega_{\bm{k}}E_{\bm{p}}}{(kp)},
\end{align*}
for $\bm{k}_{T}=0$, one can show that
\begin{align}
    \delta^{(2)}(\bm{k}'_{T}-\bm{k}_{T})\,
    \delta^{(4)}(k+p-k'-q)
    =\frac{\omega_{\bm{k}}E_{\bm{q}}}{(kq)}\,
    \delta^{(3)}(\bm{k}'-\bm{k})\,
    \delta^{(3)}(\bm{p}-\bm{q}).
    \label{CollinearIdentity}
\end{align}
This is the collinear delta-function identity.\footnote{This approach is more rigorous than the standard impact-parameter argument~\cite{Peskin:1995}.}

Applying Eq.~\eqref{CollinearIdentity} to $\mathcal{I}(\bm{P};\bm{Q};\bm{R})_{\text{I}}$ yields
\begin{align}
    \mathcal{I}(\bm{P};\bm{Q};\bm{R})_{\text{I}}
    =&\frac{\omega_{\bm{k}}E_{\bm{q}}}{(kq)}\,
    (2\pi)^{3}\delta^{(3)}(\bm{p}-\bm{q})
    \int d\mu_{\gamma}(k_{\parallel})\,|a_{\text{cl},\parallel}(k_{\parallel})|^{2}\int d^{3}\mu_{\gamma}(\bm{l})
    \nonumber\\[2pt]
    &\times(2\pi)^{4}\delta^{(4)}(l+r-k-p)
    \sum_{\lambda}
    \mathcal{M}^{(2)}_{\text{C}}(\bm{K};\bm{Q}|\bm{L};\bm{R})^{\dagger}
    \mathcal{M}^{(2)}_{\text{C}}(\bm{K};\bm{P}|\bm{L};\bm{R}),
    \label{Iincoherent}
\end{align}
where $\bm{k}=(\bm{0}_{T},k_{\parallel})$ and $\bm{K}=(\kappa_{L},\bm{k})$. Applying the
same identity to both delta functions in $\mathcal{I}(\bm{P};\bm{Q};\bm{R})_{\text{C}}$, the factorized structure is manifest:
\begin{align}
    \mathcal{I}(\bm{P};\bm{Q};\bm{R})_{\text{C}}
    &=\left[
    \int d\mu_{\gamma}(k_{\parallel})\,
    \frac{\omega_{\bm{k}}E_{\bm{q}}}{(kq)}\,
    |a_{\text{cl}}(k_{\parallel})|^{2}\,
    (2\pi)^{3}\delta^{(3)}(\bm{r}-\bm{q})\,
    \mathcal{M}^{(2)}_{\text{C}}(\bm{K};\bm{Q}|\bm{K};\bm{R})^{\dagger}
    \right]\nonumber\\[2pt]
    &\quad\times
    \left[
    \int d\mu_{\gamma}(l_{\parallel})\,
    \frac{\omega_{\bm{l}}E_{\bm{p}}}{(lp)}\,
    |a_{\text{cl}}(l_{\parallel})|^{2}\,
    (2\pi)^{3}\delta^{(3)}(\bm{r}-\bm{p})\,
    \mathcal{M}^{(2)}_{\text{C}}(\bm{L};\bm{P}|\bm{L};\bm{R})
    \right],
    \label{Icoherent}
\end{align}
where $\bm{K}=(\kappa_{L},\bm{k})$ and $\bm{L}=(\kappa_{L},\bm{l})$ with polarization sums
performed.

\paragraph{Contributions to the scattered spectrum.}
The contributions to the unpolarized scattered electron momentum spectrum from the two
forward-scattering channels are then
\begin{align}
    f^{(4)}_{\bm{q}_{f}}(\bm{q})_{\text{II}}
    =-\frac{e^{4}}{8}\,f_{\bm{p}_{i}}(\bm{q})
    \int &\frac{dk_{\parallel}}{2\pi}\,|a_{\text{cl},\parallel}(k_{\parallel})|^{2}
    \int\frac{d^{3}\bm{l}}{(2\pi)^{3}}
    \int\frac{d^{3}\bm{r}}{(2\pi)^{3}}\nonumber\\
    &\times\frac{1}{\omega_{\bm{l}}E_{\bm{r}}(kq)}(2\pi)^{4}\delta^{(4)}(k+q-l-r)
    \,\Sigma_{\text{C}}(k,q,l,r),
    \label{ElectronScatteredMomentumSpectrumIncoherentInterference}
\end{align}
where we define the Compton kinematic factor
\begin{align}
    \Sigma_{\text{C}}(k,q,l,r)
    \equiv \frac{(kr)}{(kq)}+\frac{(kq)}{(kr)}
    -2m^{2}
    \left(
    \frac{(\varepsilon^{\kappa_{L}}(\bm{k})q)}{(kq)}
    -\frac{(\varepsilon^{\kappa_{L}}(\bm{k})r)}{(kr)}
    \right)^{2},
    \label{ComptonKinematic}
\end{align}
and
\begin{align}
    f^{(4)}_{\bm{q}_{f}}(\bm{q})_{\text{IC}}
    &=-\frac{e^{4}}{4}\,\delta^{(2)}(\bm{0}_{T})\,
    f_{\bm{p}_{i}}(\bm{q})
    \left(
    \int \frac{dk_{\parallel}}{2\pi}\,
    \frac{|a_{\text{cl}}(k_{\parallel})|^{2}}{(kq)}
    \right)^{2}.
    \label{ElectronScatteredMomentumSpectrumCoherentInterference}
\end{align}
Here we have taken the continuum limit. The coherent contribution, while formally divergent due
to the $\delta^{(2)}(\bm{0}_{T})$ factor, cancels exactly against a corresponding term in the
direct calculation; its appearance here is an artifact of decomposing the total spectrum into
interference and direct contributions.

\subsection{The Direct Term}

The calculation of the direct second-order term in Eq.~\eqref{ScatteredElectronEnergySpectrum}
mirrors the structure of the preceding section, but with a crucial difference in interpretation:
whereas the interference calculation was organized by the optical theorem and Cutkosky cuts,
here we are computing a genuine squared amplitude summed over final states. The two calculations
are therefore complementary, and we will find that their coherent contributions cancel exactly
--- confirming that coherent absorption and re-emission leaves the electron spectrum unchanged.

We insert the identity for the single-electron Hilbert space tensored with the photonic Fock
space, then apply the electronic number density operator to the scattered state, giving
\begin{align}
    f^{(4)}_{\bm{Q}_{f}}(\bm{Q})_{\text{D}}
    &=\frac{e^{4}}{(2!)^{2}}\bra{\Phi_{i}}\mathcal{S}^{(2)\dagger}
    \mathcal{N}_{e^{-}}(\bm{q})\mathcal{S}^{(2)}\ket{\Phi_{i}}\nonumber\\
    &=\frac{e^{4}}{(2!)^{2}}\frac{1}{2E_{\bm{q}}}
    \sum_{n}\frac{1}{n!}
    \int d^{3n}\mu_{\gamma}(\bm{L}^{(n)})\,
    \bigl|\bra{\bm{L}^{(n)};\bm{Q};0}
    \mathcal{S}^{(2)}\ket{\Phi_{i}}\bigr|^{2}.
\end{align}
The only diagrams contributing to the change in the spectrum are again the tree-level Compton
diagrams in Fig.~(\ref{fig:compton-2nd-ord-tree}), with symmetry factor $N(2,\text{C})=2!$. Expanding the incident spectral states and
defining
\begin{align}
    \mathcal{J}(\bm{P};\bm{P}';\bm{Q})
    \equiv
    \sum_{n}\frac{1}{n!}
    \int d^{3n}\mu_{\gamma}(\bm{L}^{(n)})\,
    \bra{A_{\text{cl}};\bm{P}';0}\mathcal{S}^{(2)\dagger}_{\text{C}}
    \ket{\bm{L}^{(n)};\bm{Q};0}
    \bra{\bm{L}^{(n)};\bm{Q};0}\mathcal{S}^{(2)}_{\text{C}}
    \ket{A_{\text{cl}};\bm{P};0},
\end{align}
the direct spectrum becomes
\begin{align}
    f^{(4)}_{\bm{Q}_{f}}(\bm{Q})_{\text{D}}
    &=\frac{e^{4}}{2E_{\bm{q}}}
    \int d^{3}\lambda_{e^{-}}(\bm{P}')\,\psi_{\text{cl}}(\bm{P}')^{\dagger}
    \int d^{3}\lambda_{e^{-}}(\bm{P})\,\psi_{\text{cl}}(\bm{P})\;
    \mathcal{J}(\bm{P};\bm{P}';\bm{Q}).
    \label{ElectronScatteredMomentumSpectrumDirectPolarized}
\end{align}
The task has thus been reduced to evaluating $\mathcal{J}(\bm{P};\bm{P}';\bm{Q})$.

\paragraph{Step 1: Clearing the spectator photons.}
As in the interference calculation, the first step is to clear the spectator photons from the
coherent state expansions. The explicit reduction, most cleanly carried out in position space
before Fourier transforming to momentum space,\footnote{The position-space route is the most
direct: one works from the definition of $\mathcal{S}^{(2)}_{\text{C}}$, clears spectators
by commuting field operators through the coherent state, and only then passes to momentum
space. This makes transparent why two distinct terms arise.} again produces an incoherent and
a coherent contribution.

The parallel with the interference calculation is structurally exact, but the role of the
electron momenta is different: here $\bm{Q}$ is the fixed final momentum, while $\bm{P}$ and
$\bm{P}'$ are the initial momenta being integrated over. Using the shorthand $\mathcal{T}_{\text{C}}$ defined in
Eq.~\eqref{Tshorthand},
\begin{align}
    \mathcal{J}(\bm{P};\bm{P}';\bm{Q})
    &=\mathcal{J}(\bm{P};\bm{P}';\bm{Q})_{\text{I}}+\mathcal{J}(\bm{P};\bm{P}';\bm{Q})_{\text{C}}.
\end{align}

The term $\mathcal{J}(\bm{P};\bm{P}';\bm{Q})_{\text{I}}$ corresponds to \emph{incoherent} direct scattering. This term arises diagrammatically from the second-order tree-level Compton diagrams by modifying the Feynman rule for incident external lines to account for absorption from the classical electromagnetic field. It is given by
\begin{align}
    \mathcal{J}(\bm{P};\bm{P}';\bm{Q})_{\text{I}}
    =\int d^{3}\lambda_{\gamma}(\bm{K}')\int &d^{3}\lambda_{\gamma}(\bm{K})\,
    a_{\text{cl}}(\bm{K}')^{\dagger}a_{\text{cl}}(\bm{K})
    \int d^{3}\mu_{\gamma}(\bm{L})\nonumber\\
    &\times\mathcal{T}_{\text{C}}(\bm{K}';\bm{P}'|\bm{L};\bm{Q})^{\dagger}
    \mathcal{T}_{\text{C}}(\bm{K};\bm{P}|\bm{L};\bm{Q}).
\end{align}

The term $\mathcal{J}(\bm{P};\bm{P}';\bm{Q})_{\text{C}}$ corresponds to \emph{coherent} direct scattering. This term arises diagrammatically from the second-order tree-level Compton diagrams by modifying the Feynman rules for incident and scattered external lines to account for absorption from and emission into the classical electromagnetic field. It is given by
\begin{align}
    \mathcal{J}(\bm{P};\bm{P}';\bm{Q})_{\text{C}}
    =\int d^{3}\lambda_{\gamma}(\bm{K}')&\int d^{3}\lambda_{\gamma}(\bm{K})\,
    a_{\text{cl}}(\bm{K}')^{\dagger}a_{\text{cl}}(\bm{K})
    \int d^{3}\lambda_{\gamma}(\bm{L}')\int d^{3}\lambda_{\gamma}(\bm{L})\nonumber\\
    &\times a_{\text{cl}}(\bm{L})^{\dagger}a_{\text{cl}}(\bm{L}')
    \mathcal{T}_{\text{C}}(\bm{K}';\bm{P}'|\bm{L}';\bm{Q})^{\dagger}
    \mathcal{T}_{\text{C}}(\bm{K};\bm{P}|\bm{L};\bm{Q}).
\end{align}

\paragraph{Step 2: Collinear delta-function identity.}
Applying the collinear identity Eq.~\eqref{CollinearIdentity} to both terms, the incoherent
contribution reduces to
\begin{align}
    \mathcal{J}(\bm{P};\bm{P}';\bm{Q})_{\text{I}}
    =&\frac{\omega_{\bm{k}}E_{\bm{p}}}{(kp)}\,
    (2\pi)^{3}\delta^{(3)}(\bm{p}'-\bm{p})
    \int d\mu_{\gamma}(k_{\parallel})\,|a_{\text{cl},\parallel}(k_{\parallel})|^{2}\int d^{3}\mu_{\gamma}(\bm{l})
    \nonumber\\[2pt]
    &\times (2\pi)^{4}\delta^{(4)}(k+p-l-q)
    \sum_{\lambda}
    \mathcal{M}^{(2)}_{\text{C}}(\bm{K};\bm{P}'|\bm{L};\bm{Q})^{\dagger}
    \mathcal{M}^{(2)}_{\text{C}}(\bm{K};\bm{P}|\bm{L};\bm{Q}),
    \label{Jincoherent}
\end{align}
with $\bm{k}=(\bm{0}_{T},k_{\parallel})$ and $\bm{K}=(\kappa_{L},\bm{k})$. The coherent
contribution factorizes as
\begin{align}
    \mathcal{J}(\bm{P};\bm{P}';\bm{Q})_{\text{C}}
    &=\left[
    \int d\mu_{\gamma}(k'_{\parallel})\,
    \frac{\omega_{\bm{k}'}E_{\bm{q}}}{(k'q)}\,
    |a_{\text{cl},\parallel}(k'_{\parallel})|^{2}\,
    (2\pi)^{3}\delta^{(3)}(\bm{p}'-\bm{q})\,
    \mathcal{M}^{(2)}_{\text{C}}(\bm{K}';\bm{P}'|\bm{K}';\bm{Q})^{\dagger}
    \right]\nonumber\\[2pt]
    &\quad\times
    \left[
    \int d\mu_{\gamma}(k_{\parallel})\,
    \frac{\omega_{\bm{k}}E_{\bm{q}}}{(kq)}\,
    |a_{\text{cl},\parallel}(k_{\parallel})|^{2}\,
    (2\pi)^{3}\delta^{(3)}(\bm{p}-\bm{q})\,
    \mathcal{M}^{(2)}_{\text{C}}(\bm{K};\bm{P}|\bm{K};\bm{Q})
    \right],
    \label{Jcoherent}
\end{align}
with $\bm{K}'=(\kappa_{L},\bm{k}')$, $\bm{K}=(\kappa_{L},\bm{k})$. Comparing with
Eq.~(\ref{Iincoherent}) and Eq.~(\ref{Icoherent}), the coherent parts $\mathcal{J}(\bm{P};\bm{P}';\bm{Q})_{\text{C}}$ and
$\mathcal{I}(\bm{P};\bm{Q};\bm{R})_{\text{C}}$ are identical up to relabeling, while the incoherent parts differ in the
placement of $\bm{P}$ and $\bm{Q}$ --- reflecting the transposition of initial and final
roles between the two calculations.

\paragraph{Contributions to the scattered spectrum.}
The incoherent direct contribution to the scattered electron momentum spectrum is
\begin{align}
    f^{(4)}_{\bm{q}_{f}}(\bm{q})_{\text{DI}}
    =\frac{e^{4}}{8E_{\bm{q}}}
    \int \frac{d^{3}\bm{p}}{(2\pi)^{3}}&\,f_{\bm{p}_{i}}(\bm{p})
    \int\frac{dk_{\parallel}}{2\pi}\,|a_{\text{cl},\parallel}(k_{\parallel})|^{2}
    \int\frac{d^{3}\bm{l}}{(2\pi)^{3}}\nonumber\\
    &\times\frac{1}{\omega_{\bm{l}}(kp)}(2\pi)^{4}\delta^{(4)}(k+p-l-q)\,
    \Sigma_{\text{C}}(k,p,l,q),
    \label{ElectronScatteredMomentumSpectrumIncoherentDirect}
\end{align}
where $\Sigma_{\text{C}}$ is the Compton kinematic factor defined in Eq.~\eqref{ComptonKinematic}.
Note that Eq.~\eqref{ElectronScatteredMomentumSpectrumIncoherentDirect} and
Eq.~\eqref{ElectronScatteredMomentumSpectrumIncoherentInterference} have the same functional
form, related by the exchange $\bm{p}\leftrightarrow\bm{q}$ in the kinematic arguments of
$\Sigma_{\text{C}}$: the direct term involves the initial electron momentum in the
denominator, while the interference term involves the final.

The coherent direct contribution is
\begin{align}
    f^{(4)}_{\bm{q}_{f}}(\bm{q})_{\text{DC}}
    &=\frac{e^{4}}{4}\,\delta^{(2)}(\bm{0}_{T})\,
    f_{\bm{p}_{i}}(\bm{q})
    \left(
    \int \frac{dk_{\parallel}}{2\pi}\,
    \frac{|a_{\text{cl},\parallel}(k_{\parallel})|^{2}}{(kq)}
    \right)^{2},
    \label{ElectronScatteredMomentumSpectrumCoherentDirect}
\end{align}
which is equal and opposite to Eq.~\eqref{ElectronScatteredMomentumSpectrumCoherentInterference}.
The two $\delta^{(2)}(\bm{0}_T)$ divergences cancel exactly. This is the precise sense in which
coherent forward scattering decouples from the spectrum: processes in which the electron
absorbs and re-emits a laser photon with no net momentum transfer contribute equally to the
direct and interference terms and cancel in total. The sole physical contributions to the
scattered electron momentum spectrum therefore come from incoherent processes --- genuine
Compton events in which the electron scatters into a new momentum eigenstate.

\section{An Individual Electron Scattered by a Coherent State}

The spectral framework developed in the preceding sections assumes access to a statistical ensemble of electron beams generated by the same source with fixed specifications --- a LINAC or a laser-wakefield accelerator with fixed tuning. In practice, measuring the incident electron energy spectrum with a detector yields a histogram, i.e.\ an empirical distribution for a
single microstate. For a sufficiently large sample of electrons from a beam with negligible internal interactions, this histogram approximates the true ensemble description: through interpolation and possible deconvolution for finite detector resolution, one recovers the expected number density of the beam.

This spectral approach is highly realistic and very well physically motivated, however, it is noticeably different from the wave-packet description of electronic states that is typically encountered in quantum physics. In fact, it is a generalization thereof. This can be seen explicitly. For simplicity, assume that the incident electron spectrum is transversely localized along the scattering axis. Imagine an ideal electron beam source that may be tuned to produce electrons in a single bandwidth $\sigma$ of energies about some mean $E_{\bm{p}}$. The wave-packet description emerges for very small $\sigma\ll 1$.


The limit $\sigma\rightarrow 0$ requires care, and in it the full spectral presentation of the previous section becomes uninteresting. Both the incident and scattered spectra collapse to delta-functions. Therefore, the natural observable in this limit is not the momentum distribution but the scattered electron momentum itself. This section shows that, to order $e^{4}$, the expected scattered momentum computed from the coherent-state QED framework agrees exactly with the solution to the Landau-Lifshitz equation in the classical limit. The model presented in this chapter thus describes quantum radiation reaction, with the classical theory emerging as its leading approximation.

\subsection{Expected Momentum Shift to Order $e^4$}

The expected on-shell four-momentum of a electron sampled from a single particle spectral state is
\begin{align}
    q_{f}=\int\frac{d^{3}\bm{q}}{(2\pi)^{3}}\,q\,f_{\bm{q}_{f}}(\bm{q}).
    \label{ExpectedScatteredMomentum}
\end{align}

To $e^{4}$, the calculation will proceed as in the previous sections in this chapter. That is, we begin with 
\begin{align}
    f_{\bm{q}_{f}}(\bm{q})\approx f_{\bm{p}_{i}}(\bm{q})+f^{(4)}_{\bm{q}_{f}}(\bm{q})_{\text{DI}}+f^{(4)}_{\bm{q}_{f}}(\bm{q})_{\text{II}},
    \label{ElectronScatteredMomentumSpectrum}
\end{align}
with the additional knowledge that there is no contribution due to coherent emission.

In substituting Eq.~\eqref{ElectronScatteredMomentumSpectrum} into Eq.~\eqref{ExpectedScatteredMomentum}, we first note that the initial zeroth-order term is simply the expected incident on-shell four-momentum:
\begin{align}
    p_{i}=\int\frac{d^{3}\bm{q}}{(2\pi)^{3}}\,q\,f_{\bm{p}_{i}}(\bm{q}).
    \label{ExpectedIncidentMomentum}
\end{align}
So, we define the momentum shift $\Delta q\equiv q_{f}-p_{i}$. This is the quantity that we desire to calculate.

To compress the expressions that follow, define the Compton phase-space kernel
\begin{align}
    \mathcal{K}_{\text{C}}(k,p;l,r)\equiv \frac{\omega_{\bm{k}}}{(kp)}\,(2\pi)^{4}\delta^{(4)}(k+p-l-r)\,\Sigma_{\text{C}}(k,p,l,r).
    \label{ComptonKernel}
\end{align}

Then, substituting  Eq.~\eqref{ElectronScatteredMomentumIncoherentInterference} into Eq.~\eqref{ElectronScatteredMomentumSpectrum} and then Eq.~\eqref{ExpectedScatteredMomentum}, the momentum contribution due to the incoherent interference term is
\begin{align}
    \int \frac{d^{3}\bm{q}}{(2\pi)^{3}}\,q\,f^{(4)}_{\bm{q}_{f}}(\bm{q})_{\text{II}}
    =-e^{4}&\int d\mu_{\gamma}(k_{\parallel})\,|a_{\text{cl},\parallel}(k_{\parallel})|^{2}
    \int d^{3}\mu_{\gamma}(\bm{l})
    \int d^{3}\mu_{e^{-}}(\bm{r})\,p_{i}\,\mathcal{K}_{\text{C}}(k,p_{i};l,r),
    \label{ElectronScatteredMomentumIncoherentInterference}
\end{align}
where we have used the fact that the electron incident momentum spectrum $f_{\mathcal{\bm{p}}_{i}}$ is now assumed to be sharply localized about $\bm{p}_{i}$.

Meanwhile, substituting  Eq.~\eqref{ElectronScatteredMomentumIncoherentInterference} into Eq.~\eqref{ElectronScatteredMomentumSpectrum} and then Eq.~\eqref{ExpectedScatteredMomentum} and applying the localization condition, we arrive at a strikingly similar expression for the momentum contribution due to the incoherent direct term. It is 
\begin{align}
    \int \frac{d^{3}\bm{q}}{(2\pi)^{3}}\,q\,f^{(4)}_{\bm{q}_{f}}(\bm{q})_{\text{DI}}
    =e^{4}&\int d\mu_{\gamma}(k_{\parallel})\,|a_{\text{cl},\parallel}(k_{\parallel})|^{2}
    \int d^{3}\mu_{\gamma}(\bm{l})
    \int d^{3}\mu_{e^{-}}(\bm{q})\,q\,\mathcal{K}_{\text{C}}(k,p;l,q).
    \label{ElectronScatteredMomentumIncoherentDirect}
\end{align}

So, the direct term, being a squared amplitude, weights the final-state momentum, while the interference term, being a forward amplitude, weights the initial.

\subsection{The Rest Frame of the Incident Electron and the Classical Limit}

Recalling Eq.~\eqref{LongitudinalFourierCoefficientPlaneWave}, $\tilde{a}_{\text{cl},\parallel}=\frac{e}{m}\, a_{\text{cl},\parallel}$ is defined so that the quantity $\sqrt{\frac{\omega_{\bm{k}}}{2}}\,\tilde{a}_{\text{cl},\parallel}$ is dimensionless, independent of $\hslash$, and a Lorentz scalar. It follows that the expressions in Eq.~\eqref{ElectronScatteredMomentumIncoherentInterference} and Eq.~\eqref{ElectronScatteredMomentumIncoherentDirect} are also invariant of Lorentz boosts along the scattering axis. Combining these expressions, and boosting into the rest frame of the incident electron, yields the simplified expression
\begin{align*}
    \Delta q'\approx 4\pi \alpha m^{2}\int d\mu_{\gamma}(k_{\parallel}'&) |\tilde{a}'_{\text{cl},\parallel}(k_{\parallel}')|^{2}\int d^{3}\mu_{\gamma}(\bm{l}')\int d^{3}\mu_{e^{-}}(\bm{q}')\,(q'-p_{i}')\,\mathcal{K}_{\text{C}}(k',p_{i}';l',q'),
\end{align*}
where the primes indicate the corresponding objects are in the rest frame of the incident electron, $\hat{l}=(1,\sin{\theta_{l'}}\cos{\phi_{l'}},\sin{\theta_{l'}}\sin{\phi_{l'}},\cos{\theta_{l'}})$ is the directional four-vector of the scattered photon, and the Compton phase-space kernel reduces to
\begin{align*}
    \mathcal{K}_{\text{C}}(k',p_{i}';l',q')&=\frac{1}{m}\,(2\pi)^{4}\delta^{(4)}(k+p-l-r)\,\Sigma_{\text{C}}(k',p_{i}',l',q')\nonumber\\
    &=\frac{1}{m}\,(2\pi)^{4}\delta^{(4)}(k+p-l-r)\,\left[\frac{\omega_{\bm{l}'}}{\omega_{\bm{k}'}}+\frac{\omega_{\bm{k}'}}{\omega_{\bm{l}'}}-2(\epsilon^{\kappa_{L}}(\bm{k})'\hat{l}')^{2}\right].
\end{align*}

\paragraph{Integrating out the scattered electron.}
We integrate over the scattered electron momentum to eliminate the spatial delta-function, write the scattered photon momentum in spherical coordinates, and reintroduce factors of $\hslash$ and $c$ in preparation for taking the classical limit. Defining The four-vector
\begin{align}
    \mathcal{G}(\omega_{\bm{k}'};\omega_{\bm{l}'}, \theta_{l'}, \phi_{l'})\equiv \Sigma_{\text{C}}\,
    \begin{pmatrix}
        \omega_{\bm{k}'}-\omega_{\bm{l}'}\\
        -\omega_{\bm{l}'}\sin\theta_{l'}\cos\phi_{l'}\\
        -\omega_{\bm{l}'}\sin\theta_{l'}\sin\phi_{l'}\\
        k'_{\parallel}c-\omega_{\bm{l}'}\cos\theta_{l'}
    \end{pmatrix},
\end{align}
the momentum shift takes the form
\begin{align}
    \Delta q'&\approx\frac{\alpha m\hslash^{2}c^{2}}{4}\int \frac{dk'_{\parallel}}{2\pi}\,|\tilde{a}'_{\text{cl},\parallel}(k'_{\parallel})|^{2}\int \frac{d\omega_{\bm{l}'}\,d^{2}\Omega_{l'}}{(2\pi)^{2}}\,\frac{\omega_{\bm{l}'}}{\omega_{\bm{k}'}E_{\hslash\bm{k}'-\hslash\bm{l}'}}\,(2\pi)\delta\!\left(g(\omega_{\bm{l}'})\right)\,\mathcal{G}(\omega_{\bm{k}'};\omega_{\bm{l}'},\theta_{l'},\phi_{l'}),
    \label{DeltaqBeforeCompton}
\end{align}
where
\begin{align}
    g(\omega_{\bm{l}'})\equiv \omega_{\bm{k}'}+\frac{mc^{2}}{\hslash}-\omega_{\bm{l}'}-\frac{E_{\hslash\bm{k}'-\hslash\bm{l}'}}{\hslash}.
\end{align}

\paragraph{The Compton delta-function and the classical limit.}
Then the solution to the relation $g(\omega^{*}_{\bm{l}'})=0$ is precisely the Compton shift:
\begin{align}
    \omega^{*}_{\bm{l}'}=\frac{\omega_{\bm{k}'}}{1+\dfrac{\hslash\omega_{\bm{k}'}}{mc^{2}}(1+\cos\theta_{l'})},\qquad k'_{\parallel}<0.
\end{align}
Photons moving towards the electron are thus upshifted in frequency --- a purely quantum effect. In the classical limit $\hslash\rightarrow 0$, $\omega^*_{\bm{l}'}\rightarrow\omega_{\bm{k}'}$: the Compton recoil vanishes, and scattered photons leave the interaction region with exactly the energy they carried in.\footnote{This is more striking than it first appears. It implies that the net shift in scattered electron momentum during classical inverse Thomson scattering of a laser pulse is a purely macroscopic, cumulative effect, with no single-photon analogue.}

Furthermore, the derivative of $g$ evaluated at the frequency of a Compton-shifted photon is given by
\begin{align*}
    &\left.\frac{dg}{d\omega_{\bm{l}'}}\right|_{\omega^{*}_{\bm{l}'}}=\frac{-\hslash\omega_{\bm{k}'}-mc^{2}+\hslash k'_{\parallel}c\cos{\theta_{l'}}}{E_{\hslash\bm{k}'-\hslash\bm{l}'}}\rightarrow-1,
\end{align*}
in the classical limit as $\hslash\rightarrow0$.

Thus, the shift in scattered electron momentum in the classical limit is
\begin{align*}
    \Delta q'&\approx\frac{e^{2}}{8\pi\epsilon_{0}c^{2}}\int \frac{dk_{\parallel}'}{(2\pi)}\left|\sqrt{\frac{\hslash\omega_{\bm{k}'}}{2}}\tilde{a}_{\text{cl},\parallel}(k_{\parallel}')\right|^{2}\int \frac{d^{2}\Omega}{(2\pi)}\frac{\mathcal{G}(\omega_{\bm{k}'};\omega_{\bm{k}'}, \theta_{l'}, \phi_{l'})}{\omega_{\bm{k}'}},
\end{align*}
with
\begin{align}
    \frac{\mathcal{G}(\omega_{\bm{k}'};\omega_{\bm{k}'}, \theta_{l'}, \phi_{l'})}{\omega_{\bm{k}'}}=\left(2-2(\epsilon^{\kappa_{L}}(\bm{k})'\hat{l}')^{2}\right)\begin{pmatrix}
        0\\
        -\sin{\theta_{l'}}\cos{\phi_{l'}}\\
        -\sin{\theta_{l'}}\sin{\phi_{l'}}\\
        \text{sgn}{(k'_{\parallel})}c-\cos{\theta_{l'}}
    \end{pmatrix},
\end{align}
where the quantity $(\epsilon^{\kappa_{L}}(\bm{k})'\hat{l}')^{2}=\sin^{2}{\theta_{l'}}\cos^{2}{\phi_{l'}}$ or $(\epsilon^{\kappa_{L}}(\bm{k})'\hat{l}')^{2}=\sin^{2}{\theta_{l'}}\sin^{2}{\phi_{l'}}$, depending on the direction of the transverse polarization and $\text{sgn}$ is the sign function .

\paragraph{The solid-angle integral and Lorentz-covariant form.}
Since the laser propagates in the negative scattering-axis direction, the support of
$|\tilde{a}'_{\text{cl},\parallel}|^{2}$ is $k_{\parallel}\in(-\infty,0)$. Performing the
integral over solid angle gives
\begin{align}
    \int \frac{d^{2}\Omega_{l'}}{(2\pi)}\frac{\mathcal{G}(\omega_{\bm{k}'};\omega_{\bm{k}'}, \theta_{l'}, \phi_{l'})}{\omega_{\bm{k}'}}
    =\frac{8}{3}
    \begin{pmatrix}0\\0\\0\\-1\end{pmatrix}.
\end{align}
The column vector $(0,0,0,-1)^\top$  here is not manifestly a Lorentz four-vector. The unique way to express it as a linear combination of the available unit wave four-vector $\hat{k}'_L$ of the laser pulse and the incident electron momentum $p'_i$ is
\begin{align}
    \begin{pmatrix}0\\0\\0\\-1\end{pmatrix}
    =\frac{mc}{(\hat{k}'_{L}p'_{i})}\hat{k}'_{L}-\frac{1}{mc}p'_{i}.
\end{align}

The rest-frame momentum shift therefore reads
\begin{align}
    \Delta q'\approx-\frac{e^{2}}{3\pi mc^{2}}\int\frac{dk'_{\parallel}}{2\pi}\left|\sqrt{\frac{\hslash\omega_{\bm{k}}}{2}}\,\tilde{a}'_{\text{cl},\parallel}(k'_{\parallel})\right|^{2}\left(p'_{i}-\frac{(mc)^{2}}{(\hat{k}'_{L}p'_{i})}\hat{k}'_{L}\right).
\end{align}

\paragraph{Boosting back to the lab-frame.}
A quick calculation demonstrates that $k'_{\parallel}=\frac{(\hat{k}_{L}p_{i})}{mc}k_{\parallel}$, for $k_{\parallel}<0$. Therefore, the lab-frame expected momentum shift is
\begin{align}
    \Delta q
    \approx
    -\frac{e^{2}}{3\pi m^{2}c^{3}}
    (\hat{k}_{L}p_{i})
    \int \frac{dk_{\parallel}}{2\pi}
    \left|\sqrt{\frac{\hslash\omega_{\bm{k}}}{2}}\,
    \tilde{a}_{\text{cl},\parallel}(k_{\parallel})\right|^{2}
    \left[p_{i}-\frac{(mc)^{2}}{(\hat{k}_{L}p_{i})}\hat{k}_{L}\right].
\end{align}

\paragraph{Recovery of the Landau-Lifshitz dynamics.}
Inverting the Fourier transform of the laser envelope via
Eq.~\eqref{LongitudinalFourierCoefficientPlaneWaveSI} and substituting, we obtain the central
result of this section:
\begin{align}
    \Delta q
    \approx
    -\frac{e^{2}}{6\pi m^{2}c^{3}}
    (\hat{k}_{L}p_{i})
    \int d(\hat{k}_{L}x)\,
    \left|\frac{d\tilde{A}_{\text{cl},\parallel}(k_{L}x)}
    {d(\hat{k}_{L}x)}\right|^{2}
    \left(p_{i}-\frac{(mc)^{2}}{(\hat{k}_{L}p_{i})}\hat{k}_{L}\right),
    \label{ElectronScatteredMomentumShift}
\end{align}
where the factor of $\tfrac{1}{2}$ arises from taking the real part of the inverse Fourier transform of the Heaviside function. To order $e^{4}$, this is precisely the solution to the Landau-Lifshitz equation for an electron scattered by a linearly polarized plane-wave as previously noted in Eq.~\eqref{ElectronScatteredMomentumShiftLL}.

The significance of this result is threefold. First, it confirms that the coherent-state QED framework reproduces the standard classical limit found in the literature, establishing internal consistency. Second, it demonstrates that the full spectral equations describing quantum radiation reaction
for an electron sampled from a beam and scattered by a linearly polarized plane-wave to order $e^{4}$ are exactly Eq.~(\ref{ScatteredElectronMomentumSpectrum}), Eq.~(\ref{ElectronScatteredMomentumSpectrumIncoherentInterference}), and Eq.~(\ref{ElectronScatteredMomentumSpectrumIncoherentDirect}). Thirdly, and most importantly, it establishes that coherent-state QED — specifically, the model of electron-photon scattering in which the laser field is represented by a photonic coherent state — is the realization of the Landau-Lifshitz model of classical electrodynamics within quantum electrodynamics.

A final consideration that should be addressed is that the analogous calculation in Ref.~\cite{Ilderton:2013} in the Furry picture of QED using Volkov states also reproduces the Landau-Lifshitz dynamics to order $e^{4}$. This is not surprising as coherent state QED and background field are formally dual to each other. The main discrepancy lies in the Furry picture, where the background field is assumed to be fixed, so that conservation of energy-momentum between the electron and the background field is not fully captured therein. However, one need not assume that the field is fixed in background field QED.

\section{Macroscopic scattering by a Gaussian Plane-wave}

In this final section of chapter~5, we proceed to our ultimate task of this dissertation: deriving the scattered electron energy spectrum from the inverse Compton scattering of an electron beam by a linearly polarized Gaussian laser pulse via coherent state QED to second-order in perturbation theory. As in previous sections in this chapter, the interference and direct terms are handled separately, but in close parallel. We will show that in the ultrarelativistic limit, the model developed here is a truncation of the classical model developed in chapter~4.

\subsection{The Incident Electron Spectrum}

Modeling the incident electron beam requires care, because the full momentum-space
distribution measured in~\cite{Poder:2018} was not directly observed. The marginal energy
spectrum was measured, but the joint energy-angle distribution was fitted to a model: the incident
electron energy was sampled from the measured data, and two divergence angles were sampled
independently from an energy-dependent zero-mean Gaussian whose FWHM was taken from
experimental divergence data. This suggests a factorized joint spectrum of the form
\begin{align}
    f_{\bm{p}_{i}}(\bm{p})\,d^{3}\bm{p}
    =f_{E_{\bm{p}_{i}}}(E_{\bm{p}})\,
    f_{\bm{\theta}_{p_{i}}|E_{\bm{p}_{i}}}(\bm{\theta}_{p}|E_{\bm{p}})\,
    dE_{\bm{p}}\,d\bm{\theta}_{p},
    \label{IncidentJointSpectrum}
\end{align}
with the energy-conditional divergence angle distribution
\begin{align}
    f_{\bm{\theta}_{p_{i}}|E_{\bm{p}_{i}}}(\bm{\theta}_{p}|E_{\bm{p}})
    =\frac{2\pi}{\sigma_{\theta_{p}}^{2}(E_{\bm{p}})}
    \exp\!\left[-\frac{\theta_{p}^{2}}{2\sigma_{\theta_{p}}^{2}(E_{\bm{p}})}\right],
\end{align}
where $\bm{\theta}_{p}=\frac{\bm{p}_{T}}{p_{\parallel}}$ and $\theta_{p}=|\bm{\theta}_{p}|$.
Treating the beam as cylindrically symmetric is justified by the blowout regime of LWFA, which
produces radially symmetric accelerating fields. The divergence was measured from the transverse
beam width on the LANEX screen, perpendicular to the magnetic dispersion direction, $2\,\text{m}$
downstream of the gas cell; however,~\cite{Poder:2018} reports only the integrated value
$\text{FWHM}_{\theta_{p}}\simeq 0.70\pm 0.05\,\text{mrad}$ for $E_{\bm{p}}\gtrsim 1\,\text{GeV}$,
without providing the full spectrometer data. Lacking access to the energy-resolved divergence
profile, we marginalize over the angular dependence rather than modeling it independently,
exploiting the fact that ultrarelativistic electrons permit a clean analytical marginalization
when the laser pulse is sufficiently narrow-banded, as we will demonstrate in this section.

For this marginalization, working with the forward lightfront angles
$\bm{\chi}_{p}=\frac{\bm{p}_{T}}{p^{+}}$, where $p^{+}=E_{\bm{q}}+p_{\parallel}$, is more
natural than working with the divergence angles $\bm{\theta}_p$, both analytically and
numerically. In the ultrarelativistic regime --- which holds on the support of the incident
energy spectrum --- the two are related by $\bm{\chi}_{p}\approx\frac{1}{2}\bm{\theta}_{p}$,
so the forward lightfront angles are likewise Gaussian distributed. The incident spectrum
therefore takes the equivalent form
\begin{align}
    f_{\bm{p}_{i}}(\bm{p})\,d^{3}\bm{p}
    =f_{E_{\bm{p}_{i}}}(E_{\bm{p}})\,
    f_{\bm{\chi}_{p_{i}}|E_{\bm{p}_{i}}}(\bm{\chi}_{p}|E_{\bm{p}})\,
    dE_{\bm{p}}\,d\bm{\chi}_{p},
    \label{EnergyForwardLightfrontAngleSpectrum}
\end{align}
with
\begin{align}
    f_{\bm{\chi}_{p_{i}}|E_{\bm{p}_{i}}}(\bm{\chi}_{p}|E_{\bm{p}})
    =\frac{2\pi}{\sigma_{\chi_{p}}^{2}(E_{\bm{p}})}
    \exp\!\left[-\frac{\chi_{p}^{2}}{2\sigma_{\chi_{p}}^{2}(E_{\bm{p}})}\right],
    \qquad
    \chi_{p}=|\bm{\chi}_{p}|.
\end{align}

\subsection{From Momentum Spectrum to Energy Spectrum}

The scattered momentum spectrum computed in
Eqs.~\ref{ScatteredElectronMomentumSpectrum},~\ref{ElectronScatteredMomentumSpectrumIncoherentInterference},
and~\ref{ElectronScatteredMomentumSpectrumIncoherentDirect} is not directly the experimental
observable: detectors measure energy spectra. We therefore derive the relationship between the
two, then perform the relevant angular integrations.

Writing $Q\equiv |\bm{q}|$, the marginal distribution of scattered momentum magnitudes is
\begin{align}
    f_{|\bm{q}_{f}|}(Q)
    =Q^{2}\int\frac{d^{2}\Omega_{q}}{(2\pi)^{2}}\,f_{\bm{q}_{f}}(\bm{q}).
    \label{MarginalMomentumMagnitude}
\end{align}
Since $E_{\bm{q}}=\sqrt{m^{2}+Q^{2}}$ implies $E_{\bm{q}}\,dE_{\bm{q}}=Q\,dQ$, the
scattered energy distribution is
\begin{align}
    f_{E_{\bm{q}_{f}}}(E_{\bm{q}})
    =f_{|\bm{q}_{f}|}(Q)\frac{E_{\bm{q}}}{Q}
    =QE_{\bm{q}}\int\frac{d^{2}\Omega_{q}}{(2\pi)^{2}}\,f_{\bm{q}_{f}}(\bm{q}),
    \label{ScatteredElectronEnergySpectrum}
\end{align}
where $Q=Q(E_{\bm{q}})=\sqrt{E_{\bm{q}}^{2}-m^{2}}$. Integrating over solid angle in
Eq.~\eqref{ScatteredElectronMomentumSpectrum} via Eq.~\eqref{ScatteredElectronEnergySpectrum},
the scattered electron energy spectrum to order $e^{4}$ is
\begin{align}
    f_{E_{\bm{q}_{f}}}(E_{\bm{q}})
    \approx
    f_{E_{\bm{p}_{i}}}(E_{\bm{q}})
    +f^{(4)}_{E_{\bm{q}_{f}}}(E_{\bm{q}})_{\text{DI}}
    +f^{(4)}_{E_{\bm{q}_{f}}}(E_{\bm{q}})_{\text{II}}.
    \label{ScatteredElectronEnergySpectrumSecondOrder}
\end{align}

\subsection{Lightfront Reduction of the Interference Spectrum}

Starting from Eq.~\eqref{ElectronScatteredMomentumSpectrumIncoherentInterference} and applying
the energy spectrum relation Eq.~\eqref{ScatteredElectronEnergySpectrum}, the interference
contribution to the scattered energy spectrum is
\begin{align}
    f^{(4)}_{E_{\bm{q}_{f}}}(E_{\bm{q}})_{\text{II}}
    =-\frac{e^{4}QE_{\bm{q}}}{8}
    \int\frac{d^{2}\Omega_{q}}{(2\pi)^{2}}\,
    f_{\bm{p}_{i}}&(\bm{q})
    \int \frac{dk_{\parallel}}{2\pi}\,|a_{\text{cl},\parallel}(k_{\parallel})|^{2}
    \int\frac{d^{3}\bm{l}}{(2\pi)^{3}}
    \int\frac{d^{3}\bm{r}}{(2\pi)^{3}}\nonumber\\
    &\times
    \frac{(2\pi)^4\delta^{(4)}(k+q-l-r)}
    {\omega_{\bm{l}}E_{\bm{r}}(kq)}
    \,\Sigma_{\text{C}}(k,q,l,r).
    \label{InterferenceEnergyStart}
\end{align}
We now perform a sequence of changes of variables that reduce this nine-dimensional integral
to a numerically tractable form, using lightfront coordinates throughout. The key identities
are: $QE_{\bm{q}}\,dE_{\bm{q}}\,d^{2}\Omega_{q}=d^{3}\bm{q}$, the lightfront
phase-space measures $\omega_{\bm{l}}\,dl^{+}=l^{+}\,dl_{\parallel}$ and
$d^{3}\bm{r}=E_{\bm{r}}r^{+}\,dr^{+}\,d^{2}\bm{\chi}_{r}$, and the delta-function
factorization
\begin{align}
    \delta(\omega_{\bm{k}}+E_{\bm{q}}-\omega_{\bm{l}}-E_{\bm{r}})\,
    \delta(k_{\parallel}+q_{\parallel}-l_{\parallel}-r_{\parallel})
    =2\,\delta(q^{+}-l^{+}-r^{+})\,
    \delta(2k_{\parallel}-q^{-}+l^{-}+r^{-}),
    \label{DeltaFactorization}
\end{align}
where $q^{-}=E_{\bm{q}}-q_{\parallel}$, $l^{-}=\omega_{\bm{l}}-l_{\parallel}$, and
$r^{-}=E_{\bm{r}}-r_{\parallel}$. After integrating out the transverse delta-function
(setting $\bm{l}_{T}=\bm{q}_{T}-\bm{r}_{T}$), substituting
Eq.~\eqref{EnergyForwardLightfrontAngleSpectrum} for the incident electron spectrum and
Eq.~\eqref{LongitudinalFourierCoefficientGaussianPlaneWave} for the laser envelope, and
defining
\begin{align}
    H(k_{\parallel})
    \equiv
    \bigl[g(\hslash k_{\parallel}c-E_{L})+g(\hslash k_{\parallel}c+E_{L})\bigr]^{2},
    \label{Hdef}
\end{align}
the interference term becomes
\begin{align}
    f^{(4)}_{E_{\bm{q}_{f}}}(E_{\bm{q}})_{\text{II}}
    =-\frac{\pi m^{2} e^{2}a_{0}^{2}}{8\sigma_{E_{L}}^{2}}
    \int\frac{d^{2}\bm{\chi}_{q}}{(2\pi)^{2}}\,
    f_{E_{\bm{p}_{i}},\bm{\chi}_{p_{i}}}&(E_{\bm{q}},\bm{\chi}_{q})
    \int_{0}^{q^{+}}\frac{dr^{+}}{2\pi}
    \int\frac{d^{2}\bm{\chi}_{r}}{(2\pi)^{2}}
    \frac{r^{+}\,\Sigma_{\text{C}}\,H(E_{\bm{k}})}{q^{+}(q^{+}-r^{+})},
    \label{InterferenceLightfront}
\end{align}
where the remaining delta-functions have been used to eliminate $l^+$ and $k_\parallel$, and
the photon energy enforced by the latter is
\begin{align}
    E_{\bm{k}}
    =\tfrac{1}{2}(l^{-}+r^{-}-q^{-})
    =\frac{1}{2}\left[
    m^{2}\frac{1-\frac{r^{+}}{q^{+}}}{r^{+}}
    +\frac{r^{+}}{1-\frac{r^{+}}{q^{+}}}
    |\bm{\chi}_{q}-\bm{\chi}_{r}|^{2}
    \right],
    \label{Ek}
\end{align}
with the Compton kinematic factor evaluating to
\begin{align}
    \Sigma_{\text{C}}(k,q,l,r)
    =\frac{r^{+}}{q^{+}}+\frac{q^{+}}{r^{+}}
    -\frac{2m^{2}}{E_{\bm{k}}^{2}}
    \bigl(\bm{\varepsilon}_{T}{\cdot}(\bm{\chi}_{q}-\bm{\chi}_{r})\bigr)^{2},
    \label{SigmaCLightfront}
\end{align}
where $\bm{\varepsilon}_{T}=\bm{\varepsilon}_{T}(\hat{\bm{k}}_{L})$ is the transverse laser polarization vector.

\paragraph{Change of variables.}
Equation~\ref{InterferenceLightfront} is ready for numerical evaluation, but applying the
narrow-band approximation is complicated by the fact that the level set $E_{\bm{k}}=E_L$
has a complicated dependence on $(r^+,\bm{\chi}_r)$ in the current parametrization. We therefore introduce variables adapted to the structure of $E_{\bm{k}}$. Write
\begin{align}
    \bm{\chi}_{r}
    &=\bm{\chi}_{q}-|\bm{\chi}_{q}-\bm{\chi}_{r}|\,\hat{\bm{n}}_{\alpha},
    \qquad
    \hat{\bm{n}}_{\alpha}=(\cos\alpha,\sin\alpha),
    \label{chiCOV}
\end{align}
and then reparametrize via the dimensionless variables $s>0$ and $t\in(-1,1)$ defined by
\begin{align}
    r^{+}=\frac{q^{+}}{1+\rho s(1+t)},
    \qquad
    |\bm{\chi}_{q}-\bm{\chi}_{r}|=\frac{E_{L}}{m}s\sqrt{1-t^{2}},
    \label{stCOV}
\end{align}
where $\rho\equiv\frac{E_{L}q^{+}}{m^{2}}$ is the macroscopic Compton recoil parameter---quantifying the effect of quantum recoil for a photon at the laser carrier energy $E_L$ scattered by an electron with lightfront momentum $q^+$. Before the narrow-band approximation is applied, the naturally appearing combination is $\rho s = \frac{E_{\bm{k}}q^+}{m^2}$, which is the standard single-photon Compton recoil parameter
evaluated at the actual photon energy $E_{\bm{k}} = E_L s$ drawn from the laser spectrum. The Jacobian of this transformation is
\begin{align}
    \frac{r^{+}\,dr^{+}\,d^{2}\bm{\chi}_{r}}{q^{+}(q^{+}-r^{+})}
    =\frac{E_{L}^{2}}{m^{2}}\,\frac{s\,ds\,dt\,d\alpha}{(1+\rho s(1+t))^{2}}.
    \label{Jacobian}
\end{align}
The payoff is immediate: the expressions for $E_{\bm{k}}$ and $\Sigma_{\text{C}}$ linearize,
\begin{align}
    E_{\bm{k}}=E_{L}s,
    \qquad
    \Sigma_{\text{C}}
    =\frac{1}{1+\rho s(1+t)}+(1+\rho s(1+t))
    -2(\bm{\varepsilon}_{T}{\cdot}\hat{\bm{n}}_{\alpha})^{2}(1-t^{2}),
    \label{SimplifiedKinematics}
\end{align}
and the argument of $H$ becomes simply $H(E_Ls)$. The interference term is now
\begin{align}
    f^{(4)}_{E_{\bm{q}_{f}}}(E_{\bm{q}})_{\text{II}}
    =-\frac{e^{2}a_{0}^{2}E_{L}^{2}}{32\pi\sigma_{E_{L}}^{2}}
    \int\frac{d^{2}\bm{\chi}_{q}}{(2\pi)^{2}}\,
    f_{E_{\bm{p}_{i}},\bm{\chi}_{p_{i}}}(E_{\bm{q}},\bm{\chi}_{q})
    \int_{0}^{\infty}ds
    \int_{-1}^{1}dt
    \int_{0}^{2\pi}\frac{d\alpha}{2\pi}\,
    \frac{s\,\Sigma_{\text{C}}\,H(E_{L}s)}{(1+\rho s(1+t))^{2}}.
    \label{InterferenceST}
\end{align}

\paragraph{Analytic integration over $\alpha$ and $t$.}
The $\alpha$ and $t$ integrals can be evaluated in closed form. Define
\begin{align}
    T(\rho s)
    &\equiv
    \int_{-1}^{1}dt
    \int_{0}^{2\pi}\frac{d\alpha}{2\pi}\,
    \frac{\Sigma_{\text{C}}}{(1+\rho s(1+t))^{2}}.
    \label{Tdef}
\end{align}
This quantity encodes the Klein-Nishina kinematics for the full polychromatic coherent state.

A direct calculation gives the exact result
\begin{align}
    T(\rho s)
    &=\frac{1}{2\rho s}
    \left[2\log(1+2\rho s)+1-\frac{1}{(1+2\rho s)^{2}}\right]\nonumber\\
    &\quad-\frac{2}{(\rho s)^{2}}
    \left[\log(1+2\rho s)-1+\frac{1}{1+2\rho s}\right]\nonumber\\
    &\quad-\frac{1}{(\rho s)^{3}}
    \left[2\log(1+2\rho s)-1-2\rho s+\frac{1}{1+2\rho s}\right],
    \label{Texact}
\end{align}
which in the ultrarelativistic limit $\rho s\ll 1$ reduces to
\begin{align}
    T(\rho s)\approx \tfrac{8}{3}-\tfrac{16}{3}\rho s.
    \label{Tapprox}
\end{align}

\paragraph{Narrow-band approximation and marginalization.}
It remains to approximate the $s$-integral and marginalize the forward lightfront angles. Since
$g^2(\hslash k_\parallel c + E_L)$ is exponentially suppressed for $k_\parallel < 0$, the
function $H(E_L s) \approx g^2(E_L s - E_L)$ is effectively a single Gaussian peaked at
$s=1$ with width $\sigma_{E_L}/E_L$. In the narrow-band limit $\sigma_{E_L}\ll E_L$, the
remaining factors in the $s$-integrand are slowly varying on the scale of this Gaussian and
may be evaluated at $s=1$. Physically, this is precisely the replacement of the
polychromatic recoil parameter $\rho s$ by the macroscopic recoil parameter $\rho$: the
full distribution of single-photon Compton recoils encoded in $T(\rho s)$ collapses to a
single value $T(\rho)$ set by the carrier frequency of the laser, which is valid when
$\sigma_{E_L}$ is small enough that $T(\rho s)$ is approximately constant across the
spectral support of $H(E_L s)$. This gives
\begin{align}
    \int_0^\infty ds\; s\, T(\rho s)\, H(E_L s)
    \approx T(\rho)\int_0^\infty ds\; s\, g^2(E_L s - E_L)
    =T(\rho)\,\frac{\sqrt{\pi}\,\sigma_{E_L}}{2E_L}.
\end{align}
Substituting into Eq.~\eqref{InterferenceST} and marginalizing the forward lightfront angles
via Eq.~\eqref{EnergyForwardLightfrontAngleSpectrum}, the interference contribution to the
scattered energy spectrum most appropriate for numerical computation is
\begin{align}
    f^{(4)}_{E_{\bm{q}_{f}}}(E_{\bm{q}})_{\text{II}}
    &\approx
    -\frac{\sqrt{\pi}\,e^{2}a_{0}^{2}E_{L}}{32\pi\sigma_{E_{L}}}\,
    T(\rho)\,f_{E_{\bm{p}_{i}}}(E_{\bm{q}}).
    \label{InterferenceNarrowCompute}
\end{align}
Taking the ultrarelativistic limit and substituting Eq.~\eqref{Tapprox}, this simplifies to
\begin{align}
    f^{(4)}_{E_{\bm{q}_{f}}}(E_{\bm{q}})_{\text{II}}
    &\approx
    \left(
    -\frac{\sqrt{\pi}\,\alpha a_{0}^{2}E_{L}}{3\sigma_{E_{L}}}
    +\frac{2E_{\bm{q}}}{E_{\bm{q},\text{max}}}
    \right)f_{E_{\bm{p}_{i}}}(E_{\bm{q}}),
    \label{ElectronScatteredEnergySpectrumIncoherentInterference}
\end{align}
where $\alpha$ is the fine-structure constant, and $E_{\bm{q},\text{max}}$ the classical cutoff energy from classical model of the scattered spectrum obtained via Landau-Lifshitz dynamics.

\subsection{Lightfront Reduction of the Direct Spectrum}

The direct term is reduced by the same sequence of steps as the interference term, but the
kinematics are structurally distinct in a way that will matter for the narrow-band
approximation. Starting from Eq.~\eqref{ElectronScatteredMomentumSpectrumIncoherentDirect} and
applying Eq.~\eqref{ScatteredElectronEnergySpectrum},
\begin{align}
    f^{(4)}_{E_{\bm{q}_{f}}}(E_{\bm{q}})_{\text{DI}}
    =\frac{e^{4}Q}{8}
    \int\frac{d^{2}\Omega_{q}}{(2\pi)^{2}}
    \int \frac{d^{3}\bm{p}}{(2\pi)^{3}}\,f_{\bm{p}_{i}}(\bm{p})
    \int\frac{dk_{\parallel}}{2\pi}\,|a_{\text{cl},\parallel}(k_{\parallel})|^{2}
    \int\frac{d^{3}\bm{l}}{(2\pi)^{3}}\nonumber\\
    \times\frac{(2\pi)^{4}\delta^{(4)}(k+p-l-q)}
    {\omega_{\bm{l}}(kp)}\,
    \Sigma_{\text{C}}(k,p,l,q).
    \label{DirectEnergyStart}
\end{align}
We pass to lightfront coordinates using the same identities as in the preceding subsection,
together with the ultrarelativistic approximations
\begin{align}
    dE_{\bm{p}}\approx\tfrac{1}{2}(1+\chi_{p}^{2})\,dp^{+},
    \qquad
    d^{2}\Omega_{q}\approx\frac{4\,d^{2}\bm{\chi}_{q}}{(1+\chi_{q}^{2})^{2}},
    \label{URApprox}
\end{align}
and use Eq.~\eqref{DeltaFactorization} with the substitutions $r\to q$ and $q\to p$ appropriate
to the direct-term kinematics. After eliminating the transverse delta-function, inserting
Eq.~\eqref{LongitudinalFourierCoefficientGaussianPlaneWave}, and performing the two remaining
delta-function integrals to fix $l^+$ and $k_\parallel$, the photon energy enforced is
\begin{align}
    E_{\bm{k}}
    =\tfrac{1}{2}(l^{-}+q^{-}-p^{-})
    =\frac{1}{2}\left[
    m^{2}\frac{l^{+}}{q^{+}(l^{+}+q^{+})}
    +\frac{q^{+}(l^{+}+q^{+})}{l^{+}}
    |\bm{\chi}_{p}-\bm{\chi}_{q}|^{2}
    \right],
    \label{EkDirect}
\end{align}
with the Compton kinematic factor
\begin{align}
    \Sigma_{\text{C}}(k,p,l,q)
    =\frac{q^{+}}{l^{+}+q^{+}}+\frac{l^{+}+q^{+}}{q^{+}}
    -\frac{2m^{2}}{E_{\bm{k}}^{2}}
    \bigl(\bm{\varepsilon}_{T}{\cdot}(\bm{\chi}_{p}-\bm{\chi}_{q})\bigr)^{2},
    \label{SigmaCDirect}
\end{align}
and the incident electron energy
\begin{align}
    E_{\bm{p}}
    =\tfrac{1}{2}\left[(l^{+}+q^{+})(1+\chi_{p}^{2})+\frac{m^{2}}{l^{+}+q^{+}}\right].
    \label{EpDirect}
\end{align}
The direct term then reads
\begin{align}
    f^{(4)}_{E_{\bm{q}_{f}}}(E_{\bm{q}})_{\text{DI}}
    \approx
    \frac{\pi m^{2}e^{2}a_{0}^{2}}{16\sigma_{E_{L}}^{2}}
    \int\frac{4E_{\bm{q}}\,d^{2}\bm{\chi}_{q}}{(2\pi)^{2}(1+\chi_{q}^{2})^{2}}
    \int\frac{d^{2}\bm{\chi}_{p}}{(2\pi)^{2}}
    \int\frac{dl^{+}}{2\pi}\nonumber\\
    \times\frac{(1+\chi_{p}^{2})}{l^{+}(l^{+}+q^{+})}\,
    \Sigma_{\text{C}}\,H(E_{\bm{k}})\,
    f_{E_{\bm{p}_{i}},\bm{\chi}_{p_{i}}}(E_{\bm{p}},\bm{\chi}_{p}).
    \label{DirectLightfront}
\end{align}

\paragraph{Change of variables.}
Equation~\ref{DirectLightfront} has the same obstruction as Eq.~\eqref{InterferenceLightfront}
before the change of variables: the level set $E_{\bm{k}}=E_L$ is complicated in the current
parametrization. We introduce the analogous adapted coordinates. Write
\begin{align}
    \bm{\chi}_{p}=\bm{\chi}_{q}+|\bm{\chi}_{p}-\bm{\chi}_{q}|\,\hat{\bm{n}}_{\alpha},
    \qquad
    \hat{\bm{n}}_{\alpha}=(\cos\alpha,\sin\alpha),
\end{align}
and reparametrize via $s>0$ and $t\in(-1,1)$ defined by
\begin{align}
    l^{+}=\frac{q^{+}\rho s(1+t)}{1-\rho s(1+t)},
    \qquad
    |\bm{\chi}_{p}-\bm{\chi}_{q}|=\frac{E_{L}}{m}s\sqrt{1-t^{2}},
    \label{stCOVDirect}
\end{align}
where $\rho = \frac{E_L q^+}{m^2}$ is the macroscopic Compton recoil parameter defined in
Eq.~\eqref{stCOV}. Comparing with the interference-term change of variables
Eq.~\eqref{stCOV}, the two parametrizations
reflect the transposition of the roles of the initial and final electron roles between the two terms. The physical requirement $l^+>0$ imposes $\rho s(1+t)<1$, which in the
ultrarelativistic limit is automatically satisfied on the support of $H(E_L s)$ and allows
the $t$-limits to be extended to $(-1,1)$. The Jacobian is
\begin{align}
    \frac{(1+\chi_{p}^{2})\,dl^{+}\,d^{2}\bm{\chi}_{p}}{l^{+}(l^{+}+q^{+})}
    \approx
    \frac{E_{L}^{2}}{m^{2}}\,
    \frac{(1+\chi_{q}^{2})\,s\,ds\,dt\,d\alpha}{q^{+}},
    \label{JacobianDirect}
\end{align}
and the kinematics simplify to
\begin{align}
    E_{\bm{k}}=E_{L}s,
    \qquad
    \Sigma_{\text{C}}
    =\frac{1}{1-\rho s(1+t)}+(1-\rho s(1+t))
    -2(\bm{\varepsilon}_{T}{\cdot}\hat{\bm{n}}_{\alpha})^{2}(1-t^{2}).
    \label{SimplifiedKinematicsDirect}
\end{align}
The direct-term $\Sigma_{\text{C}}$ is related to its interference-term counterpart
Eq.~\eqref{SimplifiedKinematics} by $\rho s(1+t)\to -\rho s(1+t)$: the Compton recoil shifts
the final-state lightfront fraction upward in the direct term and downward in the interference
term, consistent with their respective roles as emission and absorption amplitudes. The
incident electron energy becomes
\begin{align}
    E_{\bm{p}}
    \approx
    \frac{1}{2}\left[
    \frac{q^{+}(1+\chi_{q}^{2})}{1-\rho s(1+t)}
    +\frac{m^{2}}{q^{+}}(1-\rho s(1+t))
    \right],
    \label{EpNewVars}
\end{align}
where we have used the fact that the deflection $|\bm{\chi}_{p}-\bm{\chi}_{q}|\,\hat{\bm{n}}_{\alpha}$ causes negligible variation of $f_{E_{\bm{p}_{i}},\bm{\chi}_{p_{i}}}$, so that
$f_{E_{\bm{p}_{i}},\bm{\chi}_{p_{i}}}(E_{\bm{p}},\bm{\chi}_{p})\approx
f_{E_{\bm{p}_{i}},\bm{\chi}_{p_{i}}}(E_{\bm{p}}(\bm{\chi}_{q}),\bm{\chi}_{q})$ on the ultrarelativistic scale. In fact, in this specific experiment, the energy variation due to deflection is far below the resolution of the detector. The direct term is then
\begin{align}
    f^{(4)}_{E_{\bm{q}_{f}}}(E_{\bm{q}})_{\text{DI}}
    \approx
    \frac{e^{2}a_{0}^{2}E_{L}^{2}}{64\pi\sigma_{E_{L}}^{2}}
    \int\frac{4E_{\bm{q}}\,d^{2}\bm{\chi}_{q}}{(2\pi)^{2}(1+\chi_{q}^{2})}
    \int_{0}^{\infty}ds
    \int_{-1}^{1}dt
    \int_{0}^{2\pi}\frac{d\alpha}{2\pi}\,
    \frac{s\,\Sigma_{\text{C}}\,H(E_{L}s)}{q^{+}}\,
    f_{E_{\bm{p}_{i}},\bm{\chi}_{p_{i}}}(E_{\bm{p}},\bm{\chi}_{q}).
    \label{DirectST}
\end{align}

\paragraph{First narrow-band approximation and the $\alpha$-integral.}
As in the interference term, $H(E_L s)\approx g^2(E_L s - E_L)$ is a Gaussian sharply
peaked at $s=1$, so the macroscopic recoil parameter $\rho$ again governs the kinematics
under the narrow-band approximation $s\to 1$. After evaluating the $\alpha$-integral and
setting $q^+\approx 2E_{\bm{q}}$, the direct term becomes
\begin{align}
    f^{(4)}_{E_{\bm{q}_{f}}}(E_{\bm{q}})_{\text{DI}}
    \approx
    \frac{\sqrt{\pi}e^{2}a_{0}^{2}E_{L}}{32\pi\sigma_{E_{L}}}
    \int_{-1}^{1}dt\,
    U(\rho,t)
    \int\frac{d^{2}\bm{\chi}_{q}}{(2\pi)^{2}(1+\chi_{q}^{2})}\,
    f_{E_{\bm{p}_{i}},\bm{\chi}_{p_{i}}}(\bm{\chi}_{q}|E_{\bm{p}})\,
    f_{E_{\bm{p}_{i}}}(E_{\bm{p}}),
    \label{DirectNarrowBand}
\end{align}
where we define the direct-term kinematic integrand
\begin{align}
    U(\rho,t)
    \equiv
    \frac{1}{1-\rho(1+t)}+1-\rho(1+t)-(1-t^{2}),
    \label{TtildeDef}
\end{align}
The sign difference in the recoil
argument --- $-\rho(1+t)$ here versus $+\rho s(1+t)$ in Eq.~(\ref{SimplifiedKinematics}) ---
is the same transposition noted above, and persists through to the final result.

\paragraph{Second narrow-band approximation.}
A further approximation is available because the angular spread of the beam, while much
larger than the laser bandwidth ($\sigma_{\bm{\chi}_p}(E_{\bm{p}})\gg\sigma_{E_L}/E_L$),
is itself small in the ultrarelativistic regime: $\sigma_{\bm{\chi}_p}(E_{\bm{p}})\ll 1$.
The conditional angular distribution $f_{\bm{\chi}_{p_i}|E_{\bm{p}_i}}$ is therefore sharply
peaked about $\bm{\chi}_q\approx 0$ on the scale of the integrand, and the factor
$(1+\chi_q^2)^{-1}$ can be set to unity. Marginalizing the angular distribution then gives
\begin{align}
    f^{(4)}_{E_{\bm{q}_{f}}}(E_{\bm{q}})_{\text{DI}}
    \approx
    \frac{\sqrt{\pi}e^{2}a_{0}^{2}E_{L}}{32\pi\sigma_{E_{L}}}
    \int_{-1}^{1}dt\;
    U(\rho,t)\;
    f_{E_{\bm{p}_{i}}}(E_{\bm{p}}),
    \label{ElectronScatteredEnergySpectrumIncoherentDirectCompute}
\end{align}
with the incident electron energy
\begin{align}
    E_{\bm{p}}
    \approx E_{\bm{q}}+\frac{q^{+}\rho(1+t)}{2}.
    \label{EpFinal}
\end{align}
Equation~\ref{ElectronScatteredEnergySpectrumIncoherentDirectCompute} is the most convenient
form for numerical evaluation, with $E_{\bm{p}}$ varying across the $t$-integral and the
incident spectrum sampled at the energy required to produce a final-state electron at $E_{\bm{q}}$.

\paragraph{Analytic approximation.}
For an analytic closed form, we Taylor-expand the incident spectrum about $E_{\bm{q}}$.
Since the energy shift $\frac{q^+\rho(1+t)}{2}$ is small when $\rho\ll 1$,
\begin{align}
    f_{E_{\bm{p}_{i}}}(E_{\bm{p}})
    \approx f_{E_{\bm{p}_{i}}}(E_{\bm{q}})
    +\frac{q^{+}\rho(1+t)}{2}f'_{E_{\bm{p}_{i}}}(E_{\bm{q}}),
\end{align}
valid whenever the spectrum is differentiable --- satisfied for any physical ensemble or test
function. Using the leading-order approximation
$U(\rho,t)\approx 1+t^{2}$ (valid for $\rho\ll 1$) and integrating over $t$,
the direct contribution takes the simple closed form
\begin{align}
    f^{(4)}_{E_{\bm{q}_{f}}}(E_{\bm{q}})_{\text{DI}}
    \approx
    \frac{\sqrt{\pi}\,\alpha a_{0}^{2}E_{L}}{3\sigma_{E_{L}}}\,
    f_{E_{\bm{p}_{i}}}(E_{\bm{q}})
    +\frac{E_{\bm{q}}^{2}}{E_{\bm{q},\text{max}}}\,
    f'_{E_{\bm{p}_{i}}}(E_{\bm{q}}),
    \label{ElectronScatteredEnergySpectrumIncoherentDirectFormal}
\end{align}
where $\alpha$ is the fine-structure constant, and $E_{\bm{q},\text{max}}$ the classical cutoff energy.

\subsection{Comparison with Classical model}

We now have two novel macroscopic frameworks for describing the scattering of charged particles by an electromagnetic field that is free from them — one grounded in classical electrodynamics via the Landau-Lifshitz equation, and one in the coherent-state QED formalism developed in this chapter. The relationship between these frameworks has already been partially established at the level of single-electron dynamics. In the context of Compton scattering, the classical limit of the coherent-state framework reproduces the Landau-Lifshitz dynamics to order $e^4$. A more discriminating comparison is possible at the spectral level, where both frameworks admit explicit derivations of the scattered electron energy spectrum for a Gaussian pulse. It is this comparison that begins to reveal a more compelling and deeper relationship.

Combining the incident spectrum with
Eqs.~\ref{ElectronScatteredEnergySpectrumIncoherentInterference}
and~\ref{ElectronScatteredEnergySpectrumIncoherentDirectFormal} yields the total scattered
energy spectrum
\begin{align}
    f_{E_{\bm{q}_{f}}}(E_{\bm{q}})
    \approx f_{E_{\bm{p}_{i}}}(E_{\bm{q}})
    +\frac{2E_{\bm{q}}}{E_{\bm{q},\text{max}}}\,f_{E_{\bm{p}_{i}}}(E_{\bm{q}})
    +\frac{E_{\bm{q}}^{2}}{E_{\bm{q},\text{max}}}\,f'_{E_{\bm{p}_{i}}}(E_{\bm{q}}).
    \label{ElectronScatteredElectronSpectrumFormal}
\end{align}
This is precisely the scattered spectrum obtained from the Landau-Lifshitz model with a
Gaussian pulse truncated at order $e^{4}$ in Eq.~\eqref{ElectronScatteredEnergySpectrumLLCED}. Remarkably, no classical
limit was taken anywhere in the derivation: the agreement is exact on the strict
ultrarelativistic scale, where the primary departure from classical behavior consists of
imperceptibly small Compton recoil shifts in the individual electron energies. This is a
precise and quantitative reaffirmation of the central thesis of this chapter: coherent-state
QED is the realization of the Landau-Lifshitz model of classical electrodynamics within quantum electrodynamics.

\begin{figure}[h!]
    \centering
    \includegraphics[width=0.8\linewidth]{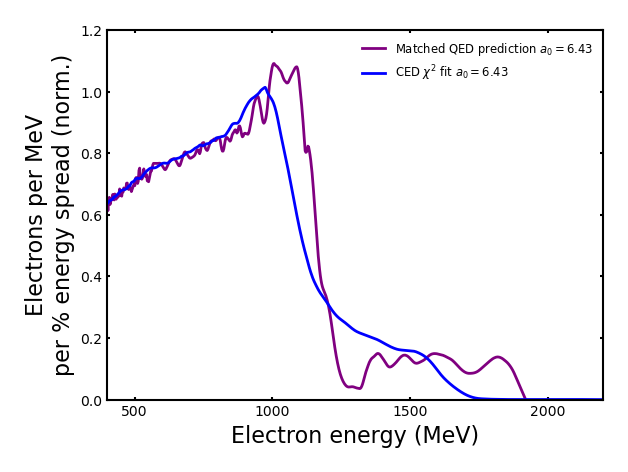}
    \caption[Matched plot of macroscopic QED model]{Matched scattered electron energy spectrum due to inverse Compton scattering by a Gaussian-shaped laser pulse with optimal $a_{0}=6.43$ derived from the Landau-Lifshitz spectral pushforward model of CED to the observed scattered spectra from the experiment in~\cite{Poder:2018}. }
    \label{fig:poderQEDmatch}
\end{figure}

The structural relationship between the two models, however, is more intricate than their
numerical agreement at order $e^4$ might suggest. The Landau-Lifshits equation is a self-consistent
correction to the Lorentz force law at second order in perturbation theory ---
order $e^2$ as a dynamical equation --- but being a differential equation, its solution and
the scattered spectrum derived from it contain non-perturbative content. They contain
contributions to all orders in $e$ simultaneously. Crucially, the classical scattered energy
spectrum in Eq.~\eqref{ScatteredElectronEnergySpectrumPushforward} has a transparent geometric interpretation: it is a pushforward of the incident
energy spectrum under the Landau-Lifshitz dynamics. That is, the map from initial to final
energy is a smooth deformation of the real line, and the scattered spectrum is obtained by
the corresponding change of variables. The spectrum is therefore unimodal, preserves the
topology of the incident distribution, and shifts monotonically with the field strength.

The coherent-state spectrum as presented here, by contrast, is manifestly perturbative. Rather than deforming
the incident distribution, it corrects it order by order: the two terms in
Eq.~\eqref{ElectronScatteredElectronSpectrumFormal} beyond the unscattered contribution are a
proportional shift and a derivative correction, each suppressed by $E_{\bm{q},\text{max}}^{-1}$.
These qualitative differences are visible even when the two models agree numerically to order
$e^4$, as shown in~\eqref{fig:poderQEDmatch}: the classical spectrum is a smooth
rigid shift of the initial distribution, while the quantum spectrum develops a characteristic
asymmetry driven by the truncation at the derivative term $f'_{E_{\bm{p}_i}}$. At higher field strengths or electron energies, where higher-order terms become considerably larger on a term-by-term comparison, one would expect that either the perturbative expansion must eventually depart from the non-perturbative push-forward, and the two descriptions will diverge, or the opposite happens, and the summation over all terms actually produces the Landau-Lifshitz dynamics as a mean at high energies.

This raises a deeper question about the ultimate relationship between these two frameworks.
The S-matrix expansion in powers of $e$ in Eq.~\eqref{SmatrixGeneral}, from which the
truncation Eq.~\eqref{SmatrixExpansion} was obtained, is in general asymptotic rather than
convergent: beyond some optimal truncation order, adding further terms may not improve the approximation. The classical spectrum, on the other hand, admits a convergent
expansion in powers of $e$ --- a consequence of its non-perturbative, push-forward structure.
The agreement between the two models to any given finite order is therefore not guaranteed to
persist: it is a non-trivial statement about the structure of the QED perturbation series.
If the agreement does hold to arbitrary finite order, this would suggest that the complete
quantum extension of the Landau-Lifshitz model --- that is, the full QED description of
radiation reaction for a beam of electrons scattered by a coherent state --- may in principle
be recovered by resumming the Feynman diagram expansion to infinite order. Whether such a
resummation is possible, and whether it converges to the Landau-Lifshitz push-forward or to
a genuinely distinct quantum distribution in the ultrarelativistic limit, is a question of considerable physical interest that lies well beyond the scope of this dissertation, but which the framework developed here
places on a precise footing for future investigation.
\chapter{Conclusion}

The interaction of a relativistic electron with an intense electromagnetic field is one of the oldest and most persistently difficult problems in theoretical physics. It has resisted complete resolution for over a century not because the underlying physics is exotic or inaccessible, but because the problem sits precisely at the intersection of two of the most successful physical theories ever constructed --- classical electrodynamics and quantum field theory --- and demands that they be reconciled at the level of the electron's own field, not merely at the level of externalized interactions. Two distinct pathologies have obstructed progress, and tracing their histories reveals how deeply intertwined the classical and quantum aspects of the problem actually are.

The first pathology is classical. It begins with the apparent simplicity of the Lorentz force and the Larmor formula: an electron in an electromagnetic field is accelerated by the field and radiates energy proportional to the square of that acceleration. The difficulty is that the electron's own radiation field acts back on the electron, and accounting for this self-consistently within classical electrodynamics proved extraordinarily resistant to resolution. Lorentz and Abraham introduced the radiation reaction force~\cite{Lorentz:1892,Abraham:1903}; Dirac placed it on rigorous field-theoretic foundations~\cite{Dirac:1938}, exposing in the process the pathological runaway solutions and pre-acceleration of the Lorentz-Abraham-Dirac equation; Landau and Lifshitz proposed the self-consistent reduction that bears their name~\cite{Landau:1951}; and finally Gralla, Harte, and Wald demonstrated rigorously, through a careful limiting procedure applied to extended charged matter, that the Landau-Lifshitz equation is the correct classical equation of motion for a radiating charged particle~\cite{Gralla:2009}. This century-long program is now complete, and its culmination --- the Landau-Lifshitz equation and its exact analytic solution in a plane-wave background --- serves as the classical foundation of this dissertation.

The second pathology is quantum. It begins with Volkov's exact solution of the Dirac equation in a plane-wave background~\cite{Volkov:1935}, and the subsequent development by Furry, Ritus, and others of a perturbative QED organized around dressed electron states in the background field rather than free vacuum states~\cite{Furry:1951,Ritus:1975,Ritus:1985}. This Furry-picture framework is technically powerful --- the electron propagates as a Volkov state, an exact solution to the Dirac equation in the plane-wave background field, so the full effect of the background on the electron's dynamics is incorporated exactly into the propagator rather than treated as a sequence of perturbative interactions --- but it is also deeply restrictive: the desire for analytic solutions to the Dirac equation demands that the background field be a function of a single lightcone coordinate, it generates highly oscillatory phase integrals that resist analytic and computational evaluation for even idealized pulse profiles, and its practical implementation requires the locally constant field approximation whose validity is uncertain in precisely the experimental regimes where quantum radiation reaction is most interesting. A century of theoretical development within this framework has produced exact formal results for monochromatic backgrounds and a sophisticated approximation scheme for general pulses, but has not produced a model with genuine predictive accuracy for measured scattered spectra --- the actual observable of any inverse Compton scattering experiment.

This dissertation presents the resolution to the second pathology. The key insight is that the restrictive structure of the Furry picture --- its dependence on a fixed classical background field threading the Dirac propagator --- is not a necessary feature of QED but an artifact of importing the background field methods of quantum mechanics into a framework where the electromagnetic field is itself a quantum object. The field operators of quantum electrodynamics provide a more natural language: a laser field is not a classical background but a photonic coherent state, which is the precise QED realization of a classical electromagnetic field without any restriction on its functional form. Organizing the perturbation theory around this coherent state, rather than around a background-field-dressed vacuum, frees the calculation from the plane-wave constraint, eliminates the oscillatory phase integrals, and can seemlessly produce scattered spectra as a primary analytic output rather than a statistical aggregate of single-particle simulations. The framework simultaneously recovers Landau-Lifshitz dynamics for an electron traversing a plane-wave electromagnetic field as its classical limit confirming that the classical and quantum descriptions are not competing alternatives but a unified hierarchy, with Landau-Lifshitz governing the mean and quantum corrections appearing as perturbative fluctuations around it.

\section{Discussion}

The results of this dissertation invite reflection on questions that extend beyond the immediate technical contributions. At the heart of these contributions is a perspective shift that is both conceptually simple and consequential: standard QED is organized around particle-particle scattering --- the collision of quantized excitations of matter and radiation fields in the perturbative vacuum. The macroscopic QED framework
presented here is organized instead around the scattering of a charged particle by an electromagnetic field --- the laser pulse is not a collection of individual photons to
be exchanged one at a time but a macroscopic quantum state of the radiation field, a photonic coherent state, and the electron beam is not a single particle but a
statistical ensemble encoded in an electronic spectral state. This is not a technical distinction but a physical one: it is the natural language for the class of experiments in which a macroscopic classical field drives quantum dynamics in a charged particle beam, and it is precisely this language that the inverse Compton scattering experiments of the past decade have been demanding without yet having a theoretical framework to
meet them. The macroscopic QED framework is that framework.

Beyond this, the framework produces a question concerning the relationship between classical and quantum radiation reaction: the Landau-Lifshitz equation, long treated as a classical approximation requiring quantum
correction, may be the exact mean of quantum radiation reaction dynamics, with quantum corrections appearing as perturbative fluctuations around it rather than as corrections to it. This claim --- speculative but grounded in the macroscopic QED derivation of this dissertation --- has significant implications for the direction of the field. If this Landau-Lifshitz-as-mean hypothesis is correct, the current experimental program, which targets non-perturbative quantum effects at high field strengths and high electron energies, is searching in a regime where quantum stochasticity is most suppressed, and the correct observable is not a threshold effect like pair production but the spectral shape of the scattered electron energy distribution around the Landau-Lifshitz mean. The macroscopic QED framework provides the first model capable of predicting this spectral shape from first principles. The experimental program capable of measuring it has not yet been designed with this question clearly in view.

\subsection{Landau-Lifshitz Dynamics and the Mean of Quantum Radiation Reaction}

The Landau-Lifshitz equation occupies an unusual position in the landscape of radiation reaction models: it is universally acknowledged as the correct self-consistent classical equation of motion for a radiating charged particle, yet it is almost universally treated in the literature as an approximation that requires quantum correction. This dissertation argues that this treatment reflects a misunderstanding of what the Landau-Lifshitz equation is and what it predicts, and that the macroscopic QED framework developed here suggests a substantially different picture of the relationship between classical and quantum radiation reaction.

\paragraph{The energy conservation misconception.}
As established in Chapter~4, the most commonly cited motivation for quantum corrections to Landau-Lifshitz --- that the equation permits the electron to radiate more energy than it initially possesses --- is incorrect. The Landau-Lifshitz equation is a self-consistent statement of momentum conservation between the electron and its ambient electromagnetic field, and the exact solution for an arbitrary plane-wave pulse is manifestly positive for all physical initial conditions and field parameters. The claim is precisely backwards: it is the Lorentz force law, not the Landau-Lifshitz equation, that admits unphysical energy trajectories in a strong background field, as seen directly from Eq.~\eqref{ElectronVelocityPlanewaveLorentz}.

\paragraph{The linear picture and its breakdown.}
The linear picture of inverse Compton scattering treats the interaction as the elastic scattering of a point electron off individual laser photons --- a particle-particle
process in which each photon exchange is independent. In this picture, the field strength determines the frequency of absorptions, and, thus, emissions, while the scattered photon energy primarily depends on the incident electron energy. This picture breaks down in the nonlinear radiation reaction regime because the electron is decelerating by emitting radiation, which feeds back into every subsequent emission event. Each emission reduces the electron's energy and momentum, and this changes the kinematics of all subsequent interactions with the field. Simultaneously, the laser field is depleted by the absorbed photons. The self-consistent coupling between the electron's evolving momentum and its subsequent emission dynamics is precisely what gives the Landau-Lifshitz equation its radiation reaction character. It is also what suggests convergence toward a well-defined mean even under stochastic quantum emissions: the damping of electron motion and the depletion of the field act as stabilizing mechanisms that drive the interaction toward a determinate outcome regardless of the stochastic character of individual emission events.

This breakdown of the linear particle-particle picture and the inadequacy of the standard QFT particle-particle framework are in fact the same problem. Both fail because
the experiment involves a macroscopic coherent field interacting with a charged particle beam rather than a sequence of individual photon exchanges between point particles. Macroscopic QED, which takes the coherent state as its organizing principle and never decomposes the
field into individual photons, is the natural and correct setting for this experiment.

\paragraph{The Landau-Lifshitz-as-mean hypothesis.}
The full Landau-Lifshitz solution given in Eq.~(\ref{ElectronVelocityPlanewaveLandauLifshitz}) effectively captures contributions corresponding to processes
at all orders in perturbative QED, making it non-perturbative. Meanwhile, the macroscopic QED calculation of Chapter~5 recovers the Landau-Lifshitz dynamics
exactly to order $e^4$, under the narrow-band and ultrarelativistic approximations. There, the classical limit of the macroscopic model to order $e^4$ is forced by the ultrarelativistic limit, reproducing Landau-Lifshitz dynamics to that order. That is, at least to $e^{4}$, in the ultrarelativistic regime for scattering by a Gaussian pulse, CED and QED make the same prediction concerning electron dynamics. This suggests the possibility that the Landau-Lifshitz dynamics are not a classical approximation to the quantum dynamics but the exact mean of those dynamics, with quantum corrections appearing as perturbative fluctuations around it, and at high electron energies, the ratio of individual electron recoil per emission event to the electron momentum is suppressed, and the energy loss due to discrete stochastic quantum emissions is indistinguishable from the loss due to continuous emissions. The self-consistent damping and field depletion of the nonlinear radiation reaction regime support this conjecture, suggesting that the stochastic recoil events due to quantum emissions, whatever their individual character, aggregate toward the smooth classical Landau-Lifshitz mean.

This conjecture must be stated with precision, however. It is scoped specifically to the scattered electron energy spectrum, because the Landau-Lifshitz equation is an equation of motion for the electron and makes no direct claim about the emitted photons or any other aspect of the interaction. Moreover, whether the hypothesis holds in general cannot be determined from the present calculation alone: the classical limit was derived under the narrow-band approximation for the Gaussian pulse and then the ultrarelativistic limit, and it is not clear whether the result is a property of those approximations specifically, of the coherent state photon statistics, or of the electron dynamics in greater generality. Critically, the hypothesis cannot be confirmed or refuted within the LCFA framework used in all existing QED simulations. As established in Chapter~4, the LCFA as implemented makes at least five simultaneous uncontrolled approximations: local constancy of the field over the photon formation length, restriction to single emission matrix elements, independence of successive emission events, single-particle dynamics with no inter-electron correlations, and the assumption that the emitted photon energy is completely independent of the electron dynamics between emission events. The Landau-Lifshitz-as-mean hypothesis is a statement about the full quantum theory, not about the LCFA approximation to it, and investigating it rigorously requires going beyond all five of these approximations simultaneously --- precisely the direction in which the macroscopic QED framework points.

This hypothesis is speculative, and the dissertation does not claim to establish it. It can be investigated in two ways. The first is theoretical: extending the macroscopic QED calculation from chapter~5 non-perturbatively to determine whether the Landau-Lifshitz dynamics continue to emerge as the mean, or whether quantum corrections shift the mean itself. The second is experimental: high-energy strong-field experiments with sufficient spectral resolution to determine whether the quantum fluctuations are suppressed or observable. If they are not suppressed, the hypothesis is falsified; if they are, it gains experimental support.

\paragraph{Where to look for quantum radiation reaction.}
If Landau-Lifshitz dynamics describe the mean of the quantum scattered electron energy distribution, quantum corrections to radiation reaction manifest as fluctuations around the Landau-Lifshitz spectral pushforward rather than as deviations of the mean from it. The electron spectrum shows the aggregate effects of all emission events throughout the interaction; the photon spectrum shows the individual events. This distinction matters because photons, unlike electrons, are exclusively quantum objects --- a classical electromagnetic field has no particle interpretation without information loss, and each individual photon emission is an irreducibly quantum event.

The appropriate observable for quantum signatures of radiation reaction is therefore the emitted photon spectrum, not the electron spectrum, and the correct regime is one where individual emission events are comparable in size to the total electron momentum --- high field strength at low electron energy --- rather than the high-energy regime where the current experimental program is focused. The current strategy of maximizing $\chi$ at high electron energies may be searching in a regime where the electron spectrum is most dominated by the classical Landau-Lifshitz mean. A future experiment optimized for spectral resolution in the emitted photon distribution at moderate electron energy, with well-characterized field strength and high collision efficiency, would be better positioned to observe the genuinely quantum character of individual radiation reaction events and to test the Landau-Lifshitz-as-mean hypothesis directly.

\section{Future Work}

The macroscopic QED framework developed in this dissertation is a proof of concept --- a demonstration that the correct language for inverse Compton scattering by a macroscopic electromagnetic field is the language of quantum states of the field rather than background field methods, and that this language yields analytic, first-principles spectral predictions without recourse to the LCFA or large-scale particle simulation. As a proof of concept it is necessarily limited in scope, and several natural and important directions extend from it.

\subsection{Extensions of the Macroscopic QED Model}

The first and most natural extension is to exploit the fact that coherent-states need only be square-integrable, and extend the calculation of Chater~5 to more general pulse profiles beyond the plane-wave Gaussian. Applying it to more realistic pulse models --- pulses with asymmetric envelopes, chirp, or, more generally, as in Ref.~\cite{Siegman:1986} --- and to flying-focus field configurations would extend the analytic spectral predictions to the full range of experimental geometries currently being pursued. The flying-focus configuration is particularly important: the spatiotemporal coupling of the flying-focus field breaks the plane-wave structure that all existing models require, and coherent-states in macroscopic QED are the only current model capable of accommodating it without additional approximation.

Secondly, the calculations of Chapter~5 were carried out to order $e^4$ in perturbation theory, which captures the leading Compton scattering contribution and is sufficient to recover the Landau-Lifshitz equation as the classical limit. The natural next step is extension to order $e^6$, at which new processes appear: double Compton scattering, in which the electron emits two photons, and radiative corrections to single Compton scattering from virtual loops. These contributions would introduce the first genuinely quantum spectral corrections to the spectral pushforward model --- the stochastic broadening and asymmetry predicted by the Landau-Lifshitz-as-mean hypothesis --- and computing them within the coherent-state framework would provide the first analytic prediction of what quantum radiation reaction fluctuations actually look like in the scattered electron energy spectrum. Whether Landau-Lifshitz dynamics continue to emerge as the mean at this order, or whether quantum corrections shift the mean itself, is a question the framework is uniquely positioned to answer.

A third and more ambitious direction concerns the non-perturbative structure of the coherent-state series. The Landau-Lifshitz solution of Eq.~\eqref{ElectronVelocityPlanewaveLandauLifshitz} is non-perturbative --- it captures contributions from all orders in $e$ simultaneously --- while the macroscopic QED calculation is organized perturbatively. The agreement between the two at order $e^4$ in the ultrarelativistic limit raises the question of whether the coherent-state series, summed to all orders, would recover a similar expression to the Landau-Lifshitz spectral pushforward exactly in this regime. If so, Landau-Lifshitz dynamics would be not merely consistent with QED to low orders but a complete resummation of the coherent-state perturbative series --- a remarkable result
connecting the non-perturbative classical dynamics to the full quantum field theory. Whether such a resummation is possible, and whether it converges to the Landau-Lifshitz result or to a genuinely distinct quantum distribution, is a deep question that the framework places on a precise mathematical footing for the first time.

\subsection{The Electron Ordering Problem}
The most fundamental open problem identified in this dissertation is the electron ordering problem: the inability of any current model and simulation to account for the correlation between each electron's longitudinal position in the bunch and the field strength it encounters as the bunch traverses a decaying laser pulse. As established in Chapter~4, this correlation cannot be recovered from the marginal energy spectrum alone, and correctly modeling it would require knowledge of the joint longitudinal-energy distribution of the bunch, which is intractable with current diagnostical tools.

The strong energy-position correlation of a LWFA beam, however, means the ordering is
not random but determined by the beam physics, and this opens a path toward better
models. Recent experimental work is already converging on this conclusion: Los and Di
Piazza~\cite{Los:2026} found it necessary to reconstruct the incident
electron energy spectrum for each shot from the non-interacting reference shots via a
neural network trained on the unscattered electron spectra, precisely because the full phase-space description of the incident beam could not be directly measured. While sophisticated, this approach remains constrained by the marginal spectrum assumption --- the neural network reconstructs the energy distribution but cannot recover the longitudinal ordering of electrons through the decaying field, for the reasons established in Chapter~4. A particle-in-cell simulation of the LWFA acceleration stage --- which produces the full phase-space distribution of the beam as a natural output of the plasma dynamics --- could provide a physically motivated beam model that replaces the assumed bunch distribution with a realistic phase-space input. Feeding this into the coherent-state scattering framework would substantially reduce the dominant source of uncertainty in the comparison with experimental data, and would make the macroscopic QED model a genuinely predictive tool rather than a benchmark. This points toward a coupled LWFA-ICS experimental and theoretical program in which beam diagnostics, acceleration-stage simulations, and scattering calculations are designed together from the outset with the ordering problem explicitly in view --- the macroscopic QED framework is the natural theoretical companion for such a program. Developing this coupled LWFA-ICS pipeline is a near-term priority of the future research program.

Beyond this near-term step, the ultimate theoretical vision would dispense with the
intermediate simulation pipeline entirely. The power of QED as a physical framework is
that it does not require detailed knowledge of intermediate states --- only the
initial and final states matter. Simultaneously, the most physically honest
model of an experiment is one that takes the actual experimental controls as its inputs
and produces the observable as its output without intermediate modeling choices. Taking
both principles together, the most complete model of a LWFA-ICS experiment would take
the gas in the gas cell and the driver and scattering lasers as inputs, and output
the scattered electron or radiation spectrum as a single expression within one unified
framework. That framework is macroscopic QED, of which the coherent-state model
developed in this dissertation is one component. The second component entails a
genuinely macroscopic quantum description of charged matter --- possibly a classical
realization complementary to the role the coherent state plays for the laser field, including the ambient electromagnetic field --- embedded as a quantum state within the interaction picture of QED. While certainly ambitious, this unified description would be conceptually simpler than all current approaches that force separate laser-plasma and ICS models to interface at an intermediate stage, because it would require no such interface at all, and more importantly because it would the direct realization of scattering an electron beam by a laser pulse within macroscopic QED. 

\subsection{Macroscopic Charged Matter within QED}
The coupled LWFA-ICS pipeline proposed in the preceding subsection is a near-term
practical step, but the deeper resolution of the ordering problem requires confronting two fundamental obstacles to incorporating a generic model of macroscopic charged matter into QED: the Pauli exclusion principle and the Dirac field itself. The photonic coherent state exists because bosons can occupy the same mode in arbitrarily large numbers, allowing a macroscopic field configuration to be realized as a single quantum state. Fermions cannot: the antisymmetry of the fermionic Fock space enforces the statistical individuality of electrons at the microscopic scale, and no fermionic coherent state in the bosonic sense can be constructed. The Dirac field compounds this not because of any deficiency as a microscopic theory of electrons, but because it is, by construction, a specific model of charged matter whose fundamental excitations are bare point particles. As discussed in Chapter~3, its macroscopic states are spectral states --- Fock space realizations of statistical ensembles of individually sampled bare point particles rather than macroscopic field configurations. This is precisely what makes the ordering problem intractable: the particle decomposition is built into the theoretical framework from the outset, and no amount of refinement within that framework can produce a description of the beam that does not ultimately resolve into individual electrons.

Any attempt to construct a macroscopic fermionic quantum state that avoids this
decomposition must confront both obstacles directly, while recovering the Dirac field
as an effective model in the appropriate limit. Whether a consistent macroscopic quantum
description of charged matter can be constructed within QFT, and what mathematical
structure it would require, is an open question this dissertation identifies but does
not resolve. It is arguably the central open problem in the broader theoretical program
initiated in Chapter~2 and complemented by the coherent-state framework of Chapter~3.

More fundamentally, this theoretical consideration brings the dissertation full circle to the generic charged matter framework of Chapter~2, where the general electrodynamic action was formulated without specifying a model of matter. That formulation in CED led, via the Gralla-Harte-Wald limiting procedure, to a rigorous derivation of classical radiation reaction and a firm physical grounding of the Landau-Lifshitz equation as an effective model of electron dynamics. This resolved the classical self-energy paradox by treating matter as a macroscopic object deformable to a point rather than assuming a point particle from the outset, and separating near-field electromagnetic structure from that of the far-field, which is the concrete classical implementation of the Wilsonian interpretation of a point particle. It is, therefore, reasonable to expect that following the analogous path in pre-QED and QED --- formulating a quantum theory of electrodynamics around a generic macroscopic description of charged matter rather than around the Dirac field --- would similarly ground the quantum theory, potentially resolving the UV and IR divergences that necessitate regularization and renormalization as artifacts of the premature point particle assumption rather than fundamental features of the theory.\footnote{This is speculative, but the parallel is precise: renormalization in QED is a post-hoc resolution of the quantum self-energy problem similar to how the Lorentz-Abraham-Dirac and later Landau-Lifshitz equations are a post-hoc resolution of the classical self-energy problem. The Galla-Harte-Wald program resolved the classical pathology providing a grounded motivation for Dirac's renormalization procedure. The analogous quantum program has not yet been attempted.}

\backmatter

\printbibliography[heading=bibintoc, title={References}]

\end{document}